\documentclass[twocolumn,10pt,english,aps,prc,showpacs,letter,longbibliography]{revtex4-1}
\usepackage{amsmath}
\usepackage[    bookmarks,
                 bookmarksopen = true,
                 bookmarksnumbered = true,
                 linktocpage,
                 colorlinks = true,
                 linkcolor = blue,
                 urlcolor  = blue,
                 citecolor = blue,
                 anchorcolor = green,
                 hyperindex = true,
                 hyperfigures]
        {hyperref}

\usepackage{enumitem}
\usepackage{graphicx}
\usepackage{esint}
\usepackage[utf8]{inputenc}
\usepackage[T1]{fontenc}

\usepackage{listings}

\DeclareFixedFont{\ttb}{T1}{txtt}{bx}{n}{8} 
\DeclareFixedFont{\ttm}{T1}{txtt}{m}{n}{8}  

\usepackage{color}
\definecolor{deepblue}{rgb}{0,0,0.5}
\definecolor{deepred}{rgb}{0.6,0,0}
\definecolor{deepgreen}{rgb}{0,0.5,0}

\newcommand\pythonstyle{\lstset{
language=Python,
basicstyle=\ttm,
otherkeywords={self},             
keywordstyle=\ttb\color{deepblue},
emph={MyClass,__init__},          
emphstyle=\ttb\color{deepred},    
stringstyle=\color{deepgreen},
frame=tb,                         
showstringspaces=false            %
}}

\lstnewenvironment{python}[1][]
{
\pythonstyle
\lstset{#1}
}
{}

\makeatletter

\usepackage{babel}

\makeatother

\begin{document}

\preprint{This line only printed with preprint option}


\title{Pseudorapidity distribution and decorrelation of anisotropic flow within CLVisc hydrodynamics}

\author{Long-Gang Pang$^{1,2,3,4}$, Hannah Petersen$^{4,5,6}$, Xin-Nian Wang$^{1,2,3}$}

\address{$^{1}$Key Laboratory of Quark \& Lepton Physics (MOE) and Institute of Particle Physics, Central China Normal University, Wuhan 430079, China}

\address{$^{2}$Physics Department, University of California, Berkeley, CA 94720, USA}

\address{$^{3}$Nuclear Science Division, Lawrence Berkeley National Laboratory, Berkeley, CA 94720, USA}

\address{$^{4}$Frankfurt Institute for Advanced Studies, Ruth-Moufang-Strasse
1, 60438 Frankfurt am Main, Germany}

\address{$^{5}$Institute for Theoretical Physics, Goethe University, Max-von-Laue-Strasse
1, 60438 Frankfurt am Main, Germany}

\address{$^{6}$GSI Helmholtzzentrum f\"ur Schwerionenforschung, Planckstr.
1, 64291 Darmstadt, Germany}

\begin{abstract}
Studies of fluctuations and correlations of soft hadrons and hard and electromagnetic probes of the dense and strongly interacting medium require event-by-event hydrodynamic simulations of high-energy heavy-ion collisions that are computing intensive. We develop a (3+1)D viscous hydrodynamic model -- CLVisc that is parallelized on Graphics Processing Unit (GPU) using Open Computing Language (OpenCL) with 60 times performance increase for space-time evolution and more than 120 times for the Cooper-Frye particlization relative to that without GPU parallelization. The model is validated with comparisons with different analytic solutions, other existing numerical solutions of hydrodynamics and experimental data on hadron spectra in high-energy heavy-ion collisions. The pseudo-rapidity dependence of anisotropic flow $v_n(\eta)$ are then computed in CLVisc with initial conditions given by the A Multi-Phase Transport (AMPT) model,  with energy density fluctuations both in the transverse plane and along the longitudinal direction. Although the magnitude of $v_n(\eta)$ and the ratios between $v_2(\eta)$ and $v_3(\eta)$ are sensitive to the effective shear viscosity over entropy density ratio $\eta_v/s$, the shape of the $v_{n}(\eta)$ distributions in $\eta$ do not depend on the value of $\eta_v/s$.
The decorrelation of $v_n$  along the pseudo-rapidity direction due to the twist and fluctuation of the event-planes in the initial parton density distributions is also studied. The decorrelation observable $r_n(\eta^a, \eta^b)$ between $v_n\{-\eta^a\}$ and $v_n\{\eta^a\}$ with the auxiliary reference window $\eta^b$ is found not sensitive to $\eta_v/s$ when there is no initial fluid velocity.  For small $\eta_v/s$, the initial fluid velocity from mini-jet partons introduces sizable splitting of $r_n(\eta^a, \eta^b)$ between the two reference rapidity windows $\eta^b \in [3, 4]$ and $\eta^b \in [4.4, 5.0]$, as has been observed in experiment.
 The implementation of CLVisc and guidelines on how to efficiently parallelize scientific programs on GPUs are also provided.
\end{abstract}



\maketitle

\section{Introduction}

Heavy-ion collisions at the Relativistic Heavy-Ion Collider (RHIC) and Large Hadron Collider (LHC) create strongly coupled QCD matter that exhibits multiple extreme properties. It is the hottest -- temperature reaching more than 100,000 times that at the core of the Sun, most vortical -- angular momentum on the order of $10^3-10^5\hbar$ \cite{STAR:2017ckg} and almost perfect fluid -- very low shear viscosity over entropy density ratio \cite{Romatschke:2007mq,Song:2007fn,Song:2010mg}, that is exposed to the strongest magnetic field ($|\mathbf{B}|=5\sim 10\ m_{\pi}^2$) \cite{Kharzeev:2007jp} ever produced in laboratory. This strongly coupled QCD matter is believed to share some of the properties of the quark-gluon-plasma epoch in the early universe.

Numerical simulations of the dynamical evolution of this strongly coupled QCD matter and comparisons with experimental data are vital to extract the physical properties of the strong interaction matter. Relativistic viscous hydrodynamics is the most successful effective theory in describing the space-time evolution of QCD matter created in high-energy heavy-ion collisions \cite{Gale:2013da,Molnar:2014zha}. Hybrid approaches that comprise hydrodynamics and hadronic transport agree with experimental data on various observables such as charged multiplicity, transverse momentum spectra and transverse momentum $p_T$-differential elliptic flow of identical particles \cite{Petersen:2014yqa} (and references therein). Event-by-event simulations with energy density fluctuations \cite{Alver:2008zza,Alver:2010gr,Teaney:2010vd,Schenke:2010rr,Qiu:2011iv,Schenke:2012wb,Holopainen:2010gz,Qin:2010pf,Schenke:2012fw,Werner:2010aa} in the initial states are indispensable to describe not only the ensemble average of odd-order harmonic flows but also their probability distributions \cite{Gale:2012rq}. New observables such as the correlation between different event plane angles \cite{Qiu:2012uy,Teaney:2013dta,Aad:2014fla,Niemi:2015qia}, different harmonic flows \cite{ALICE:2016kpq} and $p_T$-differential harmonic flows \cite{Qian:2017ier} can provide more rigorous constraints on medium properties such as the shear viscosity to entropy density ratio, but also require efficient algorithms to reach sufficient statistics in a reasonable amount of CPU time.
Furthermore, (3+1)D event-by-event hydrodynamics is also necessary to understand the longitudinal structure of the collective flow. The initial state fluctuations along the longitudinal direction have been built in many models \cite{Petersen:2011fp,Cheng:2011hz,Xiao:2012uw,Pang:2014pxa,Adil:2005bb,Adil:2005qn,Bozek:2010vz,Dumitru:2011vk}. Observables \cite{Borghini:2002hm,Bzdak:2012tp,ATLAS:2015kla,Monnai:2015sca,Bozek:2015tca,Huo:2013qma,Jia:2014ysa,Csernai:2014cwa,Khachatryan:2015oea,Bozek:2017qir,Aaboud:2017tql} have been designed to either constrain the longitudinal structure in the initial state or to determine other QGP properties using the multiplicity or anisotropic flow correlations along the longitudinal direction. Taking into account the asymmetry between forward and backward going participants, the non-central heavy-ion collisions not only produce strong angular momentum, strong magnetic field but also global and local vorticity \cite{Kharzeev:2007jp} and hyperon polarization \cite{Liang:2004ph}.

The space-time evolution of high-energy heavy-ion collisions from event-by-event relativistic hydrodynamics also provide critical background information for thermal photon, di-lepton emission, heavy flavor transport and jet energy loss studies when they are produced in or traverse the fluctuating hot and dense medium. For studies of thermal photon and di-lepton production \cite{Liu:2011dk,Xu:2014ada,Shen:2013vja}, the emission rates are computed with the local temperature and fluid velocity at each space-time point from event-by-event (3+1)D viscous hydrodynamics, which is quite computing intensive. In the simultaneous simulations of parton shower propagation and bulk medium evolution, the bottle neck in the numerical simulations is also the relativistic hydrodynamic evolution of the medium in each time step of the parton shower propagation as shown in CoLBT-Hydro \cite{Chen:2017zte} and the forthcoming JetScape \cite{Cao:2017zih}.
Big data analyses in relativistic heavy-ion collisions using machine learning \cite{Pratt:2015zsa,Bernhard:2015hxa,Bernhard:2016tnd} and deep learning techniques \cite{Pang:2016vdc} demand huge amount of data from event-by-event hydrodynamic simulations with up to $O(10^7)$ events across a high dimensional parameter space. These studies will all benefit from a fast numerical solver for the (3+1)D relativistic hydrodynamics.

In order to reduce the running time of one single simulation, Message Passing Interface (MPI) library has been used in MUSIC \cite{Schenke:2010nt,Schenke:2010rr,Paquet:2015lta} to parallelize the (3+1)D viscous hydrodynamic program by communicating between multiple CPUs. The communication costs between CPUs on different nodes are usually heavy comparing to the workload of the numerical computations. On the other hand, a Graphics Processing Unit (GPU) has a huge amount of processing elements (>2500) on one single computing device, which makes it quite popular to accelerate numerical computations via massive parallelization. The SHASTA algorithm is first parallelized on heterogeneous devices using OpenCL to simulate the QGP expansion by solving the (3+1)D ideal hydrodynamic equations \cite{Gerhard:2012uf}. The (3+1)D viscous hydrodynamics for simulations of heavy-ion collisions has been parallelized on GPU using both OpenCL (CLVisc \cite{Pang:2014ipa}) and Cuda (GPU-VH \cite{Bazow:2016yra}).
In this paper and its appendix, we provide a detailed description of the parallelization of hydrodynamic evolution, hyper-surface finding and spectra calculation in CLVisc hydrodynamic model. OpenCL has the benefit that the same code can run on heterogeneous computing devices (CPUs, GPUs, FPGAs and Intel Phi). However, the basic concepts and optimization principles are the same for both OpenCL and Cuda. The acronym CLVisc refers to both CCNU (Central China Normal University) and LBNL (Lawrence Berkeley National Laboratory) viscous hydrodynamic model and OpenCL GPU parallelization that is used.

After providing validations of CLVisc through comparisons with several analytic solutions to the viscous hydrodynamics and experimental data on bulk hadron spectra in high-energy heavy-ion collisions, we apply the CLVisc to the study of pseudo-rapidity distribution and fluctuation of anisotropic flow with event-by-event initial conditions from
A Multi-Phase Transport (AMPT) model  \cite{Lin:2004en}. We compute the pseudo-rapidity dependence of the anisotropic flows $v_n(\eta)$ and $r_n(\eta^a, \eta^b)$ which represents the de-correlation between $v_n\{-\eta^a\}$ and $v_n\{\eta^a\}$ with the auxiliary reference window $\eta^b$. Effects of shear viscosity and initial fluid velocity on these longitudinal observables are also investigated for the first time with CLVisc. 

This paper is organized as follows: in Sec. \ref{sec:hydro_eq}, we rewrite the hydrodynamic equations in a specific way to simplify the numerical implementation. In Sec. \ref{sec:numerical_detail}, we describe in detail how the relativistic hydrodynamic equations are solved numerically in CLVisc with GPU parallelization. In Sec. \ref{sec:particalization}, we introduce the GPU parallelized smooth particle spectra calculation and the fast Monte-Carlo sampler to sample four-momenta of particles from freeze-out hyper-surface. 
In Sec. \ref{sec:code_varification}, we verify our numerical code with a variety of analytical solutions and numerical results from other implementations. 
Comparisons with experimental data on hadron spectra and anisotropic flow are given in Sec.~\ref{sec:data}. 
In Secs.~\ref{sec:vneta} and \ref{sec:rneta} we discuss the pseudo-rapidity distribution, correlation and fluctuation of anisotropic flow.
In the Appendix, we provide a detailed description of the structure and GPU parallelization of the  algorithm to solve the hydrodynamics equations, two methods to sample Juttner, Fermi-Dirac and Bose-Einstein distributions efficiently and assess the performance of GPU parallelization.

\section{Hydrodynamic equations}
\label{sec:hydro_eq}
Let us start by recapitulating the exact form of the relativistic hydrodynamic equations that are solved within CLVisc. 
The second-order hydrodynamic equations are simply given by

\begin{eqnarray}
    \nabla_{\mu} T^{\mu\nu} &=&  0, \\
    \nabla_{\mu} N^{\mu} &=&  0, \label{eqn:Nmu} 
\end{eqnarray}
with the energy-momentum tensor $T^{\mu\nu}=\varepsilon u^{\mu}u^{\nu}
-(p+\Pi) \Delta^{\mu\nu} + \pi^{\mu\nu}$, where $\varepsilon$ is the
energy density, $p$ the pressure, $u^{\mu}$ the fluid four-velocity
normalized as $u^{\mu}u_{\mu}=1$ and $\Delta^{\mu\nu}=g^{\mu\nu}-u^{\mu}u^{\nu}$ the projection operator which is orthogonal to the fluid velocity,
and the net charge current $N^{\mu}=nu^{\mu}+d^{\mu}$ where 
$d^{\mu}$ is the charge diffusion current. The shear stress tensor
$\pi^{\mu\nu}$ and the bulk pressure $\Pi$ represent the deviation from
ideal hydrodynamics and local equilibrium. We choose to work in Landau
frame, which yields the traceless ($\pi_{\mu}^{\mu}=0$) and transverse
$(u_{\mu}\pi^{\mu\nu}=0)$ shear stress tensor. By projecting along the
fluid velocity $u^{\mu}$ direction, we simply get $u_{\mu}T^{\mu\nu}=\varepsilon u^{\nu}$.

The bulk pressure $\Pi$ and the shear stress tensor $\pi^{\mu\nu}$ satisfy the equations \cite{Baier:2007ix},
\begin{eqnarray}
    \Pi &=&  -\zeta \theta  - \tau_{\Pi} \left[u^{\lambda} \nabla_{\lambda} \Pi + \frac{4}{3}\Pi\theta \right] \label{eqn:Pi} \\
    \pi^{\mu\nu} &=&  \eta_{v} \sigma^{\mu\nu} - \tau_{\pi} \left[ \Delta_{\alpha}^{\mu}\Delta_{\beta}^{\nu} u^{\lambda} \nabla_{\lambda} \pi^{\alpha\beta} + \frac{4}{3}\pi^{\mu\nu} \theta \right] \nonumber \\
    & & - \lambda_1 \pi^{\langle \mu}_{\lambda} \pi^{\nu\rangle\lambda} - \lambda_2 \pi^{\langle \mu}_{\lambda} \Omega^{\nu\rangle\lambda}- \lambda_3 \Omega^{\langle \mu}_{\lambda} \Omega^{\nu\rangle\lambda}, \label{eqn:pimn}
\end{eqnarray}
with the expansion rate $\theta$, symmetric shear tensor $\sigma^{\mu\nu}$ and the antisymmetric vorticity tensor $\Omega^{\mu\nu}$ defined as
\begin{eqnarray}
    \theta & \equiv & \nabla_{\mu} u^{\mu}, \nonumber \\
    \sigma^{\mu\nu} & \equiv & 2 \nabla^{\langle \mu}u^{\nu\rangle} \equiv2 \Delta^{\mu\nu\alpha\beta} \nabla_{\alpha} u_{\beta}, \nonumber \\
    \Omega^{\mu\nu} & \equiv & \frac{1}{2}\Delta^{\mu\alpha} \Delta^{\nu\beta} (\nabla_{\alpha} u_{\beta} - \nabla_{\beta} u_{\alpha}), \nonumber \\
    \Delta^{\mu\nu\alpha\beta} & \equiv & \frac{1}{2}(\Delta^{\mu\alpha}\Delta^{\nu\beta} + \Delta^{\mu\beta} \Delta^{\nu\alpha}) - \frac{1}{3}\Delta^{\mu\nu} \Delta^{\alpha\beta},
    \label{eqn:projector}
\end{eqnarray}
where $\Delta^{\mu\nu\alpha\beta}$ is the double projection operator that makes the resulting contracted tensor symmetric, traceless and orthogonal to the fluid velocity $u^{\mu}$.
In Eqs. (\ref{eqn:Pi}) and (\ref{eqn:pimn}), the $\tau_{\Pi}, \tau_{\pi}, \lambda_1, \lambda_2, \lambda_3$ are five independent second-order transport coefficients.
Nonzero relaxation times $\tau_{\Pi}$ and $\tau_{\pi}$ in the second-order
Israel-Stewart (IS) equations solve the causality problem of the first-order
Navier-Stokes equations.
In the current calculation we set $\tau_{\pi}=5\eta_v/(Ts)$ \cite{Song:2008si} and $\tau_{\Pi}=5\zeta/(Ts)$, where $T$ is the temperature,
$s$ the entropy density, $\eta_v$ the shear viscous coefficient,
and $\zeta$ the bulk viscous coefficient.

The time-like fluid four-velocity in Cartesian coordinates
$x^{\mu}=(t, x, y, z)$ is defined as,
\begin{equation}
    u^{,\mu} \equiv \frac{dx^{\mu}}{d\sigma} \equiv u^{0}(1, v_{x}^{,}, v_y^{,}, v_z^{,})
    \label{eq:umu_cartesian}
\end{equation}
where $\sigma=\sqrt{t^2 - x^2 - y^2 - z^2}$ and spatial components of the
fluid velocity are defined as $v_i^{,} = u^{,i}/u^{0}$.
We work in Milne coordinates $X^{\mu} = (\tau, x, y, \eta_s)$, in which $\tau=\sqrt{t^2 - z^2}$ is the proper time and
$\eta_s = \frac{1}{2}\ln \frac{t+z}{t-z}$ the space-time rapidity.
The fluid four-velocity in these coordinates is,
\begin{eqnarray}
    u^{\mu} &\equiv& \frac{dX^{\mu}}{d\sigma} = \frac{dX^{\mu}}{dx^{\nu}}
    \frac{dx^{\nu}}{d\sigma} = \frac{dX^{\mu}}{dx^{\nu}} u^{,\nu} \nonumber\\
    & = & \left(
    \begin{array}{l}
        u^{0}\cosh \eta_s - u^{,z}\sinh \eta_s \\
        \vec{u}_{\perp}^{,} \\
         \frac{1}{\tau}(-u^{0}\sinh \eta_s + u^{,z}\cosh \eta_s)
    \end{array}
    \right) \equiv u^{\tau} \left(
    \begin{array}{l}
         1 \\
          \vec{v}_{\perp} \\
          \frac{v_{\eta_s}}{\tau}
    \end{array}
    \right)
    \label{eqn:umu}
\end{eqnarray}
where $v_{\perp}$ and $v_{\eta_s}$ are defined as,
\begin{eqnarray}
    \vec{v}_{\perp} &=&  \vec{v}_{\perp}^{,} \cosh(y_v)/\cosh(y_v - \eta_s) \\
    v_{\eta_s} &=&  \tanh(y_v - \eta_s)
    \label{eqn:veta}
\end{eqnarray}
and $y_v$ denotes the rapidity of the longitudinal fluid velocity
as given by $v_z^{,}=\tanh y_v$, $u^{\tau}=1/\sqrt{1-v_{\perp}^2-v_{\eta_s}^2}$ and $u^{\eta_s}=u^{\tau} v_{\eta_s}/\tau$.
In the Bjorken scaling scenario where the energy density is uniform along $\eta_s$
direction, we simply get $v_{\eta_s}=0$ and $y_v = \eta_s$, which implies
$v_z = z/t$. In full 3D expansion, $v_{\eta_s}$ denotes the relative
fluid velocity at coordinate $\mathrm{(t, x, y, z)}$,
in a reference frame which is moving at the speed of $v_z = z/t$.

From the invariant line element 
$ ds^2 = g_{\mu\nu} dX^{\mu}dX^{\nu} = d\tau^2 - dx^2 - dy^2 - \tau^2 d\eta_s^2 $
we get the metric tensor in Milne coordinates,
\begin{eqnarray}
    g_{\mu\nu} &=& \mathrm{diag}(1, -1, -1, -\tau^2) \\
    g^{\mu\nu} &=& \mathrm{diag}(1, -1, -1, -1/\tau^2)
\end{eqnarray}
The Christoffel symbols are explicitly solved as a function of the metric tensor,
$\Gamma_{kl}^{i} = \frac{1}{2} g^{im}(\partial_l g_{mk} + \partial_k g_{ml} - \partial_m g_{kl})$,
and contain three nonzero components,
\begin{eqnarray}
    \Gamma_{\eta_s\eta_s}^{\tau} &=&  \tau, \quad\;
    \Gamma_{\tau\eta_s}^{\eta_s} = \Gamma_{\eta_s\tau}^{\eta_s} = 1/\tau,
    \label{eq:christoffel}
\end{eqnarray}
which are used in the covariant derivative operation $\nabla^{\mu}$
for all vectors and tensors in the hydrodynamics equations
and IS equations,
\begin{eqnarray}
    \nabla_{b} \lambda^{a} &\equiv& \partial_b \lambda^a + \Gamma_{bc}^a \lambda^c \\
    \nabla_{c} \lambda^{ab} &\equiv& \partial_c \lambda^{ab} + \Gamma_{cd}^a \lambda^{db} + \Gamma_{cd}^b \lambda^{ad}
    \label{eq:covariant}
\end{eqnarray}
For example, there are $3$ terms in $\nabla_{\mu}u^{\nu}$ which are 
different from their ordinary derivatives,
\begin{eqnarray}
\nabla_{\tau} u^{\eta_s} &=&  \partial_{\tau} u^{\eta_s} + \frac{1}{\tau}u^{\eta_s},\\
\nabla_{\eta_s} u^{\tau} &=&  \partial_{\eta_s} u^{\tau} + \tau u^{\eta_s}, \\
\nabla_{\eta_s} u^{\eta_s} &=&  \partial_{\eta_s} u^{\eta_s} + \frac{1}{\tau} u^{\tau}, \label{eq:cov_dmun}
\end{eqnarray}

The $\partial_{\tau} \lambda + \lambda/\tau$ terms from covariant
derivatives are combined as $\frac{1}{\tau}\partial_{\tau} (\tau \lambda)$,
to reduce the numerical error when $\tau$ is small.
The new independent variables are thus defined as $\tilde{\lambda}=\tau \lambda$.
In this way, we define $\tilde{T}^{\mu\nu}, \tilde{N}^{\mu}, \tilde{\pi}^{\mu\nu},\tilde{u}^{\mu}, \tilde{\partial}_{\mu}$ and  $\tilde{g}^{\mu\nu}$
as,
\begin{eqnarray}
    \tilde{T}^{\mu\nu} & = & \left \{
\begin{array}{ll}
   \tau T^{\mu\nu}      &\mathrm{for\ \mu\neq\eta_s\ and\ \nu\neq\eta_s}\\
   \tau^2 T^{\mu\eta_s} &\mathrm{for\ \mu\neq\eta_s}\\
   \tau^{3} T^{\eta_s\eta_s}& \mathrm{otherwise}
\end{array} \right.\\
\tilde{N}^{\mu} & = & \left \{
\begin{array}{ll}
    \tau N^{\mu}      &\mathrm{for\ \mu\neq\eta_s}\\
    \tau^2 N^{\eta_s} &\mathrm{for\ \mu = \eta_s}
\end{array} \right.\\
\tilde{\pi}^{\mu\nu} & = & \left \{
\begin{array}{ll}
   \pi^{\mu\nu} &\mathrm{for\ \mu\neq\eta_s\ and\ \nu\neq\eta_s}\\
   \tau \pi^{\mu\eta_s} &\mathrm{for\ \mu\neq\eta_s}\\
   \tau^2 \pi^{\eta_s\eta_s} & \mathrm{otherwise}
\end{array}\right. \\
\tilde{u}^{\mu} & = & (u^{\tau},u^{x},u^{y}, \tau u^{\eta_s})\\
\tilde{\partial}_{\mu} & = & (\partial_{\tau}, \partial_{x},\partial_{y},\partial_{\eta_s}/\tau)\\
\tilde{g}^{\mu\nu} & = & \tilde{g}_{\mu\nu}=\mathrm{diag}(1,-1,-1,-1)
\end{eqnarray}
One benefit of these substitutions is that all the components
in the same vector or tensor have the same dimension.
This technique is widely used in all well-known (2+1)D or (3+1)D
relativistic hydrodynamic codes for heavy-ion collisions
\cite{Hirano:2001eu,Schenke:2010rr,Pang:2012he,Shen:2014vra,Karpenko:2013wva}.
However, the Christoffel symbols calculated from $\tilde{g}_{\mu\nu}$ satisfy $\tilde{\Gamma}_{kl}^{i}=0$.
Neither $\tilde{\Gamma}_{kl}^{i}$ nor $\Gamma_{kl}^i$ constitute the proper new
covariant derivatives to leave the hydrodynamic equations and IS equations unchanged.
Those three covariant derivatives in the new system become,
\begin{eqnarray}
    \tilde{\nabla}_{\tau} \tilde{u}^{\eta_s} &=& \tilde{\partial}_{\tau}\tilde{u}^{\eta_s} \\
    \tilde{\nabla}_{\eta_s} \tilde{u}^{\tau} &=& \tilde{\partial}_{\eta_s}\tilde{u}^{\tau} + \frac{1}{\tau}\tilde{u}^{\eta_s} \\
    \tilde{\nabla}_{\eta_s} \tilde{u}^{\eta_s} &=& \tilde{\partial}_{\eta_s}\tilde{u}^{\eta_s} + \frac{1}{\tau}\tilde{u}^{\tau}
    \label{eq:dtuz}
\end{eqnarray}
From now on, Christoffel symbols will not appear in the equations to avoid possible typos.
Using the new covariant derivatives $\tilde{\nabla}_{\mu}\tilde{u}^{\nu}$, the hydrodynamic equations and IS
equations are expanded in the following way to simplify the explanation of the numerical
implementation in the next section,
\begin{eqnarray}
    \tilde{\partial}_{\tau} \tilde{T}^{\tau\nu} + \tilde{\partial}_{i} \tilde{T}^{i \nu} &=&  S_{T}^{\nu} \label{eqn:Ttilde}\\
    \tilde{\partial}_{\tau} \tilde{N}^{\tau} + \tilde{\partial}_{i} \tilde{N}^{i} &=&  S_{N} \label{eqn:Ntilde}\\
    \tilde{\partial}_{\tau} (\tilde{u}^{\tau}\tilde{\pi}^{\mu\nu}) + \tilde{\partial}_{i} (\tilde{u}^{i}\tilde{\pi}^{\mu\nu}) &=&  S_{\pi}^{\mu\nu} \label{eqn:pitilde}\\
 \tilde{\partial}_{\tau} (\tilde{u}^{\tau}\Pi) + \tilde{\partial}_{i} (\tilde{u}^{i}\Pi) &=&  S_{\Pi} \label{eqn:Pitilde}
\end{eqnarray}
where the source terms are,
\begin{eqnarray}
S_{T}^{\nu} &=&  (\frac{1}{\tau}\tilde{T}^{\eta_s\eta_s}, 0, 0, \frac{1}{\tau}\tilde{T}^{\tau\eta_s})^T, \\
S_{N} &=& 0, \\
S_{\pi}^{\mu\nu} &=& -\frac{\tilde{\pi}^{\mu\nu} - \eta_v \tilde{\sigma}^{\mu\nu}}{\tau_{\pi}} - \frac{1}{3} \tilde{\pi}^{\mu\nu} \tilde{\theta} \nonumber\\
& & - \tilde{g}_{\alpha\beta}(\tilde{u}^{\mu} \tilde{\pi}^{\nu\beta} +
                              \tilde{u}^{\nu} \tilde{\pi}^{\mu\beta})
                              \tilde{D}\tilde{u}^{\alpha} + \tilde{\pi}^{\mu\nu}\frac{\tilde{u}^{\tau}}{\tau} \nonumber \\
& & - \frac{1}{\tau_{\pi}} \left[\lambda_1 \tilde{\pi}^{\langle \mu}_{\lambda} \tilde{\pi}^{\nu\rangle\lambda} + \lambda_2 \tilde{\pi}^{\langle \mu}_{\lambda} \tilde{\Omega}^{\nu\rangle\lambda} + \lambda_3 \tilde{\Omega}^{\langle \mu}_{\lambda} \tilde{\Omega}^{\nu\rangle\lambda} \right] \nonumber \\
                              & & + I^{\mu\nu},\\
S_{\Pi} &=& -\frac{\Pi - \zeta \tilde{\theta}}{\tau_{\Pi}} - \frac{1}{3} \Pi \tilde{\theta}
,
\end{eqnarray}
where $\tilde{\theta}=\tilde{\partial}_{\mu}\tilde{u}^{\mu}+\tilde{u}^{\tau}/\tau$ is the expansion rate, $\tilde{D}=\tilde{u}^{\lambda}\tilde{\nabla}_{\lambda}$ the comoving derivatives.
The $I^{\mu\nu}$ are source terms from Christoffel symbols which are given in Ref.~\cite{Karpenko:2013wva},
\begin{align}
    & I^{\tau\tau} = 2\tilde{u}^{\eta_s}\tilde{\pi}^{\tau\eta_s}/\tau,
    & I^{\tau x} = \tilde{u}^{\eta_s} \tilde{\pi}^{\eta_s x} /\tau,\\
    & I^{\tau y} = \tilde{u}^{\eta_s} \tilde{\pi}^{\eta_s y} /\tau,
    & I^{\tau\eta_s} = \tilde{u}^{\eta_s}(\tilde{\pi}^{\tau\tau}+\tilde{\pi}^{\eta_s\eta_s})/\tau, \\
    & I^{\eta_s x} = \tilde{u}^{\eta_s} \tilde{\pi}^{\tau x} /\tau,
    & I^{\eta_s y} = \tilde{u}^{\eta_s} \tilde{\pi}^{\tau y} /\tau,\\
    & I^{\eta_s \eta_s} = 2\tilde{u}^{\eta_s} \tilde{\pi}^{\tau \eta_s} /\tau,
    & I^{x y} = I^{xy} = I^{yy} = 0,
    \label{eqn:Ipi}
\end{align}

\section{Numerical implementation}
\label{sec:numerical_detail}
The task of the numerical algorithm is to obtain the time evolution of the
energy density $\varepsilon$ and fluid four-velocity $u^{\mu}$ by
solving partial differential equations (\ref{eqn:Ttilde}), (\ref{eqn:Ntilde}), (\ref{eqn:pitilde}) and (\ref{eqn:Pitilde}).
These equations have the common form,
\begin{equation}
    \partial_{\tau} Q + \partial_x F^x + \partial_y F^y + \partial_{\eta_s} F^{\eta_s} = S
    \label{eq:common_form}
\end{equation}
where $Q$ is the conservative variable, $F^{x,y,\eta_s}$ the flux along $x,y,\eta_s$ directions and $S$ the source term.
We use a second-order central scheme Kurganov-Tadmor (KT) algorithm \cite{KT_algorithm} for the convective part $\partial_{\tau} Q + \partial_i F^i = 0$ in Eq. (\ref{eq:common_form}).
\begin{eqnarray}
    \frac{d\bar{Q}}{d\tau} &=& -\frac{H^x_{i+1/2,j,k}-H^x_{i-1/2,j,k}}{dx} \nonumber \\
     & & -\frac{H^y_{i,j+1/2,k}-H^y_{i,j-1/2,k}}{dy} \nonumber \\
     & & -\frac{H^{\eta_s}_{i,j,k+1/2}-H^{\eta_s}_{i,j,k-1/2}}{\tau d\eta_s} \nonumber \\
     &\equiv & S_{KT}
    \label{eq:kt}
\end{eqnarray}
where $\bar{Q}$ stands for the mean value of Q in one cell, $S_{KT}$ stands for source terms from flux in KT algorithm.
The KT algorithm is a finite volume algorithm which has a very clear physical meaning--the change of conserved quantities in a finite volume equals
to the flux entering minus the flux leaving this volume.
Take the $x$ direction as an example, the flux leaving this volume is,
\begin{eqnarray}
    H^{x}_{i+1/2} &=&  \frac{F^x(Q_{i+1/2}^r) + F^x(Q_{i+1/2}^l)}{2} \\
    & & - c_{i+1/2}\frac{Q_{i+1/2}^{r} - Q_{i+1/2}^{l}}{2}
    \label{eq:flux_out}
\end{eqnarray}
where
\begin{eqnarray}
    Q_{i+1/2}^{r} &=& \bar{Q}_{i+1} - (\partial_x Q)_{i+1} \frac{dx}{2} \\
    Q_{i+1/2}^{l} &=& \bar{Q}_{i} + (\partial_x Q)_{i} \frac{dx}{2}
\end{eqnarray}
and $c_{i+1/2}$ is the maximum propagating speed of the local collective signal given in Ref.~\cite{Schenke:2010nt}.
Notice that five nodes $(i-2, i-1, i, i+1, i+2)$ are needed to update the hydrodynamic cell at $i$ for the one-dimensional case. 
In (3+1)D hydrodynamics, another $4$ nodes $(j-2, j-1, j+1, j+2)$ along
the $y$ and $4$ nodes $(k-2, k-1, k+1, k+2)$ along the $\eta_s$ direction are needed.
The KT algorithm is widely used in relativistic hydrodynamic simulations of heavy-ion collisions \cite{Schenke:2010nt,Pang:2014ipa,Bazow:2016yra},
after being introduced to the field of high-energy physics by the McGill group \cite{Schenke:2010nt}.
Some higher order KT algorithms use more nodes in the off-diagonal direction to achieve a higher precision.
However, the simplicity of the 2nd order central scheme makes it much easier to parallelize on GPU.
The equations are further simplified by moving the KT source terms to the right hand side,
\begin{eqnarray}
    \tilde{\partial}_{\tau}\tilde{T}^{\tau\mu} &=& S_{T,tot}^{\mu} \label{eqn:T_ode}\\
    \tilde{\partial}_{\tau}\tilde{N}^{\tau} &=& S_{N,tot}^{\mu} \label{eqn:N_ode}\\
    \tilde{\partial}_{\tau}(\tilde{u}^{\tau}\tilde{\pi}^{\mu\nu}) &=& S_{\pi,tot}^{\mu\nu} \label{eqn:pi_ode} \\
    \tilde{\partial}_{\tau}(\tilde{u}^{\tau}\tilde{\Pi}) &=& S_{\Pi,tot} \label{eqn:pi_ode} 
\end{eqnarray}
where $S_{*,tot} = S_{*} + S_{\mathrm{KT}}$.
The upper index $\mu$ in the vector and $\mu,\nu$ in the tensor are neglected in the following notation for simplicity.
\begin{eqnarray}
u^{*n+1}\pi^{'n+1} & = & u^{n}\pi^{n}+h S_{\pi,tot}(\varepsilon^{n},u^{n},u^{*n+1},\pi^{n})\\
T^{'n+1} & = & T^{n}+ h S_{T,tot}(\varepsilon^{n},u^{n},\pi^{n})\\
T_{\mathrm{ideal}}^{'n+1} & = & T^{'n+1}-\pi^{'n+1}\rightarrow\varepsilon^{'n+1},u^{'n+1}\\
u^{'n+1}\pi^{n+1} & = & u^{n}\pi^{n}+\frac{h}{2}\left[S_{\pi,tot}(\varepsilon^{n},u^{n},u^{*n+1},\pi^{n})\right. \nonumber \\
& & +S_{\pi,tot}(\varepsilon^{'n+1},u^{'n+1},u^{n},\pi^{'n+1}) \left.\right]\\
T^{n+1} & = & T^{n}+\frac{h}{2}\left[ S_{T,tot}(\varepsilon^{n},u^{n},\pi^{n}) \right. \nonumber\\
 &  & +S_{T,tot}(\varepsilon^{'n+1},u^{'n+1},\pi^{n+1}) \left.\right]\\
T_{\mathrm{ideal}}^{n+1} & = & T^{n+1}-\pi^{n+1}\rightarrow\varepsilon^{n+1},u^{n+1}
\end{eqnarray}
where $h$ is the time spacing.
From this flow chart the difficulty in solving 2nd order viscous hydrodynamics becomes clear.
In order to update $\pi^{\mu\nu}$ to time step $n+1$, one needs information of fluid velocity $u^{n+1}$.
However, $u^{n+1}$ can only be determined through $T_{\mathrm{ideal}}^{\mu\nu} = T_{visc}^{\mu\nu} - \pi^{\mu\nu}$,
assuming that $\pi^{\mu\nu}$ at time step $n+1$ are already known.
Implicitly solving $T^{\mu\nu}$, $\pi^{\mu\nu}$ together with root-finding is a possible solution,
however, very complex. The two step Runge-Kutta method is good at solving this problem,
since the first step is a prediction step, it does not ask for exact solution.
We first predict $\pi^{'n+1}$, by extrapolating the fluid velocity to $n+1$ step using $u^{*n+1} = 2u^{n} - u^{n-1}$, and
then get some predicted values for $\varepsilon$ and $u^{\mu}$.
Afterwards, we update $\pi^{n+1}, \Pi^{n+1}, N^{n+1}$ and $T^{n+1}$ using the averaged source terms in 2 steps.
For the first time step where $u^{n-1}$ is not known,
ideal hydrodynamics is employed to estimate $u^{*1}$.
Notice that the bulk viscosity and net baryon density are set to $0$ in the current version.

CLVisc has been applied with a various set of initial energy-momentum tensors for the initial stage of high-energy heavy-ion collisions.
The first model is the optical Glauber model \cite{Miller:2007ri} which can reproduce the charged multiplicity, transverse momentum spectra and elliptic flow $v_2$ of heavy-ion collisions. The second model is Trento \cite{Bernhard:2016tnd} developed by the Duke group which parameterizes MC-Glauber \cite{Hirano:2005xf,Miller:2007ri}, MC-KLN \cite{Kharzeev:2000ph,Kharzeev:2002ei,Hirano:2004en,Drescher:2006ca}, IP-Glasma \cite{Lappi:2006xc,Schenke:2012wb,Schenke:2012fw} and EKRT \cite{Eskola:1999fc,Paatelainen:2012at,Eskola:2017imo} initial conditions. It can additionally describe higher order anisotropic flow $v_n$ due to the inclusion of entropy/energy density fluctuations in the transverse plane. Since Trento is very flexible and successful, this is used as the default for the public version of CLVisc. To verify that bulk observables are well described the corresponding results are presented in Sec.~\ref{sec:data}. The third model is A-Multi-Phase-Transport (AMPT) model \cite{Lin:2004en} which includes further fluctuations along the space-time rapidity and of the initial fluid velocity \cite{Pang:2012he}.
Due to the longitudinal fluctuations and the asymmetric distribution of forward and backward going participants in heavy-ion collisions, CLVisc with AMPT initial conditions can describe the twisting of event planes along the longitudinal direction \cite{Pang:2014pxa,Pang:2015zrq},
di-hadron correlation as a function of rapidity and azimuthal angle differences \cite{Pang:2013pma}. It is also used to describe
the rich vortical structure of the QGP fluid during the expansion and the global and local polarization of hyperons \cite{Pang:2016igs} in non-central heavy-ion collisions. Due to the longitudinal dynamics incorporated in the AMPT initial conditions, they are going to be used for all the results of this work shown in Secs.~\ref{sec:vneta} and \ref{sec:rneta}. 

\begin{figure}[!htp]
    \begin{center}
        \includegraphics[width=0.5\textwidth]{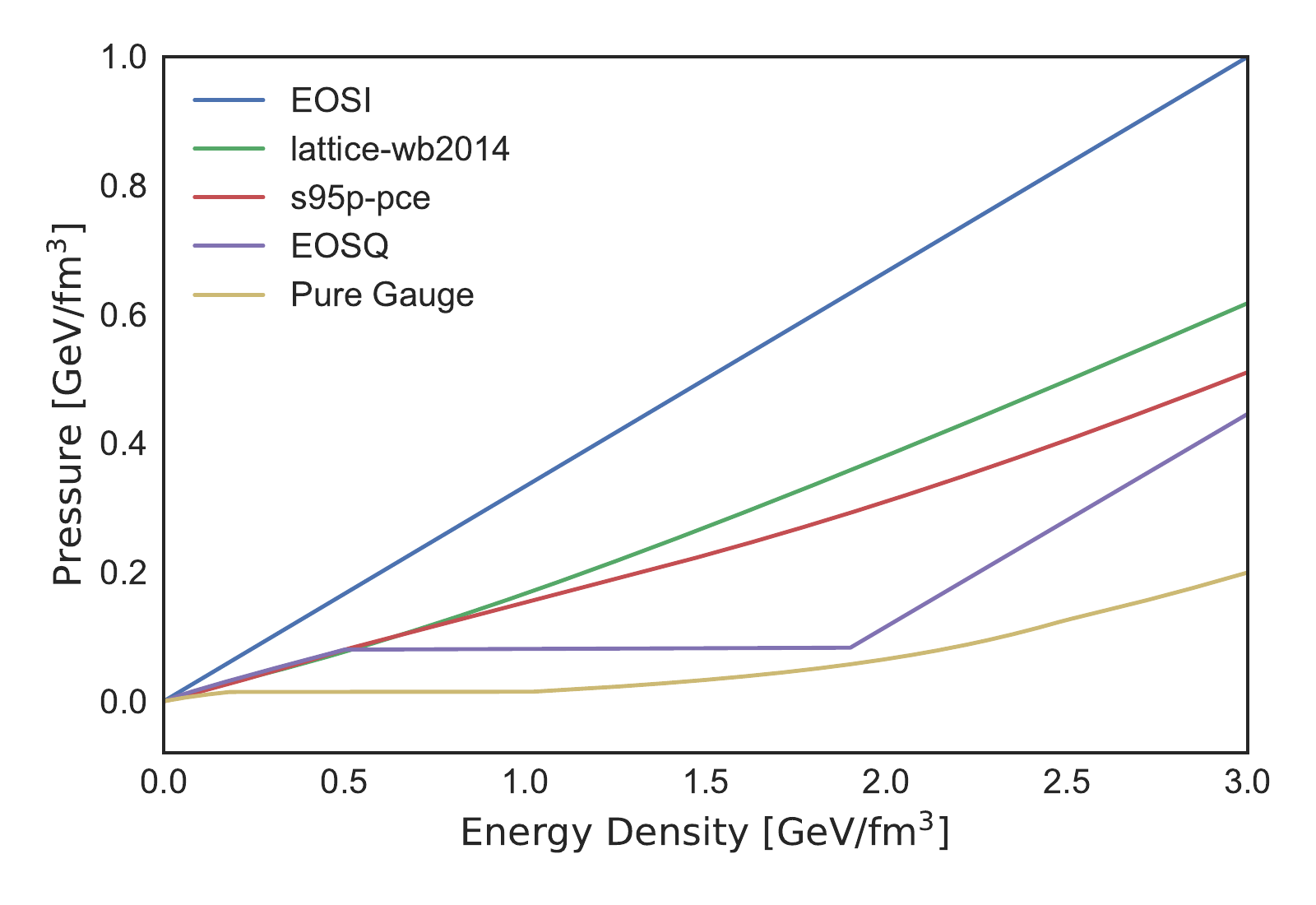}
    \end{center}
    \caption{(color online) Pressure as a function of energy density for 5 different equations of state. They are denoted as EOSI, lattice-wb2014, s95p-pce, EOSQ and pure gauge from top to down.}
    \label{fig:eos}
\end{figure}

There are 5 options for the equation of state (EoS) in CLVisc as shown in Fig.~\ref{fig:eos}:
\begin{description}[align=left]
    \item [EOSI] The simplest EoS -- ideal gas EoS where pressure is $1/3$ of energy density.
    \item [lattice-wb2014] The recent lattice QCD calculations from Wuppertal-Budapest group, whose trace anomaly differ from s95p lattice results by a large margin for the temperature range $180 - 320$ MeV \cite{Borsanyi:2013bia}.
    \item [s95p-pce] The default s95p partial chemical equilibrium EoS \cite{Huovinen:2009yb} used in this paper is given by lattice QCD EoS at high energy density and hadronic resonance gas (HRG) EoS at low energy density with a smooth crossover in between using interpolation.
    \item [EOSQ] Employs a first order phase transition between QGP and HRG \cite{Sollfrank:1996hd}.
    \item [pure gauge] Pure gauge EoS with a first order phase transition given by gluodynamics without (anti)quarks \cite{Boyd:1995zg,Borsanyi:2012ve,Vovchenko:2016mtf}.
\end{description}

\section{Freeze-out and particlization}
\label{sec:particalization}

We use the Cooper-Frye formula \cite{Cooper:1974mv} to calculate the momentum distribution of particle $i$
on the freeze-out hypersurface, 
\begin{equation}
    \frac{dN_i}{dYp_T dp_T d\phi} = \frac{g_i}{(2\pi)^3} \int p^{\mu}d\Sigma_{\mu} f_{\rm eq}(1 + \delta f)
    \label{eq:cooper_frye}
\end{equation}
where $d\Sigma_{\mu}$ is a freeze-out hyper-surface element determined by the constant
freeze-out temperature $T_{f}$ or constant freeze-out energy density $\varepsilon_{f}$.
Particles passing through the freeze-out hyper-surface elements are assumed to obey
Fermi/Bose distributions at temperature $T_{f}$ with the non-equilibrium correction $\delta f$,

\begin{eqnarray}
    f_{\rm eq} &=& \frac{1}{\exp\left[(p\cdot u - \mu_i)/T_{\mathrm{frz}} \right] \pm 1 } \\
    \delta f &=& (1 \mp f_{\rm eq})\frac{p_{\mu} p_{\nu} \pi^{\mu\nu}}{2 T_{\mathrm{frz}}^2 (\varepsilon + P)}
    \label{eqn:feq}
\end{eqnarray}
where $\pm$ is for fermion/bosons, respectively, $\mu_i$ the effective chemical potential in the  partial chemical equilibrium EoS to fix the particle ratio when the temperature is below the chemical freeze-out temperature. $\mu_i$ is set to $0$ for chemical equilibrium EoS.

Two methods are used to compute the particle spectra on the freeze-out hyper-surface.
The first method (called 'smooth') is to carry out the numerical integration over the freeze-out hyper-surface and obtain 
smooth particle spectra in $N_Y \times N_{pt} \times N_{\phi} = 41 \times 15 \times 48$
tabulated $(Y, p_T, \phi)$ bins. $p_T$ and $\phi$ are chosen to be Gaussian Quadrature nodes 
to simplify the calculation of $p_T$ or $\phi$ integrated spectra. Hadron spectra from resonance decays are
also computed via integration. In practice, there are millions of small freeze-out hyper-surface elements $d\Sigma_{\mu}$,
that make the spectra calculation quite CPU time consuming.
This module is parallelized on GPU and the implementation details are described in the Appendix.

The second method for computing final hadron spectra is Monte Carlo sampling based on Eq.~(\ref{eq:cooper_frye}) (dubbed 'MC sampling').
This method is similar to Monte Carlo event generators and the sampled particles can be redirected to hadron cascade models
like UrQMD \cite{Bass:1998ca,Bleicher:1999xi,Petersen:2008dd}, JAM \cite{Nara:1999dz} and SMASH \cite{Weil:2016zrk} to simulate 
hadronic rescattering and resonance decays. In the present work we do not employ a hadronic afterburner, but force the sampled resonances to decay to stable particles immediately after they are produced. This setup saves CPU time and allows for an efficient calculation of correlation observables and provides a baseline calculation for future more quantitative work including hadronic rescattering. By comparing with this baseline one can distinguish the effect of hadronic scattering from resonance decays only.

Since the particle number is Lorentz invariant, particles and their energy-momentum are sampled in the comoving frame of fluid,
and then boosted back to the collision frame via Lorentz transformation with the fluid velocity $u^{\mu}$. This is possible, if the proper weights are taken into account. 
The total number of hadrons produced from the freeze-out hyper-surface is $N= n \times u\cdot d\Sigma$,
where $u\cdot d\Sigma$ is the invariant volume and $n = \sum_i n_i$ is the thermal density of all hadrons in the co-moving frame.
For systems without bulk viscosity and net charge current (net baryon, net electric charge or net strangeness),
the thermal density of hadron type $i$ is fixed for a given freeze-out temperature.
In this case, the thermal densities $n_i$ for all hadron species are computed a priori and tabulated for efficiency.
For systems with non-zero net charge current and bulk viscosity, the thermal densities are different for hyper-surface
elements that have different net charge and bulk viscosity.
In that case, the thermal density $n_i$ must be computed locally for each hyper-surface element which is rather computing intensive,
and also demands parallelization on GPUs.
The present Monte Carlo particlization obeys global conservation laws in one ensemble of sampled events.
If the code is used to compute the net baryon fluctuations or charge correlation, one has to consider global conservation laws 
in each single event \cite{Schwarz:2017bdg}.

The thermal density $n_i$ in the co-moving frame is computed numerically by one-dimensional integration,

\begin{equation}
    n_i = \frac{g_{s}}{(2\pi^2)} \int_{0}^{100T} \frac{p^2 dp}{\exp\left[ (\sqrt{p^2 + m_i^2} - \mu_i)/T \right] \pm 1}
    \label{eq:thermal_density_i}
\end{equation}
where $g_{s}$ is the spin-degeneracy, $T$ is the temperature, $p$ is the momentum magnitude, $m_i$ is the mass of hadron type $i$,
$\mu_i$ is the chemical potential, $\pm 1$ is for baryons and mesons, respectively.

The total number of hadrons computed from one freeze-out hyper-surface element $d\Sigma_j$ is $\lambda_j = n u\cdot d\Sigma_j$,
where $n = \sum_i n_i$ is the summation of thermal density over all hadrons.
$\lambda_j$ is a very small float number that gives the mean number of hadrons produced from $d\Sigma_j$ in multiple independent samplings.
This probability for the hadron multiplicity in the $j$th hyper-surface element is assumed to follow a Poisson distribution,
\begin{equation}
    P_j(k) = e^{-\lambda_j} \frac{\lambda_j^k}{k!}
    \label{eq:poisson}
\end{equation}
where $k$ is an integer that indicates the hadron multiplicity in one sampling.
We draw $k$ from this Poisson distribution and determine the particle type for each of these $k$ hadrons through a discrete distribution whose probabilities are given by $n_i/\sum_i n_i$.

Once the total number of hadrons and their species are determined for one hyper-surface element,  the magnitude of their momenta in the local rest frame can be sampled.
Since the total number of hadrons from the hyper-surface element is Lorentz invariant, one can compute $dN$ from,
\begin{eqnarray}
    dN &=& \frac{g_i}{(2\pi)^3}\int \frac{d^3 p^*}{p^{*0}} \int p^{*\mu} d\Sigma^*_{\mu} f_0 (1 + \delta f) \nonumber \\
&=& \frac{g_i}{2\pi^2} \int \int |\mathbf{p}^*|^2 d|\mathbf{p}^*| d\Sigma^*_{0} f_0 \nonumber \\
&=& \frac{g_i}{2\pi^2} \int u^{\mu}d\Sigma_{\mu} \int d|\mathbf{p}^*|\times |\mathbf{p}^*|^2 f_0
    \label{eqn:dn_lrf}
\end{eqnarray}
where we have used the properties that the $p^{*i}$ is integrated over $(-\infty, \infty)$ for $i=(1, 2, 3)$ and the integration of $\delta f$ (shear viscosity only) also vanishes.
It is straight forward to sample the magnitude of the momentum $|\mathbf{p}^*|$ from $|\mathbf{p}^*|^2 f_0(|\mathbf{p}^*|, \mu, T, \lambda)$ where $\mu$ is chemical potential,
$T$ is freeze-out temperature and $\lambda= \pm 1$ for Fermi-Dirac and Boson-Einstein distribution, respectively.  See \ref{subsec:draw_mom} for details.

Once $|\mathbf{p}^{*}|$ is determined, $f_0$ and $p^{*0}=\sqrt{|\mathbf{p}^{*}|^2 + m^2}$ can be treated as constants when sampling the direction of the momentum in the co-moving frame.
The momentum directions are determined by rejection sampling with acceptance rate $r_{\mathrm{ideal}}$ and $r_{\mathrm{visc}}$, where
\begin{equation}
    r_{\mathrm{ideal}} = \frac{p^*\cdot d\Sigma^*}{p^{*0}\left(d\Sigma^{0*} + \sqrt{|d\mathbf{\Sigma}^*|^2}\right)} \le 1
    \label{eq:r_ideal}
\end{equation}
with $p^*=(p^{*0}, |\mathbf{p}^{*}|\sin\theta \cos\phi,  |\mathbf{p}^{*}|\sin\theta \sin\phi,  |\mathbf{p}^{*}|\cos\theta)$ the four-momentum determined by $|\mathbf{p}^{*}|$, the hadron mass, the polar angle $\theta$ and the azimuthal angle $\phi$. The $d\Sigma^*$ is the hyper-surface element in the co-moving frame.

For viscous hydrodynamics, there is an additional acceptance rate that depends on the direction of the momentum,
\begin{equation}
    r_{\mathrm{visc}} = \frac{A + (1\mp f_0)p_{\mu}^*p_{\nu}^*\pi^{\mu\nu*}}{A + |1\mp f_0|\times|p_{\mu}^*p_{\nu}^*\pi^{*\mu\nu}|_{\mathrm{max}}}
    \label{eq:r_visc}
\end{equation}
where $A=2T^2(\epsilon+P)$ is positive on the freeze-out hyper-surface.
Since $p^{*0}$ and $f_0$ are constants for a given $|\mathbf{p}^*|$, the easiest way to get $| p_{\mu}^*p_{\nu}^*\pi^{*\mu\nu} |_{\mathrm{max}}$ is as follows,
\begin{equation}
    |p_{\mu}^*p_{\nu}^*\pi^{*\mu\nu}|  \le \sum_{\mu \nu} | p_{\mu}^* p_{\nu}^* \pi^{*\mu\nu} |  \le  (p^{*0})^2 \sum_{\mu \nu}  | \pi^{*\mu\nu} |  
\end{equation}

\begin{figure}[!htp]
    \includegraphics[width=0.5\textwidth]{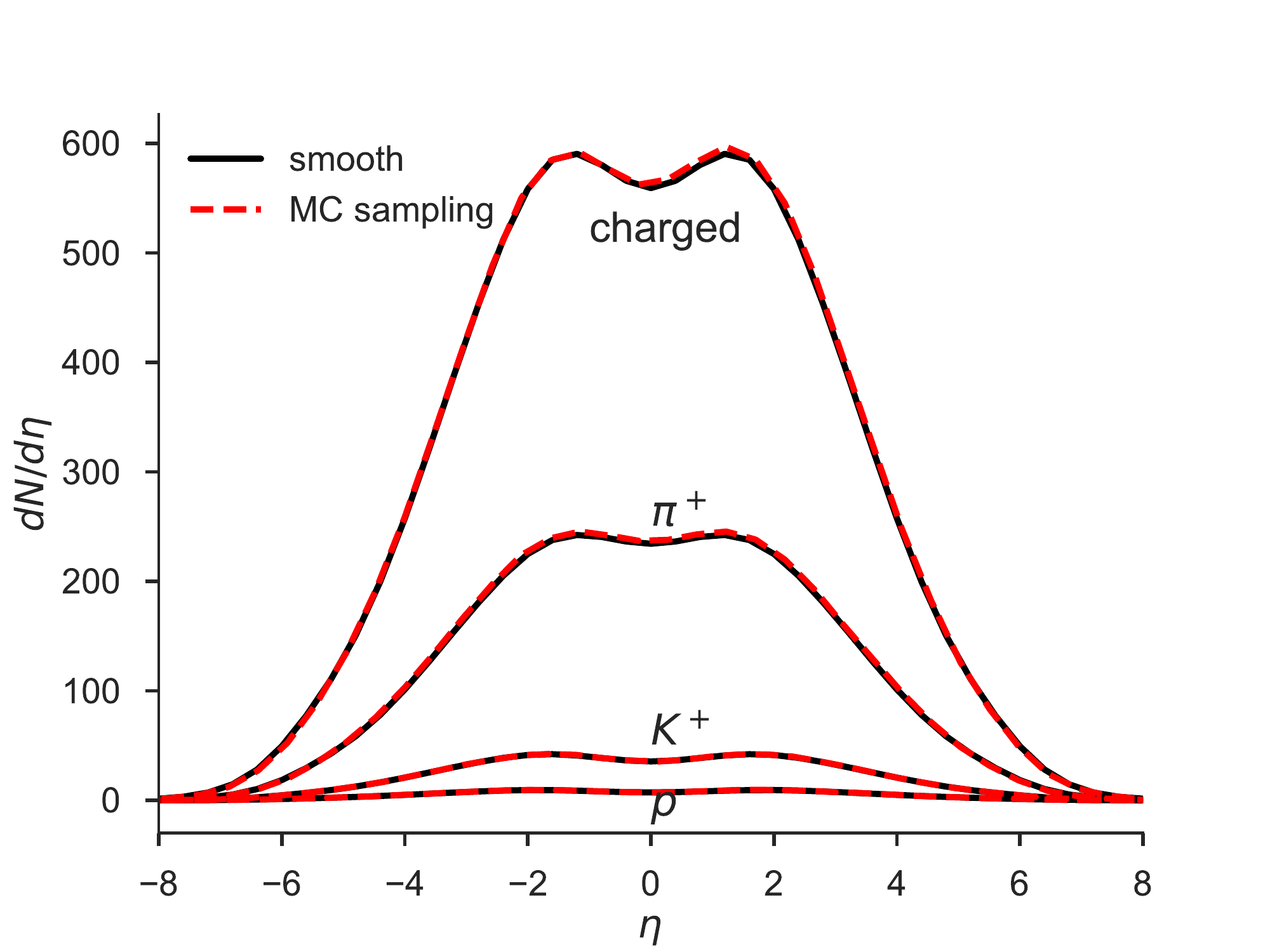}
    \protect\protect\caption{(color online) Pseudo-rapidity distributions for charged hadrons and identified particles $\pi^+$, $K^+$ and proton from smooth particle spectra (black solid line) with integral resonance decay and Monte Carlo sampling (red dashed line) with forced resonance decay. The hydrodynamic evolution is given by CLVisc with optical Glauber initial condition at impact-parameter $b=2.4$ fm, with initial time $\tau_0 = 0.4$ fm, the maximum energy density in most central collisions $\epsilon_{\mathrm{max}}=55\, \rm{ GeV/fm}^3$ and lattice QCD EoS from Wuppertal-Budapest 2014 computation. 
    \label{fig:mc_vs_sm_dNdEta}}
\end{figure}

\begin{figure}[!htp]
    \includegraphics[width=0.5\textwidth]{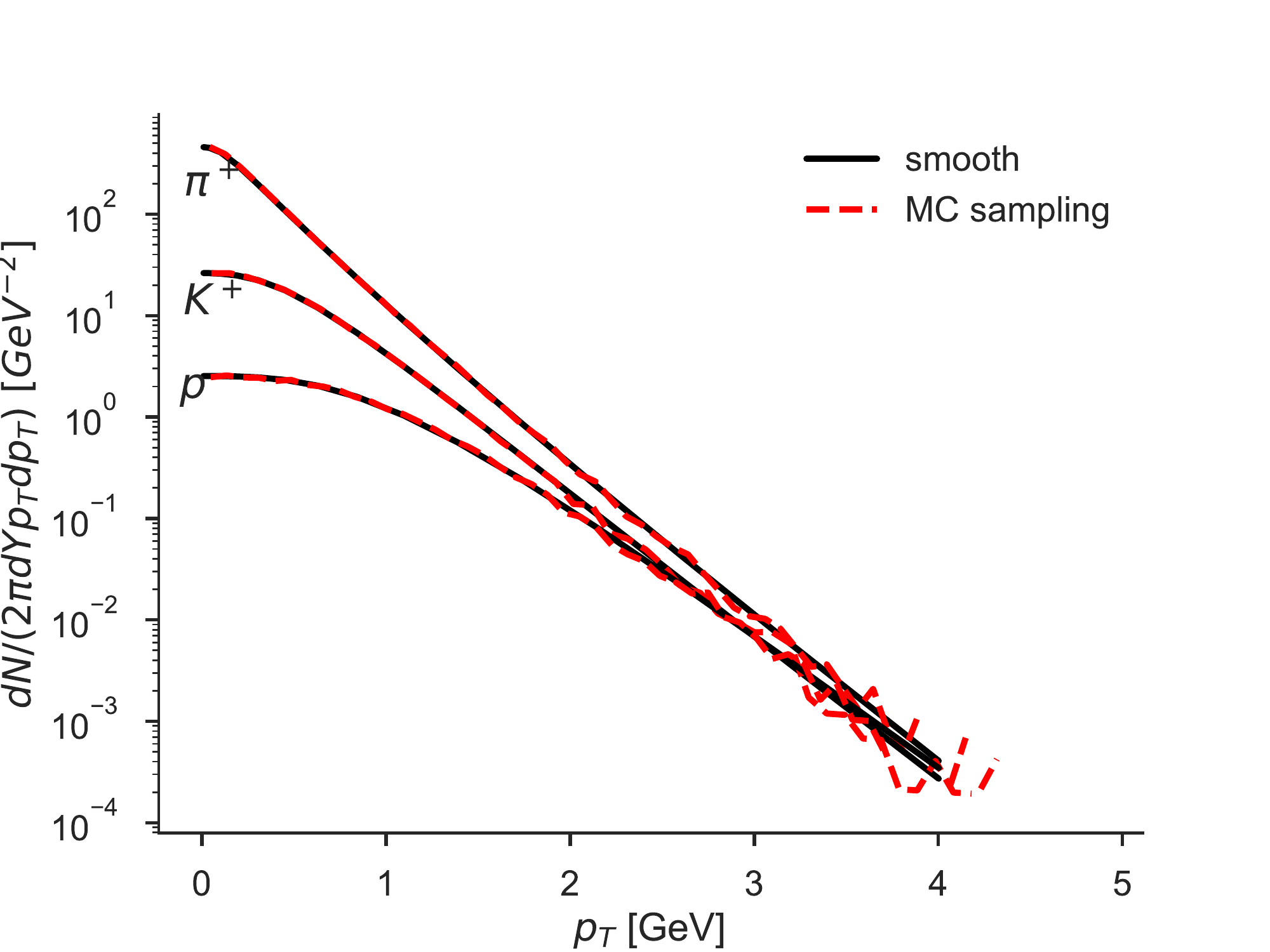}
    \protect\protect\caption{(color online) The transverse momentum distribution for identified particles $\pi^+$, $K^+$ and proton from smooth particle spectra (black solid line) with integral resonance decay and Monte Carlo sampling (red dashed line) with forced resonance decay. The hydrodynamic evolution is the same as in Fig.~\ref{fig:mc_vs_sm_dNdEta}.
    \label{fig:mc_vs_sm_ptspec}}
\end{figure}

One problem in the smooth resonance decay is that the numerical integrations over the phase space of parent hadrons are difficult to verify. The Monte Carlo sampling and decay program, however, can be tested easily. Given the freeze-out temperature, the thermal density of each hadron species before resonance decay is easily computed from numerical integration as shown in Eq.~(\ref{eq:thermal_density_i}). Given the density of each hadron and the tree-structure in the decay table, one can compute the ratio of $\pi^+$ density before and after resonance decay. We have verified that results from Monte Carlo sampling and decay agree with the analytical solution. It is straight forward to check the accuracy of the GPU parallelized smooth spectra and resonance decay via integration by comparing the particle yield and transverse momentum distribution with the Monte Carlo sampling and force decay method.

As shown in Fig.~\ref{fig:mc_vs_sm_dNdEta} and Fig.~\ref{fig:mc_vs_sm_ptspec}, the yields and the momentum distribution of charged and identified particles from the Monte Carlo sampling agree with the smooth particle spectra via integration from Cooper-Frye formula. 
These hydrodynamic simulations use optical Glauber initial condition with the impact-parameter $b=2.4$ fm,  initial time  $\tau_0=0.4$ fm, maximum energy density at the center of the overlap region $\epsilon_{\rm max}=55$ GeV/fm$^3$, $\eta_v/s=0$ and lattice QCD EoS (lattice-wb2014) based on the Wuppertal-Budapest 2014 results.

\section{Comparisons with analytical solutions and other numerical solutions}
\label{sec:code_varification}

To ensure the numerical accuracy of the GPU parallelized CLVisc code, we validate it by comparing the numerical results with both
analytical solutions of the hydrodynamic equations and numerical solutions from other independently developed codes.

For the first validation, analytical solutions are based on simple assumptions. The Bjorken solution, for example, assumes that 
the energy density distribution is uniform in $(x,y,\eta_s)$ coordinates. Under this assumption, pressure gradients
along $x,y$ and $\eta_s$ vanish, fluid velocity $v_x = v_y = v_{\eta_s}=0$, all the nonvanishing terms that affect the time evolution 
in hydrodynamic equations come from nonzero Christoffel symbols. This solution therefore can be used to check whether the Christoffel symbols
are correctly implemented and to quantify numerical errors accumulated during many time steps of evolution. On the other hand this solution can not be used to
check the accuracy of spatial derivatives.

The cross check between different codes on the other hand works for arbitrary initial configurations. However, comparisons of numerical results from different
codes with the same initial configurations, cannot directly validate one model over the other or judge which implementation results in smaller numerical errors.
Below we will compare results from CLVisc with the Riemann, Bjorken and Gubser solution for 2nd order viscous hydrodynamics and the viscous hydrodynamic code VISH2+1 developed by the Ohio State University (OSU) group.

\subsection{Riemann solution}

The Riemann solution considers fluid expansion with a step-like initial energy density distribution. 
It tests the performance of the numerical hydrodynamic simulations in regions with sharp gradients (e.g. the shock wave front) \cite{Akamatsu:2013wyk,Bouras:2009nn,Bouras:2010hm}.
The initial condition is specified as
\begin{eqnarray}
    \varepsilon(t=0, z) &=& \left\{
        \begin{array}{ll}
            \varepsilon_0, & \mathrm{z \le 0}\\
            0,              & \mathrm{z \ge 0}\\
        \end{array} \right. \\
    v_z(t=0, z) &=& \left\{
        \begin{array}{ll}
            0,             & \mathrm{z \le 0}\\
            1,              & \mathrm{z \ge 0}\\
        \end{array} \right.
    \label{eqn:riemann}
\end{eqnarray}
where the initial fluid velocity at $z>0$ is set to $1$.
In relativistic hydrodynamics, the
Riemann solution describes how the QGP expands into vacuum.
In the non-relativistic case, the Riemann solution is used to study dam breaking.
The solution is a function of the similarity variable $\zeta\equiv z/t$. Because of
causality, nothing changes in the $|\zeta|>1$ region.
For $-1<\zeta<1$, the solution is a simple rarefaction wave which is given by \cite{Rischke:1995ir},
\begin{eqnarray}
    \frac{\varepsilon(\zeta)}{\varepsilon_0} &=& \left\{
        \begin{array}{ll}
            1, & \mathrm{-1 \le \zeta \le -c_s}\\
            \left[ \frac{1-c_s}{1+c_s}\frac{1-\zeta}{1+\zeta}\right]^{(1+c_s^2)/2c_s}, & \mathrm{-c_s \le \zeta \le 1}\\
        \end{array} \right. \\
        v_z(\zeta) &=& \tanh \left[ -\frac{c_s}{1+c_s^2} \ln\left(\frac{\varepsilon}{\varepsilon_0} \right) \right] .
    \label{eqn:riemann}
\end{eqnarray}

\begin{figure}[!htp]
    \includegraphics[width=0.5\textwidth]{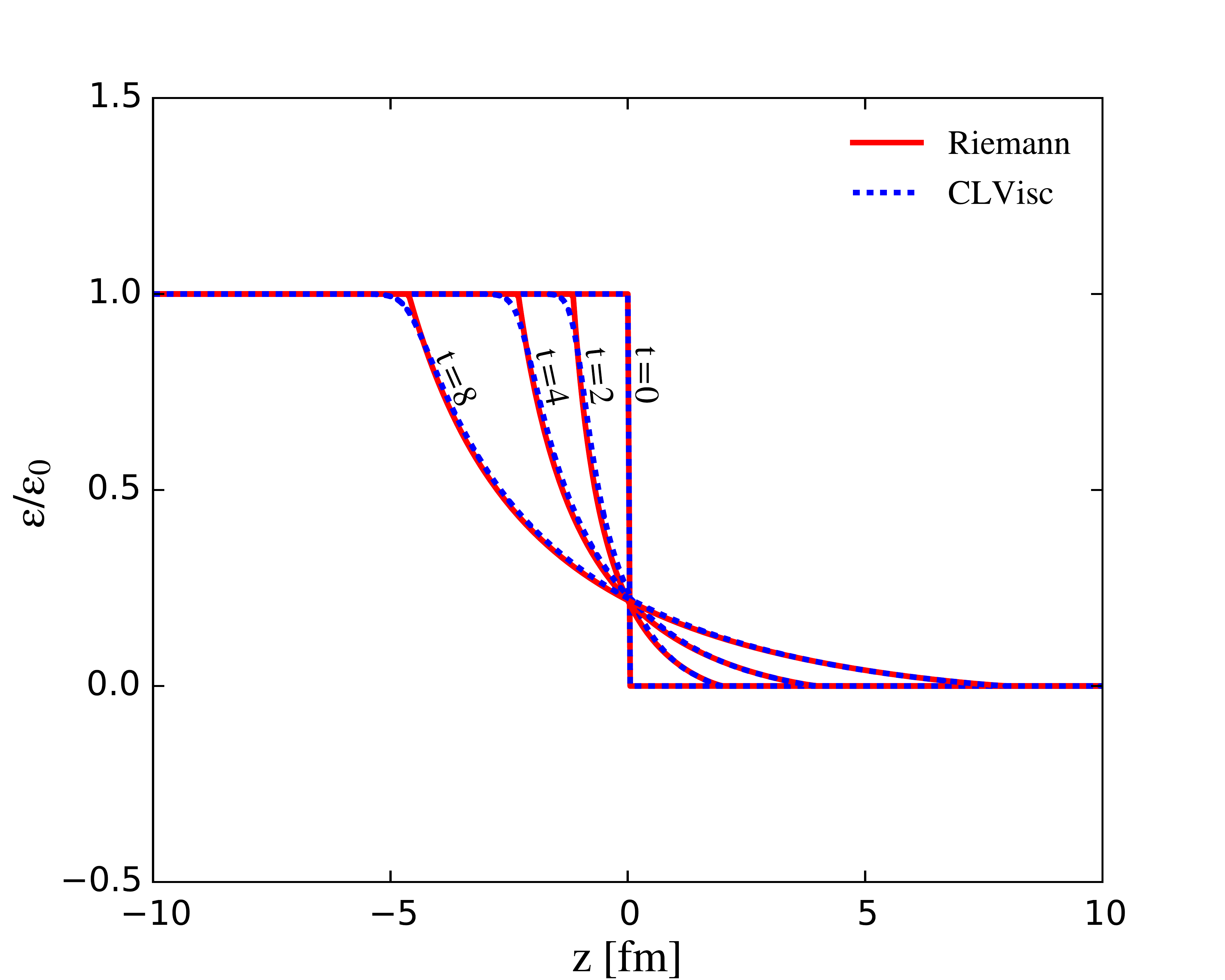}
    \protect\protect\caption{(color online) The comparison between CLVisc and Riemann solution for energy density evolution as a function of time.
       \label{fig:riemann_ed}}
\end{figure}

\begin{figure}[!htp]
    \includegraphics[width=0.5\textwidth]{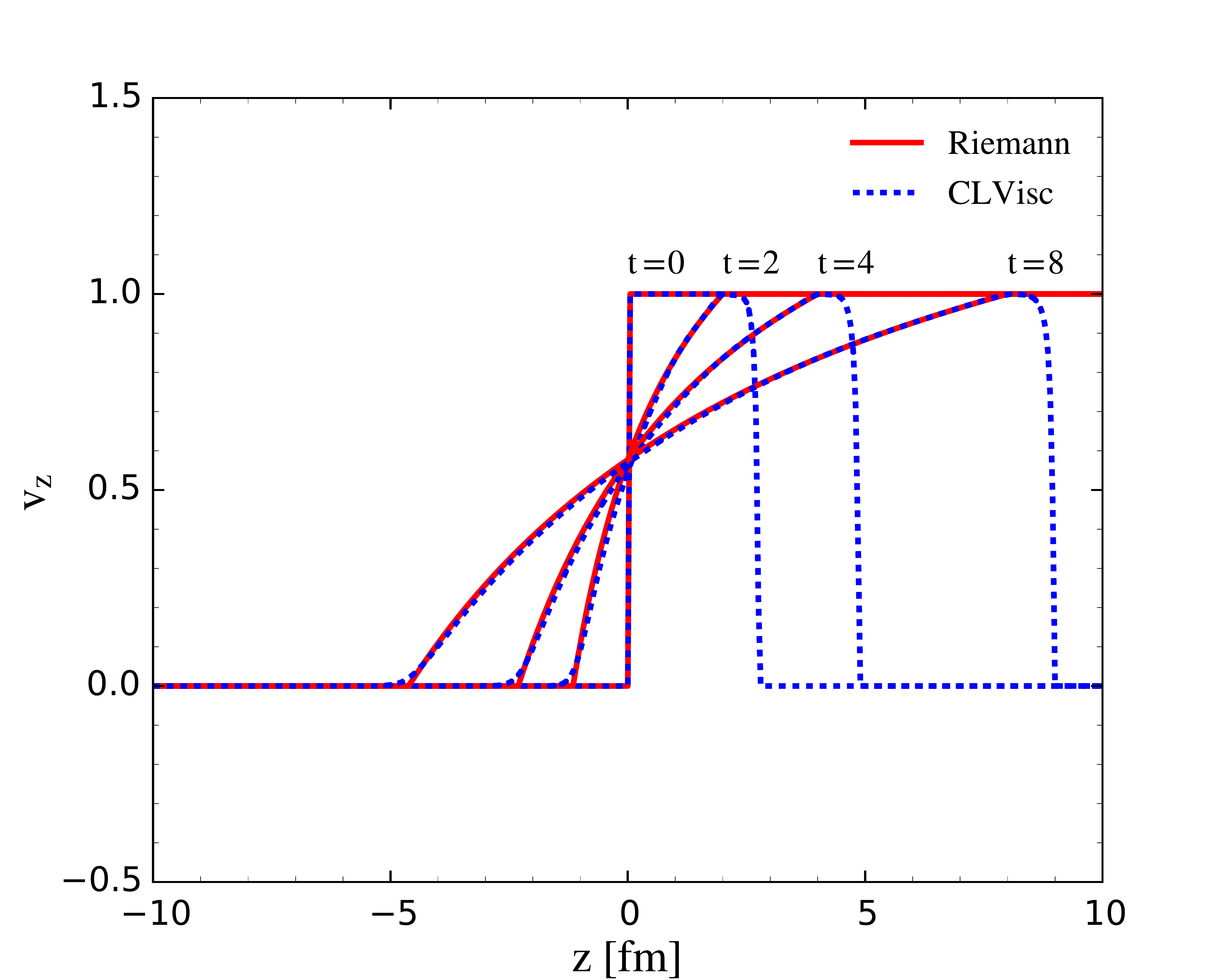}
    \protect\protect\caption{(color online) The comparison between CLVisc and Riemann solution for fluid velocity evolution as a function of time.
       \label{fig:riemann_vz}}
\end{figure}
To compare to the Riemann solution, the ideal gas EoS (EOSI) is used where the speed of sound $c_s^2=1/3$ in CLVisc simulations.
All the Christoffel symbols are set to $0$ to return to $(t,x,y,z)$ coordinates.
The energy density is set to constant in the transverse direction.
CLVisc solves the Riemann problem precisely for the energy density evolution as shown in Fig. \ref{fig:riemann_ed}.
For the fluid velocity profile, there is a quick drop-off in the light cone region ($z=t$)
which is caused by the numerical cutoff used in the simulations.
In high-energy heavy-ion collisions, an energy density cut-off
$\varepsilon=10^{-7} $ GeV/fm$^3$ is reasonably safe comparing with the typical freeze-out
energy density $\varepsilon\sim 0.1 $ GeV/fm$^3$, when the hydrodynamic evolution stops.
The physics processes at such low energy density region around and after the freeze-out should be described by hadronic 
transport models instead of hydrodynamics. 
By setting $\varepsilon=0$, when the energy density is smaller than the cutoff,
an artificial shock wave is formed at the edge of the expanding fireball.
The Riemann solution test verifies that this artificial cutoff does not lead to sizable difference for the
region where we apply hydrodynamics.

\subsection{Bjorken solution}

The Bjorken solution assumes uniform distribution in the transverse direction and in spatial rapidity $\eta_s$ in Milne coordinates,
which gives rise to $v_x=v_y=v_{\eta_s}=0$. This solution derived in \cite{Bjorken:1982qr} is used extensively to model the longitudinal expansion dynamics in high-energy heavy-ion collisions, where a pleateau in the rapidity profile is observed in final state particle spectra. It is applied in otherwise 2+1 dimensional hydrodynamic models or in analytic calculations. 
However, the energy density still decreases with time due to 
nonzero longitudinal fluid velocity $v_z=z/t$ in $(t,x,y,z)$ coordinates. 
The nonzero components of shear stress tensors are $\pi^{xx}=\pi^{yy}=-\tau^2 \pi^{\eta_s\eta_s} = \frac{4\eta_v}{3\tau}$.
With all the spatial gradients vanishing under this assumption,
the hydrodynamic equations are simplified to,
\begin{equation}
    \frac{\partial \varepsilon}{\partial \tau} + \frac{\varepsilon+P+\tau^2 \pi^{\eta_s\eta_s}}{\tau} = 0
    \label{eq:bjorken}
\end{equation}

\begin{figure}[!htp]
\includegraphics[width=0.5\textwidth]{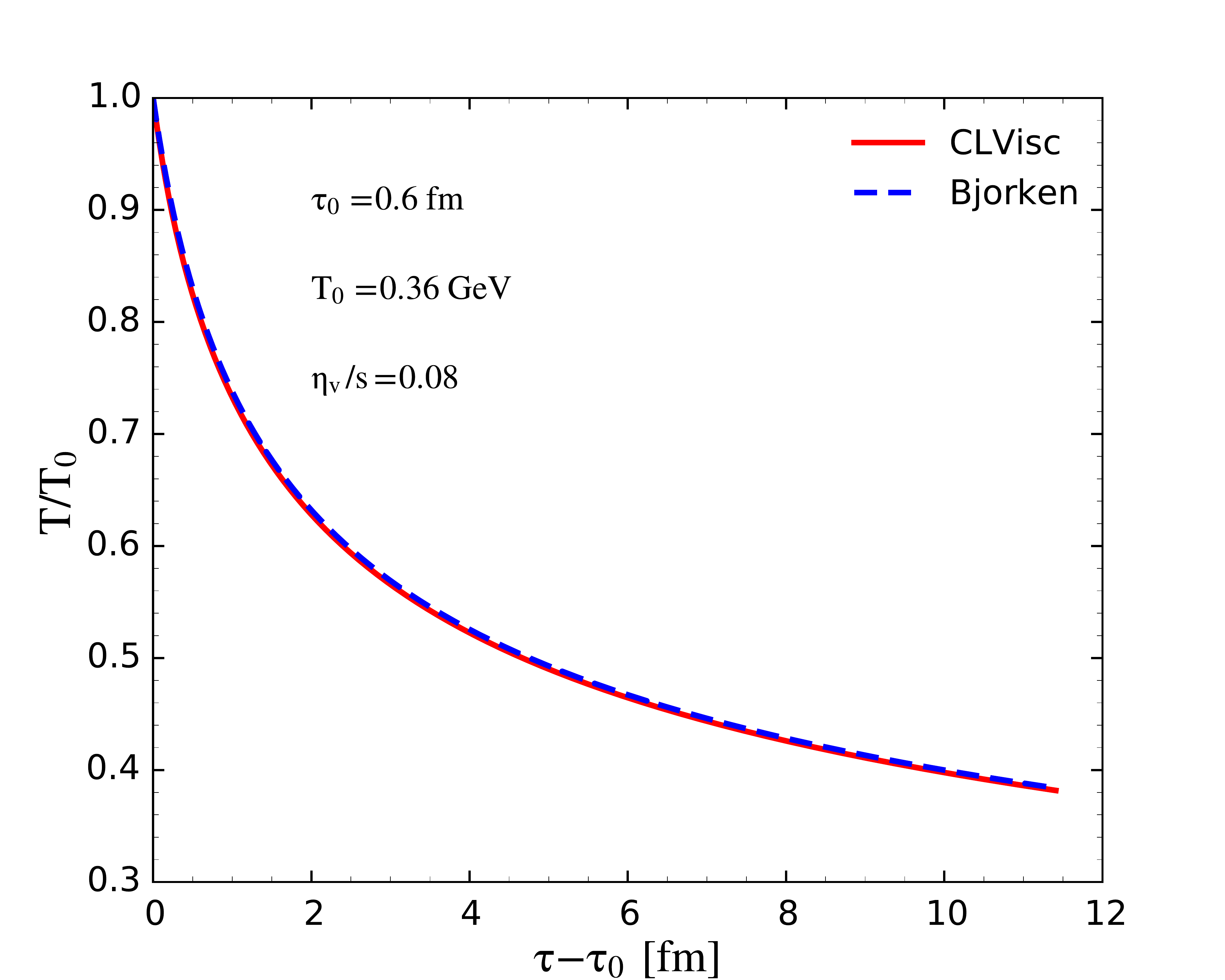} 
\protect\protect\caption{(color online) The comparison between CLVisc and Bjorken solution for viscous hydrodynamics}
\label{fig-bjorken}
\end{figure}

For the ideal gas EoS where $\varepsilon=3P$ and $T\propto \varepsilon^{1/4}$,
we have the solution,
\begin{equation}
    \frac{T}{T_0} = \left( \frac{\tau_0}{\tau} \right)^{1/3}\left[ 1+\frac{2\eta_v}{3sT\tau_0} \left( 1-\left( \frac{\tau_0}{\tau}\right)^{2/3} \right) \right],
    \label{eq:T_bjorken}
\end{equation}
where $T$ and $T_0$ are temperature at proper time $\tau$ and $\tau_0$,  respectively.  Shown in Fig.~\ref{fig-bjorken} is the numerical solution from
CLVisc (solid) compared to the above Bjorken analytic solution with the same initial temperature, time and shear viscosity to entropy ratio. 

\subsection{Gubser solution for 2nd order viscous hydrodynamics}
The Bjorken solution assumes homogeneous distribution of energy density
in $(\tau, x,y, \eta_s)$ coordinates at any given time $\tau$ which leads to $u^{\mu}=(1,0,0,0)$. This solution, 
however, gives rise to nonzero longitudinal fluid velocity $v_z=z/t$ when transformed back to $(t,x,y,z)$ coordinates.
The same philosophy is used in the Gubser solution for the 2nd order viscous hydrodynamics \cite{Pang:2014ipa},
where we perform a conformal/Weyl transformation to the coordinate system following Gubser \cite{Gubser:2010ze},
\begin{equation}
    d\hat{s}^2 \equiv \frac{ds^2}{\tau^2} = d\rho^2 - \cosh^2\rho
        (d\theta^2 + \sin^2\theta d\phi^2) - d\eta_s^2,
    \label{eq:weyl_transform}
\end{equation}
which indicates that the Minkowski space is conformal to $dS_3\times R$ with,
\begin{equation}
    \sinh \rho = - \frac{L^2 - \tau^2 + x^2_{\perp}}{2L\tau},\quad\;
    \tan \theta = \frac{2Lx_{\perp}}{L^2 + \tau^2 - x_{\perp}^2},
    \label{eq:conformal}
\end{equation}
where $L$ can be interpreted as the radius of the $dS_3$ space or the
typical size of a relativistic heavy-ion collisions.
Hereafter in this section, dynamical variables in the new coordinate system $\hat{x}^{\mu} = (\rho,\theta,\phi,\eta_s)$ will carry a hat to avoid confusion.
Assuming the energy density distribution is uniform in this $\hat{x}^{\mu}$ coordinates, one simply gets $\hat{u}^{\mu}=(1,0,0,0)$.
When $\hat{\eta}_v \hat{\lambda}_{1}^2 = 3\hat{\tau}_{\pi}$, we find a very simple analytical solution,
\begin{eqnarray}
    \hat{\varepsilon} &\propto& \left( \frac{1}{\cosh\rho} \right)^{\frac{8}{3}-\frac{2}{\hat{\lambda}_1}},\quad\; \hat{u}^{\mu} = (1, 0, 0, 0), \\
    C &=& -2 A = -2 B = \frac{2}{\hat{\lambda}_1}\hat{\varepsilon}.
    \label{eqn:visc_gubser_before_weyl}
\end{eqnarray}
where $C\equiv \hat{\pi}^{\eta_s\eta_s}$, $A\equiv \hat{\pi}^{\theta\theta}\cosh^2\rho $ and $B\equiv \hat{\pi}^{\phi\phi}\cosh^2\rho\sin^2\theta$. After Weyl rescaling, we can get back to the $(\tau,x,y,\eta_s)$ space and obtain,
\begin{eqnarray}
    \varepsilon &=&  \frac{\hat{\varepsilon}}{\tau^4}, \\ 
    \vec{v}_{\perp} &=&  \frac{-2\tau \vec{x}_{\perp}}{L^2+\tau^2+x_{\perp}^2}, \\
    \pi_{\mu\nu} &=& \frac{1}{\tau^2} \frac{\partial \hat{x}^{\alpha}}{\partial x^{\mu}} \frac{\partial \hat{x}^{\beta}}{\partial x^{\nu}}\hat{\pi}_{\alpha\beta}.
    \label{eqn:gubser_visc}
\end{eqnarray}
Notice that the dimensionless transport coefficients are defined as $\hat{\eta}_v = \eta_v/\varepsilon^{3/4}$, $\hat{\tau}_{\pi}=\tau_{\pi}\varepsilon^{1/4}$, $\hat{\lambda}_1 = \lambda_1 \varepsilon$. 
The conditional solution is nontrivial since there are three different 
transport coefficients and many non-vanishing $\pi^{\mu\nu}$ components.
Since the energy density distribution is not uniform in the transverse plane of $(\tau,x,y,\eta_s)$ coordinates, the spatial gradients along $x$ and $y$ are nontrivial. This solution is very good at verifying the numerical capability of any 2nd order viscous hydrodynamics code.

\begin{figure}
    \includegraphics[width=0.5\textwidth]{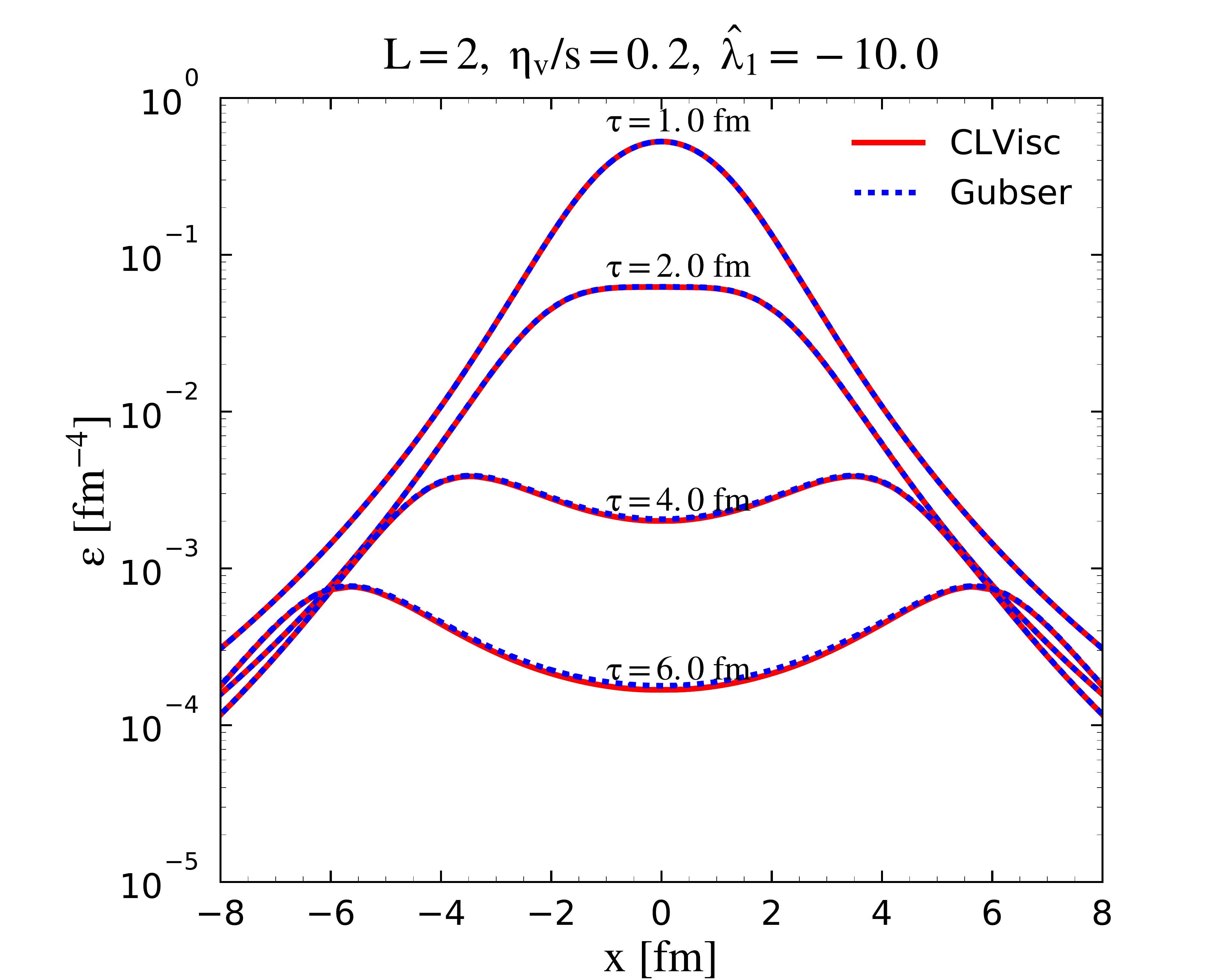}
    \caption{(color online) The time evolution of energy density distribution from CLVisc numerical results (solid)  and Gubser analytical solution (dashed) for 2nd order viscous hydrodynamics. \label{fig:gubser_ed}}
\end{figure}

\begin{figure}
    \includegraphics[width=0.5\textwidth]{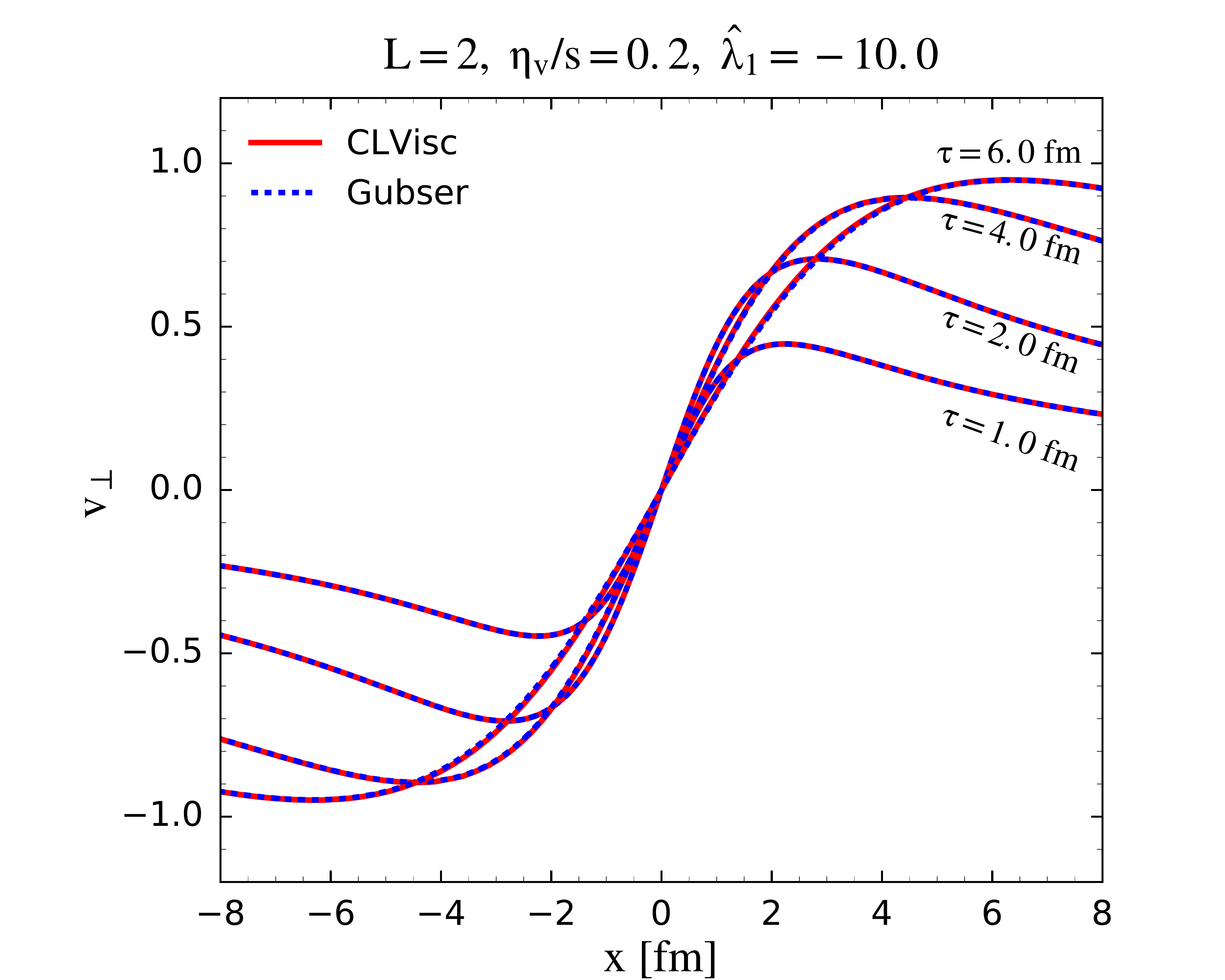}
    \caption{(color online) The time evolution of transverse fluid velocity from CLVisc numerical results (solid)  and Gubser analytical solution (dashed) for 2nd order viscous hydrodynamics.\label{fig:gubser_vr}}
\end{figure}

The parameters we used for the comparison in this section are $L=2$, $\eta_v/s=0.2$ and $\hat{\lambda}_1=-10$.
The relaxation time $\hat{\tau}_{\pi}$ is calculated from the constraint equation $\hat{\eta}_v \hat{\lambda}_{1}^2 = 3\hat{\tau}_{\pi}$.
Notice that we can still cover the whole parameter space for $\eta_v/s$ and $\hat{\lambda}_{1}$,
to investigate the stability of the code in different limits.
In practice, $\hat{\lambda}_1 = \frac{\hat{\varepsilon}}{\hat{\pi}^{\mu\nu}} >> 1$ is required for consistency and stability.
When $\hat{\lambda}_1 \rightarrow \infty$, the hydrodynamic equations recover the ideal fluid solution.
As shown in Figs. \ref{fig:gubser_ed} and \ref{fig:gubser_vr}, with $\hat{\lambda}_1=-10$, CLVisc reproduces very accurately 
the energy density and transverse fluid velocity evolution given by the Gubser solution.
Another interesting property of this 2nd order Gubser solution is that the fluid velocity is the same as that for ideal hydrodynamics, since it is fixed by conformal transformation.

\begin{figure}
    \includegraphics[width=0.5\textwidth]{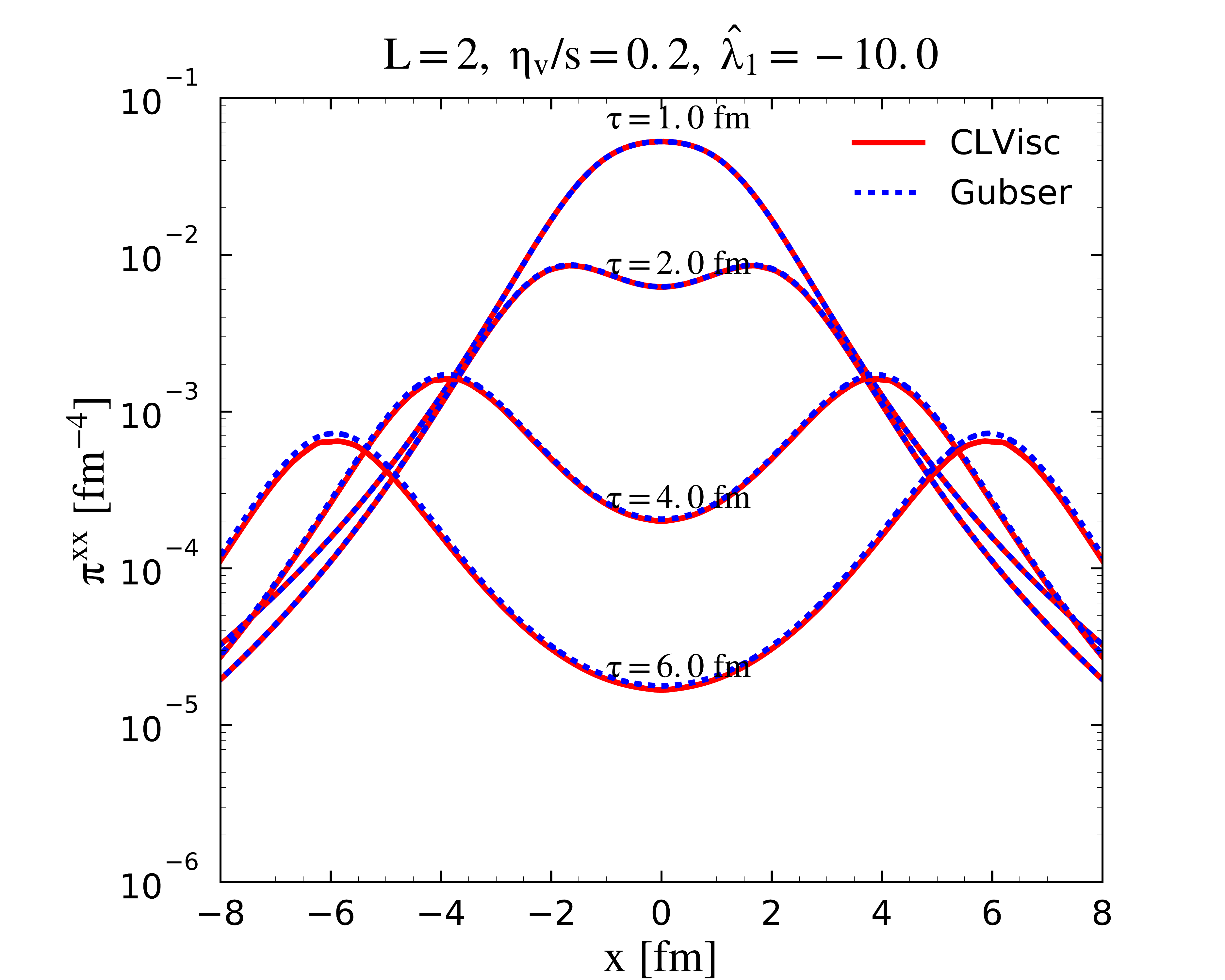}
    \caption{(color online) The time evolution of $\pi^{xx}$ from CLVisc numerical results and Gubser analytical solution for 2nd order viscous hydrodynamics.}
    \label{fig:pixx}
\end{figure}

\begin{figure}
    \includegraphics[width=0.5\textwidth]{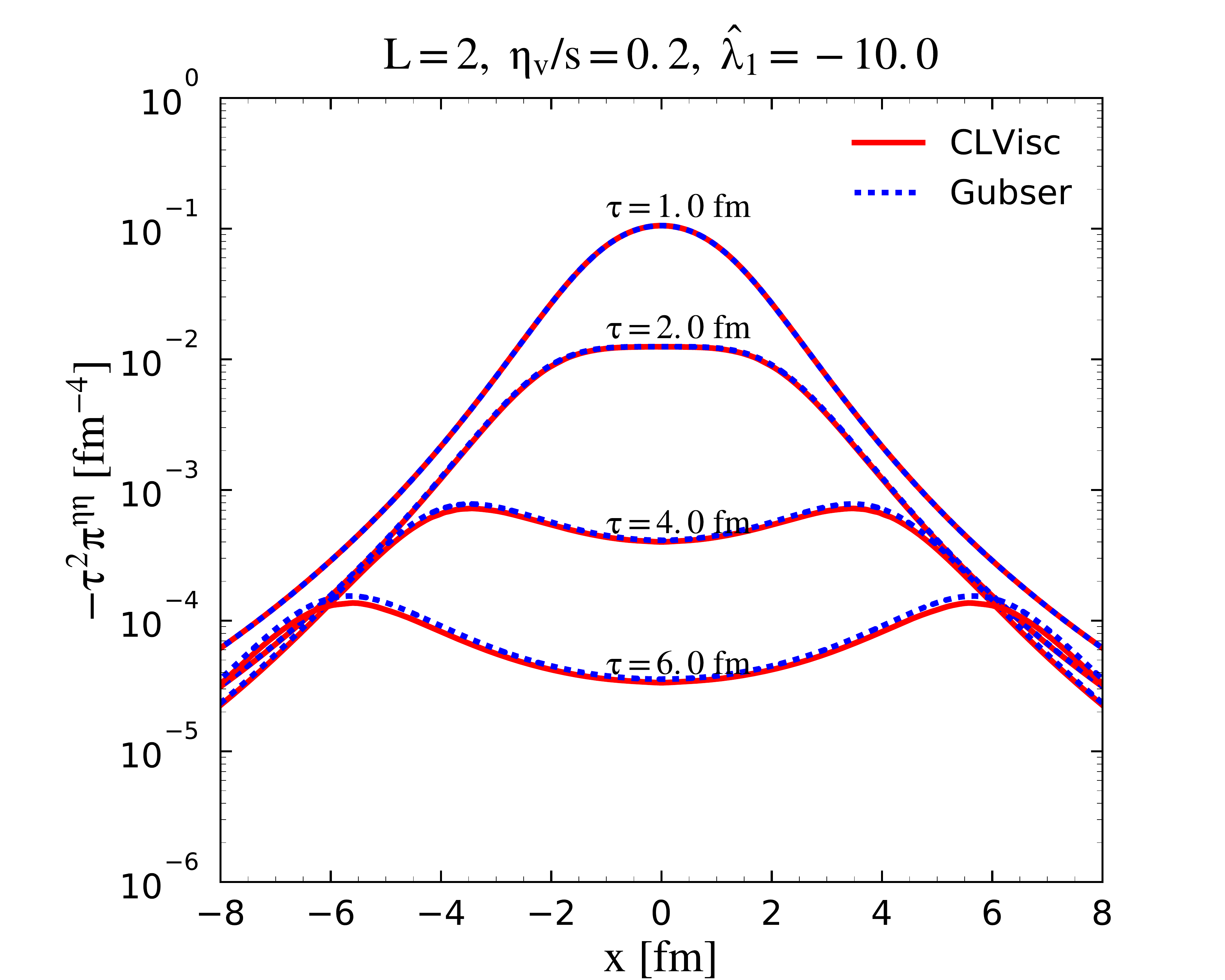}
   \caption{(color online) The time evolution of $-\tau^2\pi^{\eta_s\eta_s}$ from CLVisc numerical results and Gubser analytical solution for 2nd order viscous hydrodynamics.}
    \label{fig:pihh}
\end{figure}

In principle $\hat{\lambda}_{1}$ can be either positive or negative.
In heavy-ion collisions, one gets negative $\pi^{\eta_s\eta_s}$ in Bjorken scaling. Therefore 
we choose a negative $\hat{\lambda}_1$ for positive $\pi^{xx}, \pi^{yy}$ and negative $\pi^{\eta_s\eta_s}$.
As a result, $-\tau^2 \pi^{\eta_s\eta_s} $ is roughly two times $\pi^{xx}$ and $\pi^{yy}$,
which preserve the traceless property together with a small but nonzero $\pi^{\tau\tau}$ in this solution.

As shown in Figs.~\ref{fig:pixx} and \ref{fig:pihh}, there are tiny deviations between the analytical solution and the CLVisc relativistic hydrodynamic simulations,
on the shoulders ($x=\pm 6$) of $\pi^{xx}$ and $-\tau^2 \pi^{\eta_s\eta_s}$ at a late time $\tau=6$ fm.
It is expected that the deviation could be larger at even later time due to the accumulated numerical error.
At present, this tiny deviation is acceptable since the energy density drops much faster in Gubser expansion,
than Bjorken expansion or realistic time evolutions of QGP in heavy-ion collisions.

We have collected these analytical solutions and put them in a python package \textbf{gubser}. The package is uploaded to the Python Package Index website,
and can be downloaded and installed on a local machine using \textbf{pip install --user gubser}.
More analytical solutions \cite{Biro:2000nj,Csorgo:2003ry,Nagy:2007xn,Borshch:2007uf,Beuf:2008vd,Lin:2009kv,Peschanski:2009tg,Csorgo:2013ksa,Wong:2014sda,Hatta:2014upa,Hatta:2014gga,Csanad:2014dpa,Hatta:2015ldk,Pu:2016rdq} from the community are welcomed to be added to the package.

\subsection{Comparison with VISH2+1}

\begin{figure}
    \includegraphics[width=0.5\textwidth]{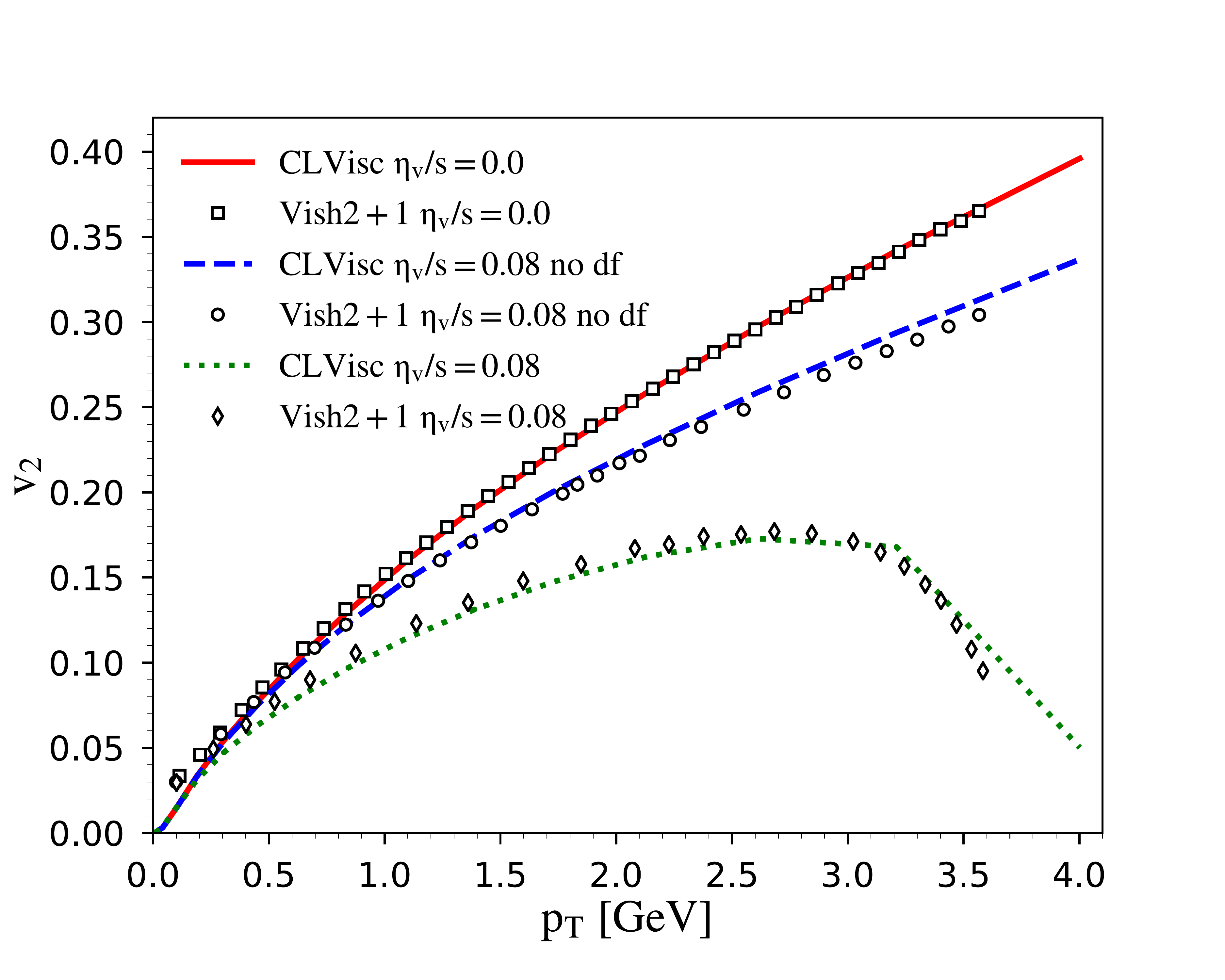}
    \protect\protect\caption{(color online) Comparison between CLVisc (symbol points) and VISH2+1 (lines) results for elliptic flow of direct $\pi^+$ in Au+Au collisions
    at $\sqrt{s_{NN}}=200$ GeV with the optical Glauber initial condition at impact-parameter $b=7$ fm and with different values of shear viscosity to entropy ratio. Results
    without the viscous correction $\delta f$ to the local phase-space distributions [Eq.~(\ref{eqn:feq})] are also shown. 
     \label{fig:cmp_hc_v2}}
\end{figure}

\begin{figure}
    \includegraphics[width=0.5\textwidth]{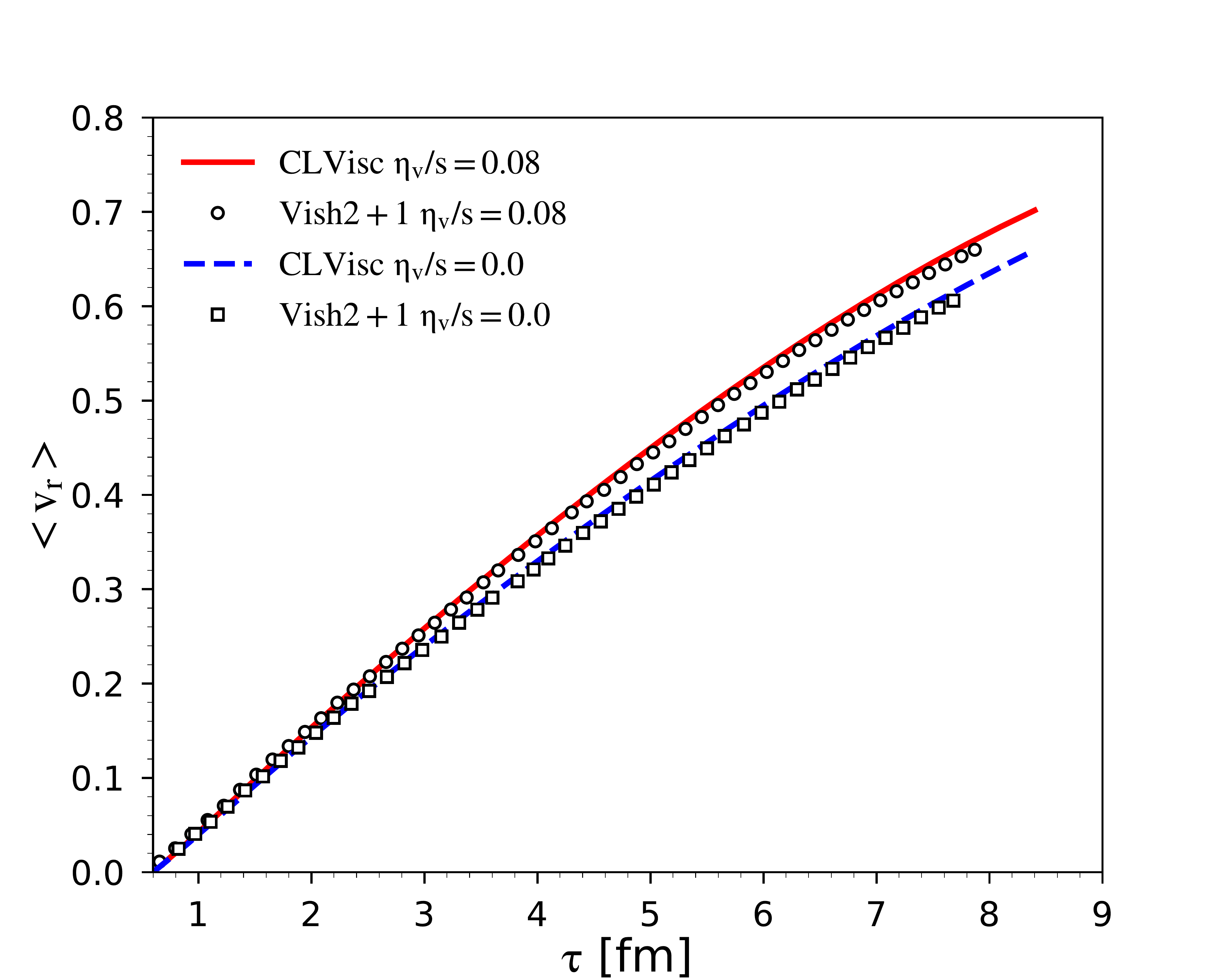}
    \protect\protect\caption{(color online) Comparison between CLVisc (symbol points) and VISH2+1 (lines) results for mean transverse fluid velocity $\langle v_r\rangle$ in Au+Au collisions  at $\sqrt{s_{NN}}=200$ GeV with the optical Glauber initial condition at impact-parameter $b=7$ fm and with different values of shear viscosity to entropy density ratio.
    \label{fig:cmp_hc_vr}}
\end{figure}

\begin{figure}
    \includegraphics[width=0.5\textwidth]{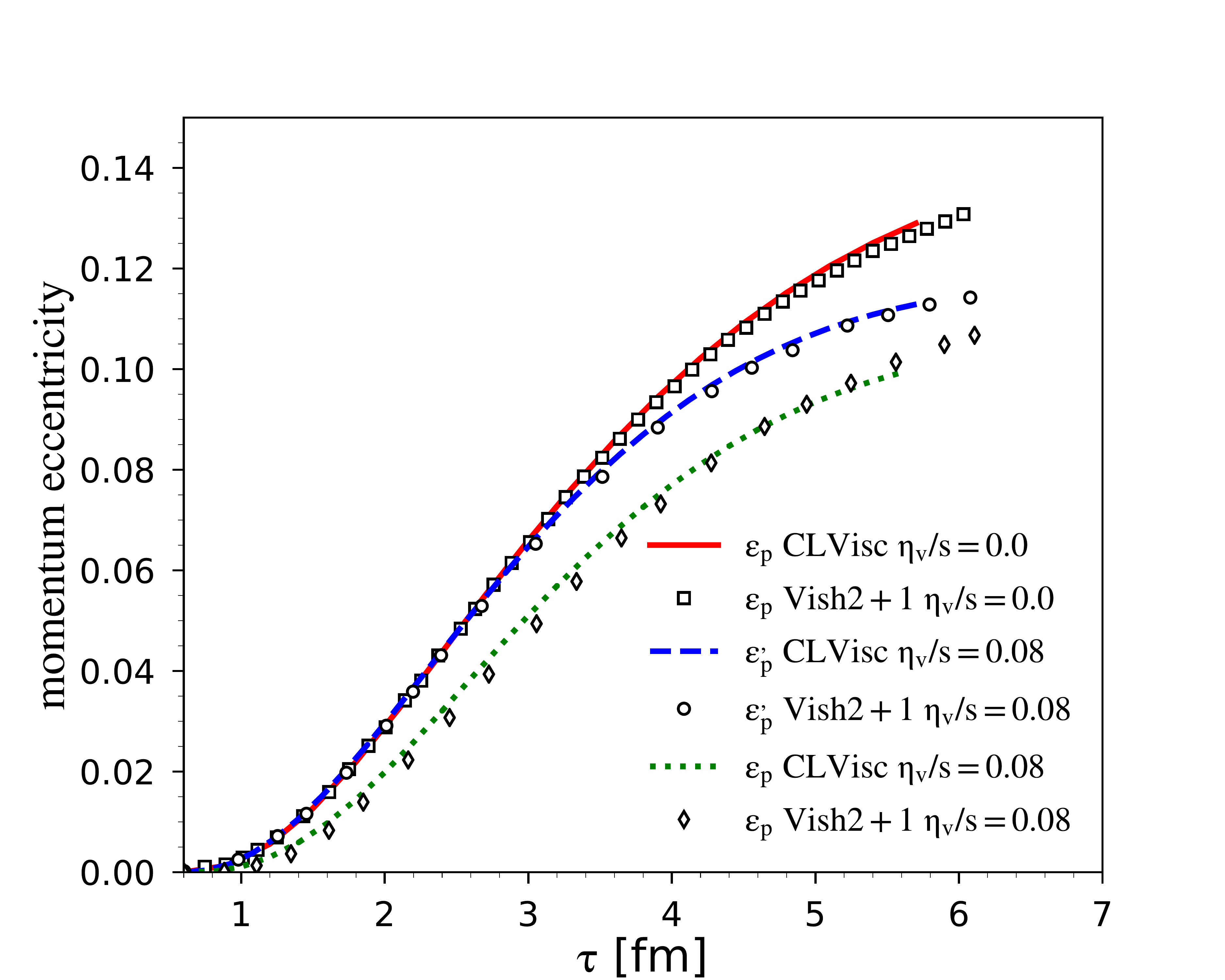}
    \protect\protect\caption{(color online) Comparison between CLVisc (symbols points) and VISH2+1 (lines) results for momentum eccentricity in Au+Au collisions  at $\sqrt{s_{NN}}=200$ GeV with the optical Glauber initial condition at impact-parameter $b=7$ fm and with different values of shear viscosity to entropy density ratio.
    \label{fig:cmp_hc_eccp}}
\end{figure}

We now compare the numerical solutions from CLVisc with VISH2+1 viscous hydrodynamic model developed by the OSU group, which is a (2+1)D viscous
hydrodynamic model assuming Bjorken scaling in the longitudinal direction. The configurations and hydrodynamic results from VISH2+1 can be found in TechQM website \url{https://wiki.bnl.gov/TECHQM/index.php/Momentum_anisotropies}. We use the same initial conditions and model parameters in the simulations for comparison. Shown in Fig.~\ref{fig:cmp_hc_v2} are results for the $p_T$ differential elliptic flow $v_2$,
in Fig.~\ref{fig:cmp_hc_vr} the mean transverse fluid velocity $ \langle v_r \rangle $ and in Fig.~\ref{fig:cmp_hc_eccp}  the momentum eccentricity 
from CLVisc (symbol points) as compared to results from VISH2+1 viscous hydro (lines), for Au+Au collisions at $\sqrt{s_{NN}}=200$ GeV at  impact parameter $b=7$ fm
with the optical Glauber initial condition. They agree with each other to a reasonable precision.

From this extensive comparison to available analytical solutions and other numerical solution of relativistic hydrodynamics, we conclude that CLVisc is performing competitively well. 

\section{Hadron spectra and anisotropic flow}

\label{sec:data}

In this section, we compare CLVisc results for hadron spectra and anisotropic flow in heavy-ion collisions to experimental data at both RHIC and LHC energies.
We use the Trento Monte Carlo model with the default option of the IP-Glasma approximator for fluctuating initial conditions in event-by-event hydrodynamic simulations. Since the public version of CLVisc uses Trento as the default initial state configuration the results in this Section provide a reference baseline for future users as well as for further calculations within CLVisc. The Trento Monte Carlo model assumes fluctuations in the transverse plane with a spatial-rapidity-dependent envelop in the longitudinal direction. Therefore, we switch to AMPT initial conditions for the later sections of this manuscript that include also longitidunal initial dynamics. The centrality range is determined by the event-by-event distributions of the total entropy. Initial conditions with top $5\%$ highest total entropies are chosen as $0-5\%$ collisions and so on. The partial chemical equilibrium EoS  s95p-pce \cite{Huovinen:2009yb} is used in the hydrodynamic simulations. The other model parameters for Au+Au $\sqrt{s_{NN}}=200$ GeV, Pb+Pb $\sqrt{s_{NN}}=2.76$ TeV and $\sqrt{s_{NN}}=5.02$ TeV collisions are listed in Tab. \ref{tab:default_parameters},

\begin{table}[htp]
\centering
\begin{ruledtabular}
\begin{tabular}{lllllll}
   system      & $\tau_0$ fm &  norm  & $T_{f}$ MeV &  $\eta_v/s$  &  $\eta_{w}$ & $\sigma_{\eta}$ \\
Au+Au 200 GeV &  0.6         &  57    &  100-137      &   0.15     &  1.3        &   1.5  \\
Pb+Pb 2760 GeV & 0.6         &  128   &  100-137      &   0.15     &  2.0        &   1.8  \\
Pb+Pb 5020 GeV & 0.6         &  151   &  100-137      &   0.15     &  2.2        &   1.8  \\
\end{tabular}
\end{ruledtabular}
\caption{Default parameters for event-by-event hydrodynamics using Trento initial conditions. The normalization is fitted to the hadron multiplicity in the
central rapidity region in the most central heavy-ion collisions.}
\label{tab:default_parameters}
\end{table}
Where $\eta_w$ and $\sigma_{\eta}$ are used to parameterize the initial state longitudinal profile using the following function
\begin{equation}
    H(\eta_s) = \exp\left[ -\frac{(\eta_s - \eta_w)^2}{2\sigma_{\eta}^2} \theta(\eta_s - \eta_w) \right]
    \label{eq:h_eta}
\end{equation}


\subsection{Au+Au at $\sqrt{s_{\rm NN}}$ 200 GeV collisions}
\begin{figure}[!htp]
    \includegraphics[width=0.5\textwidth]{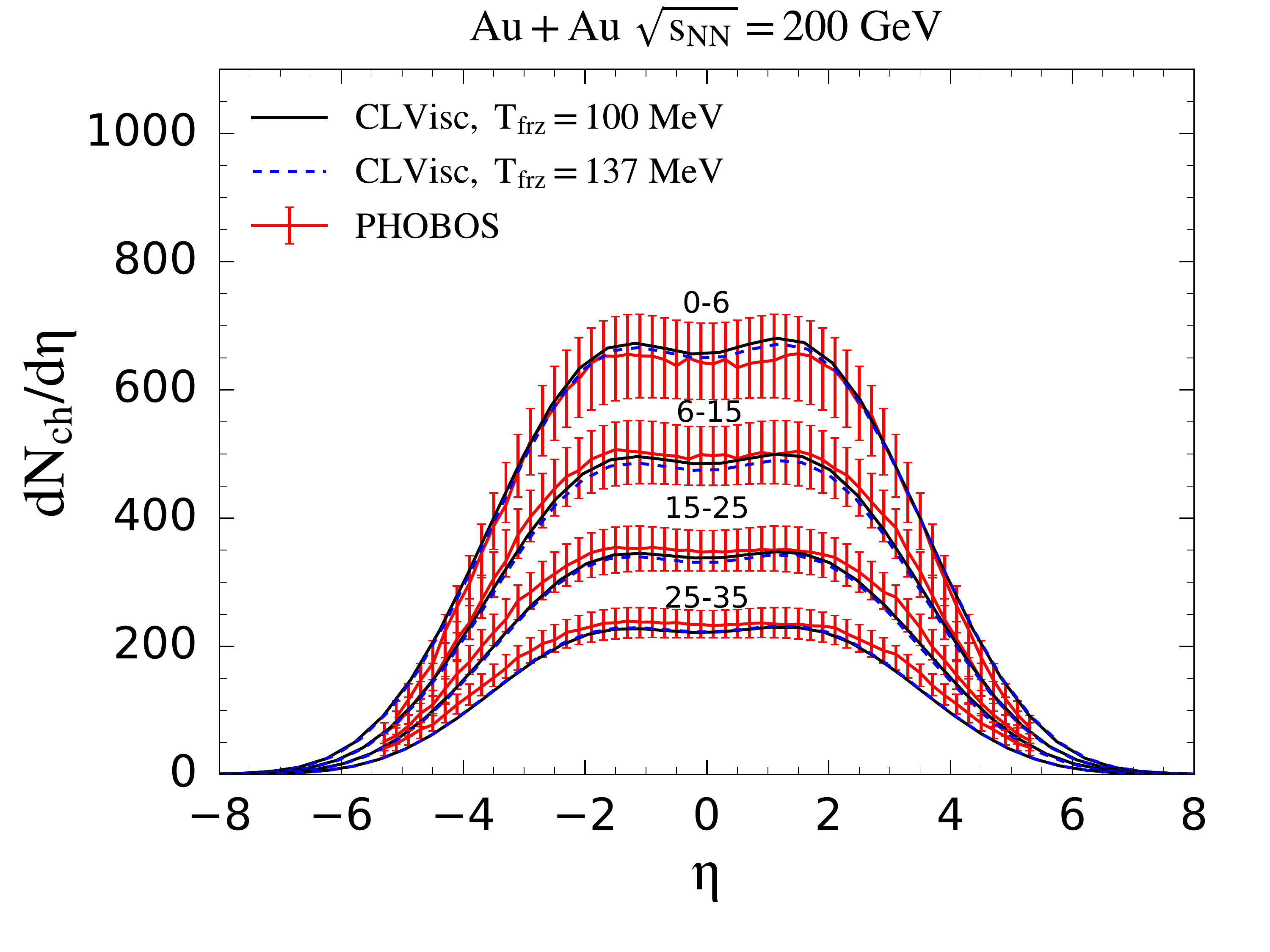}
    \protect\protect\caption{(color online) Pseudo-rapidity distribution for charged hadrons in Au+Au collisions at $\sqrt{s_{NN}}=200$ GeV with centrality range $0-6\%$, $6-15\%$, $15-25\%$ and $25-35\%$, from CLVisc with freeze-out temperature 100 MeV (solid-lines) and 137 MeV (dashed lines) as compared with RHIC experimental data by PHOBOS collaboration \cite{Abbas:2013bpa}.   \label{fig:auau200_dNdEta}}
\end{figure}

\begin{figure}[!htp]
    \includegraphics[width=0.5\textwidth]{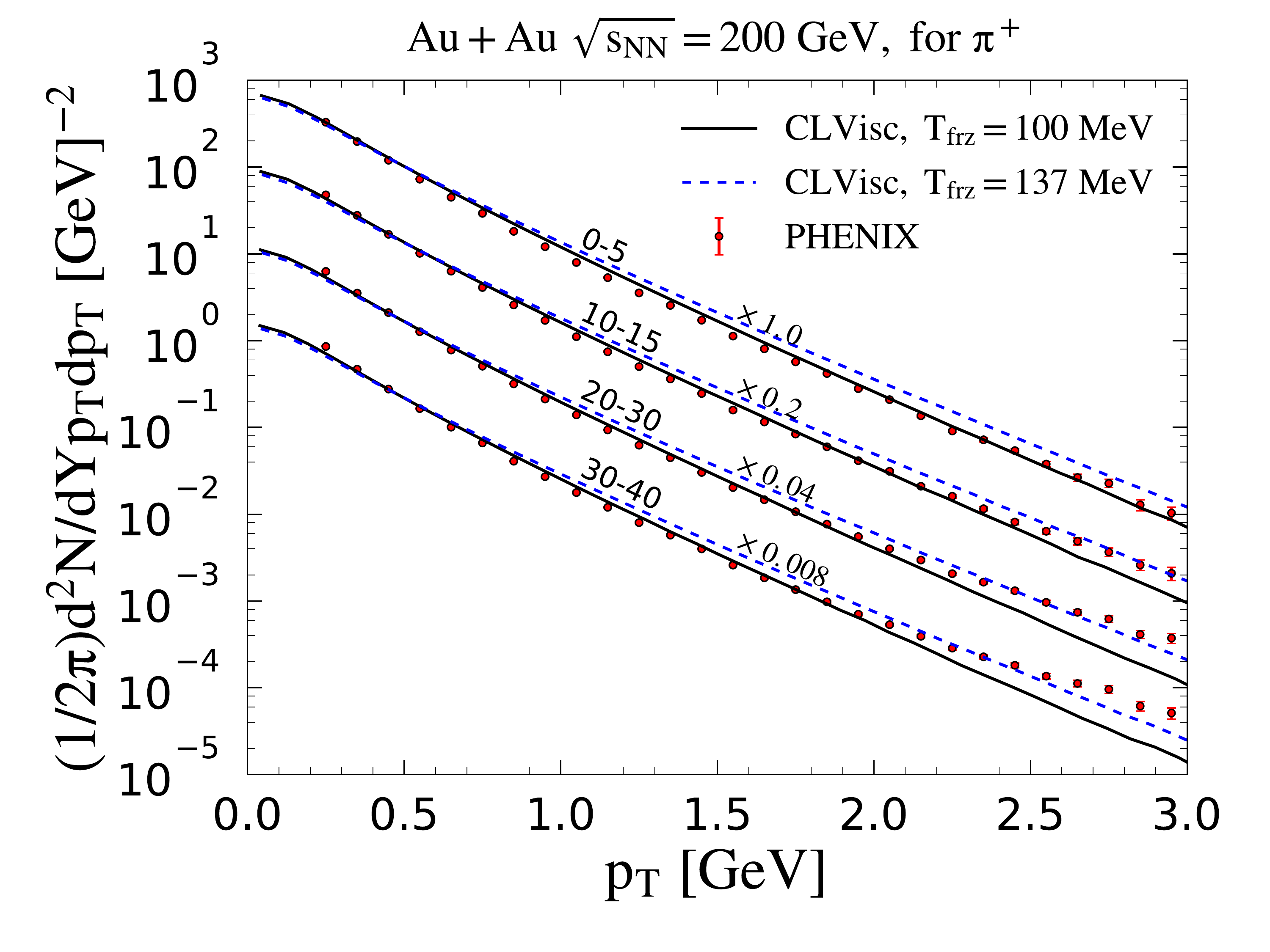}
    \protect\protect\caption{(color online) Invariant yield of $\pi^+$ in Au+Au collisions at $\sqrt{s_{NN}}=200$ GeV with centrality range $0-5\%$, $10-15\%$, $20-30\%$ and $30-40\%$, from CLVisc with freeze-out temperature 100 MeV (solid-lines) and 137 MeV (dashed lines) as compared with RHIC experimental data by PHENIX collaboration.   \label{fig:auau200_ptspec_pion}}
\end{figure}

Shown in Figs.~\ref{fig:auau200_dNdEta} and \ref{fig:auau200_ptspec_pion}
are the pseudo-rapidity distributions for charged hadrons and the transverse momentum spectra for identified particles $\pi^+$. We focus on pion transverse momentum spectra in this section since for pure relativistic hydrodynamic results without considering hadronic after-burner, the transverse momentum spectra of kaon and proton are not expected to agree with experimental data.

We use a constant $\eta_v/s$ in the current CLVisc simulations. It has been shown that the linear relationship between initial entropy and final charged multiplicity breaks down in viscous hydrodynamics with a temperature-dependent $\eta_v/s$ \cite{Niemi:2015qia}. In future studies using Bayesian analysis with temperature-dependent $\eta_v/s$, the centrality classes should be defined by the final state multiplicities after hydrodynamic evolution.

Notice that the pseudo-rapidity distributions for charged hadrons does not change much, if the freeze-out temperature $T_{\mathrm{frz}}$
changes from $137$ MeV to $100$ MeV in CLVisc with partial chemical equilibrium EoS, and the same group of $\tau_0$, normalization factor and $\eta_v/s$.
However, the slope of the pion transverse momentum spectra becomes slightly steeper and describes low $p_\mathrm{T}$ experimental data better with $T_{\mathrm{frz}}=100$ MeV than $T_{\mathrm{frz}}=137$ MeV.
At the same time, the $p_\mathrm{T}$ differential anisotropic flow increases approximately $10\%$ when $T_{\mathrm{frz}}$ is decreased from $137$ MeV to $100$ MeV 
which agrees with the observation in \cite{Song:2010aq}.
In order to get the best global fit to many different observables, a Bayesian analysis \cite{Pratt:2015zsa,Bernhard:2015hxa,Bernhard:2016tnd} has to be employed
to explore the huge parameter space. Mini-jets and their thermalization will also play a role in the transverse momentum spectra at high $p_T>2$ GeV/$c$.

%

\subsection{Pb+Pb at $\sqrt{s_{\rm NN}}$ = 2760 GeV collisions}

\begin{figure}[!htp]
    \includegraphics[width=0.5\textwidth]{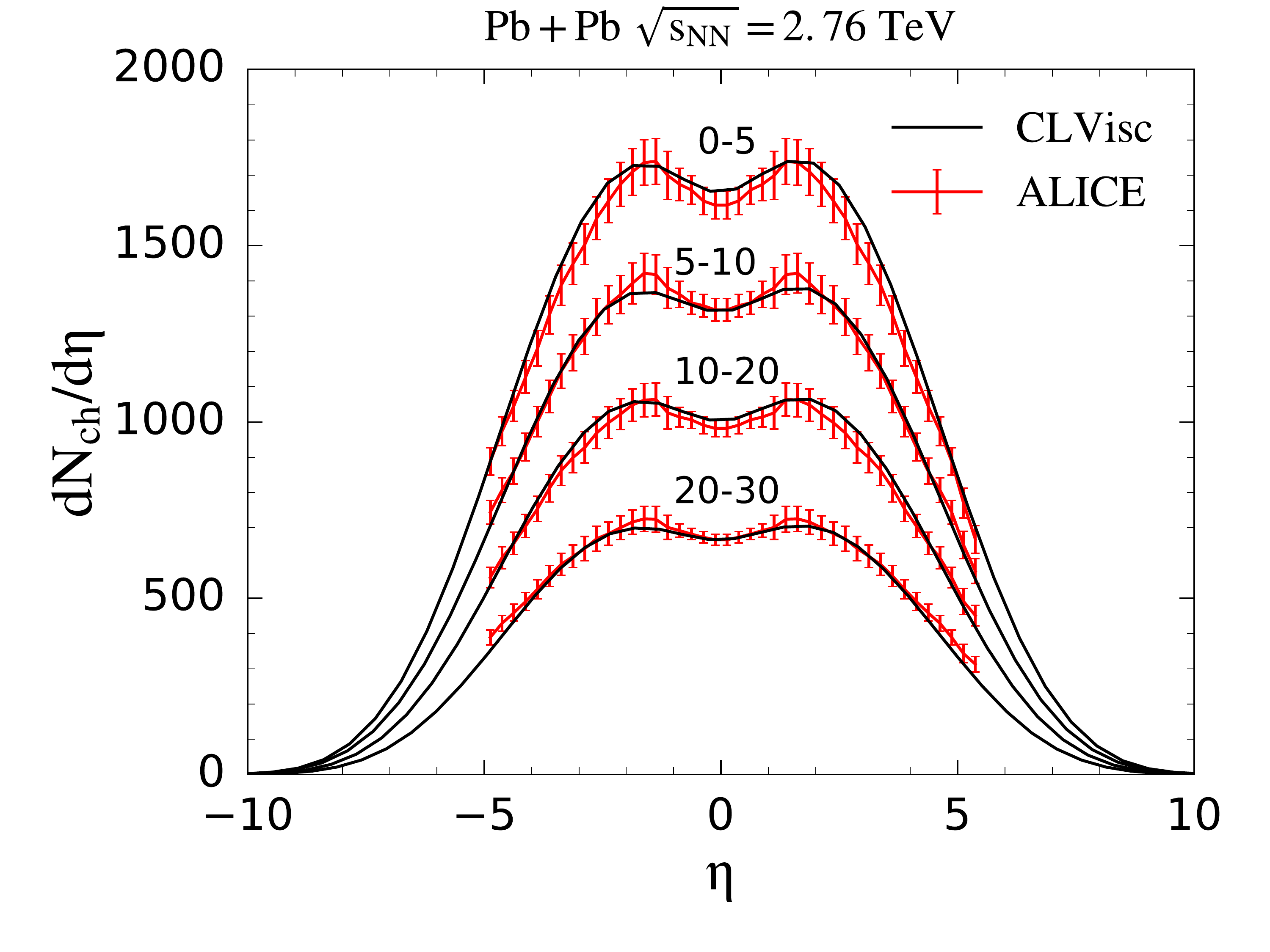}
    \protect\protect\caption{(color online) Pseudo-rapidity distribution for charged hadrons in Pb+Pb collisions at $\sqrt{s_{NN}}=2.76$ TeV with centrality range $0-5\%$, $5-10\%$, $10-20\%$ and $20-30\%$, from CLVisc (solid-lines) and LHC experimental data by ALICE collaboration \cite{Abbas:2013bpa}.   \label{fig:pbpb2760_dNdEta}}
\end{figure}

Shown in Fig. \ref{fig:pbpb2760_dNdEta} are pseudo-rapidity distributions for charged hadrons in Pb+Pb collisions at $\sqrt{s_{NN}}=2.76$ TeV for 4 different centralities -- $0-5\%$, $5-10\%$, $10-20\%$ and $20-30\%$. The centrality dependence of the event-averaged charged multiplicity is determined by event-by-event distributions of initial total entropy. A freeze-out temperature of $T_{\mathrm{frz}}=100$ MeV is used in the CLVisc simulations. Nice agreement with experimental data on the pseudo-rapidity distribution of charged particles is found over a wide range of centralities.

\begin{figure}[!htp]
    \includegraphics[width=0.5\textwidth]{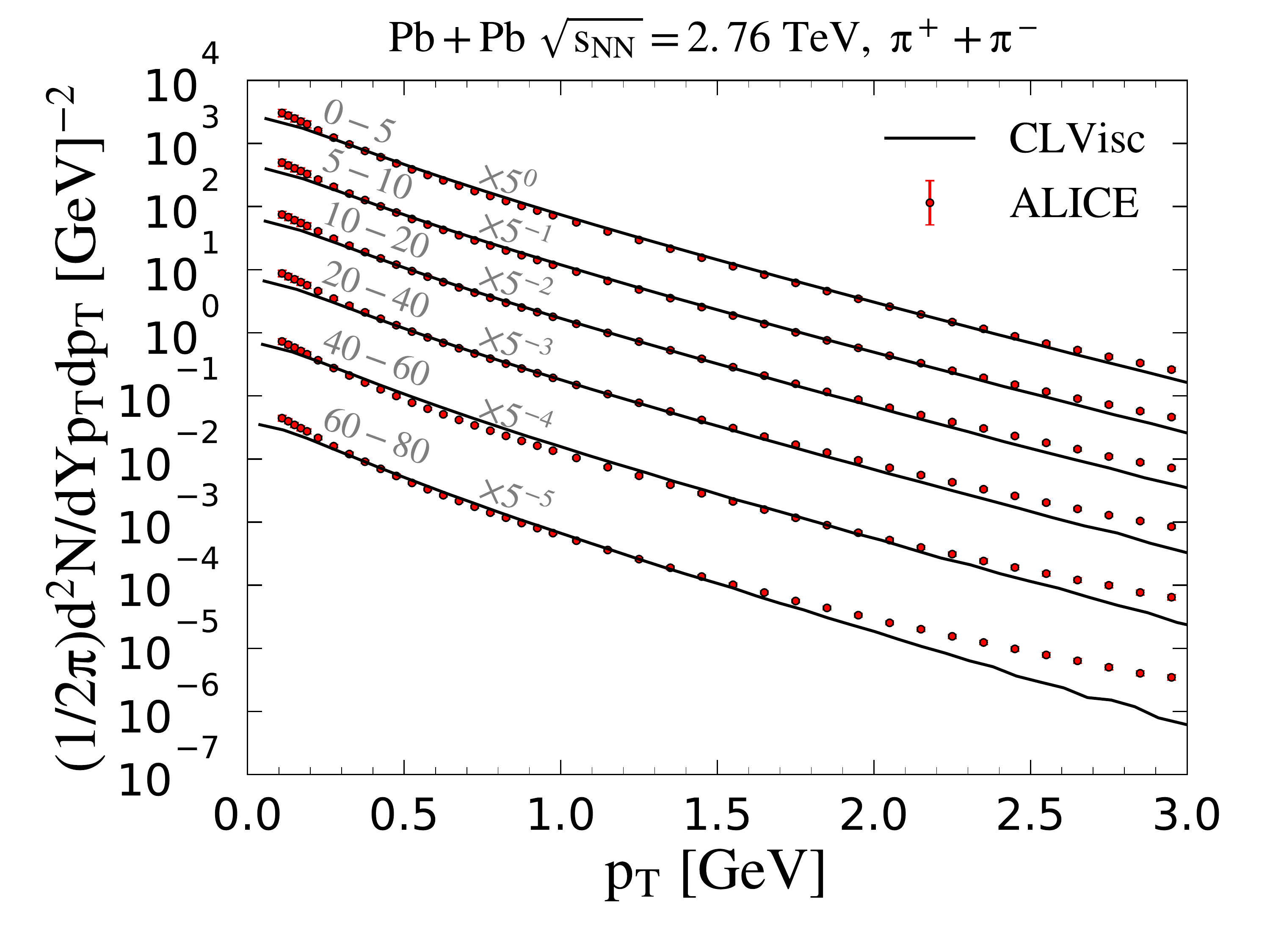}
    \protect\protect\caption{(color online) $p_T$ spectra of charged pions for Pb+Pb $\sqrt{s_{NN}}=2.76$ TeV collisions at centrality range $0-5\%$, $5-10\%$,  $10-20\%$,  $20-40\%$,  $40-60\%$,  $60-80\%$,  from CLVisc (solid-lines) and LHC experimental data by ALICE collaboration \cite{Adam:2015kca}.
    \label{fig:pbpb2760_ptspec_pion}}
\end{figure}

%

Shown in Fig. \ref{fig:pbpb2760_ptspec_pion}, 
is the transverse momentum spectra for charged pions, 
in 6 different centralities of collisions, which agree with experimental data well. The hydrodynamic simulations always underestimate low $p_T$ pions as compared to the experimental data at LHC. This problem is not solved up to date, but may be partially explained by the missing finite widths of resonances \cite{Huovinen:2016xxq} in the current hadronization modules.

\subsection{Higher order harmonic flow in Pb+Pb at $\sqrt{s_{\rm NN}}$ = 2760 GeV collisions}

\begin{figure*}[!htp]
    \includegraphics[width=0.32\textwidth]{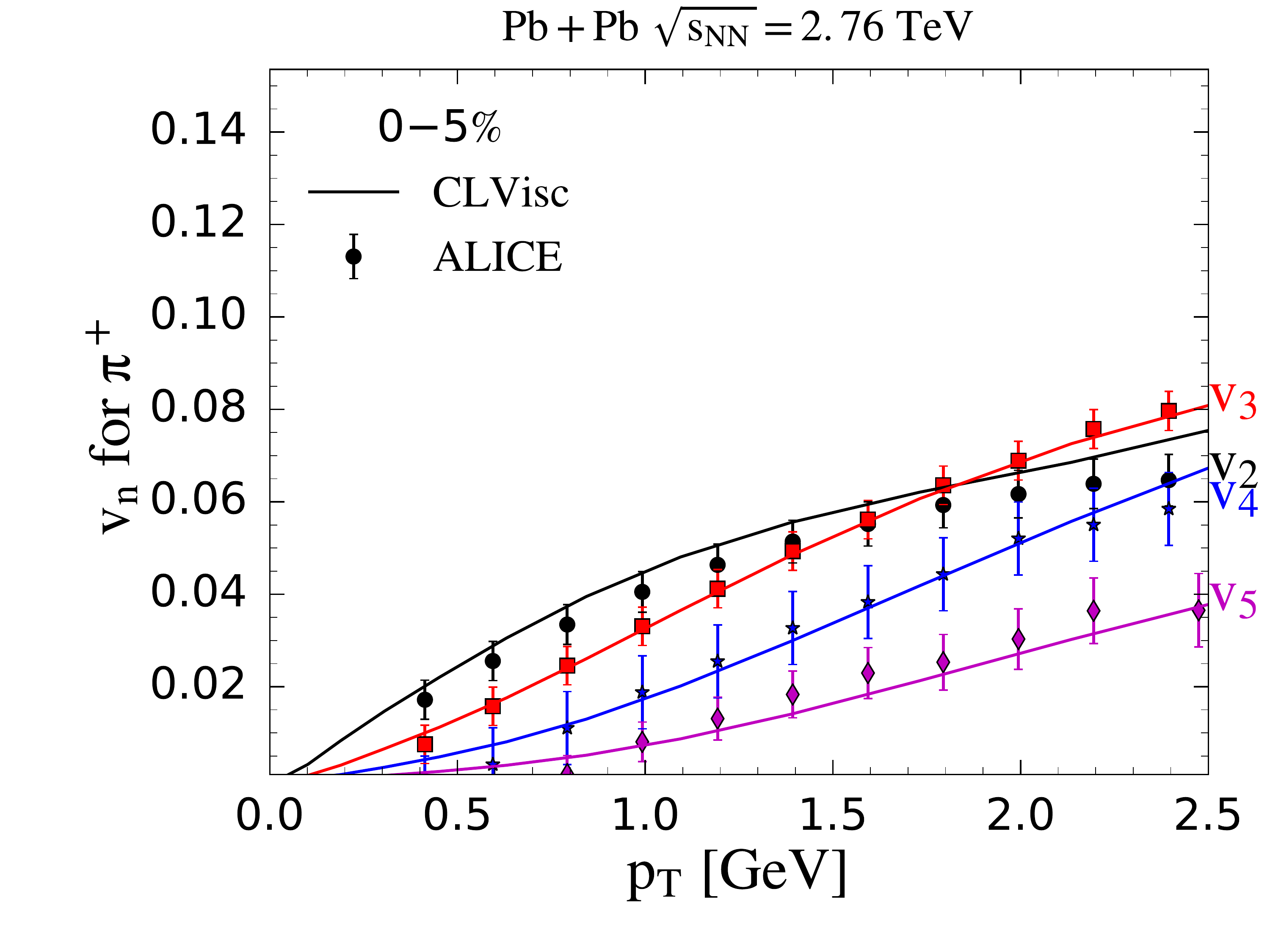} \includegraphics[width=0.32\textwidth]{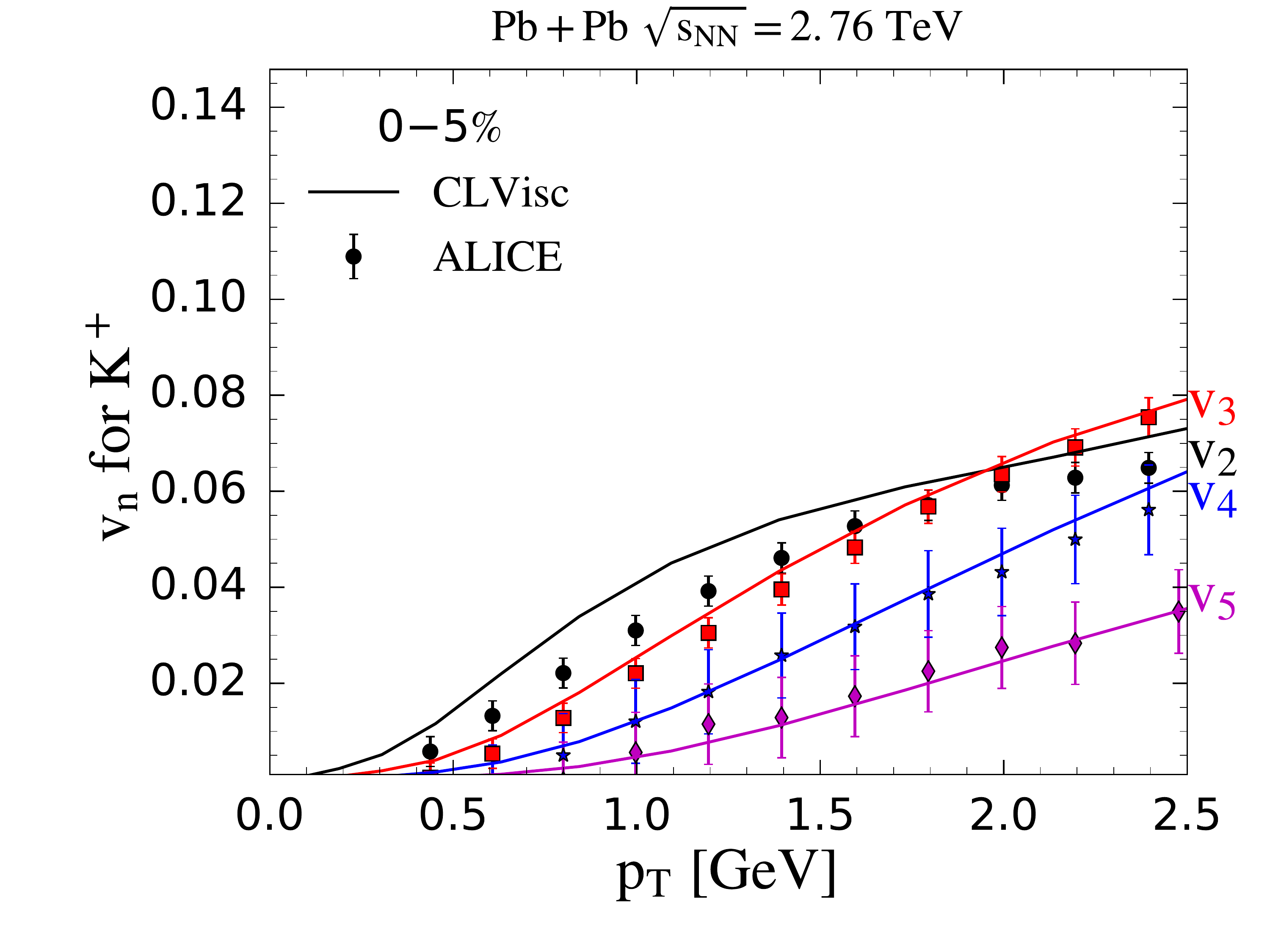} \includegraphics[width=0.32\textwidth]{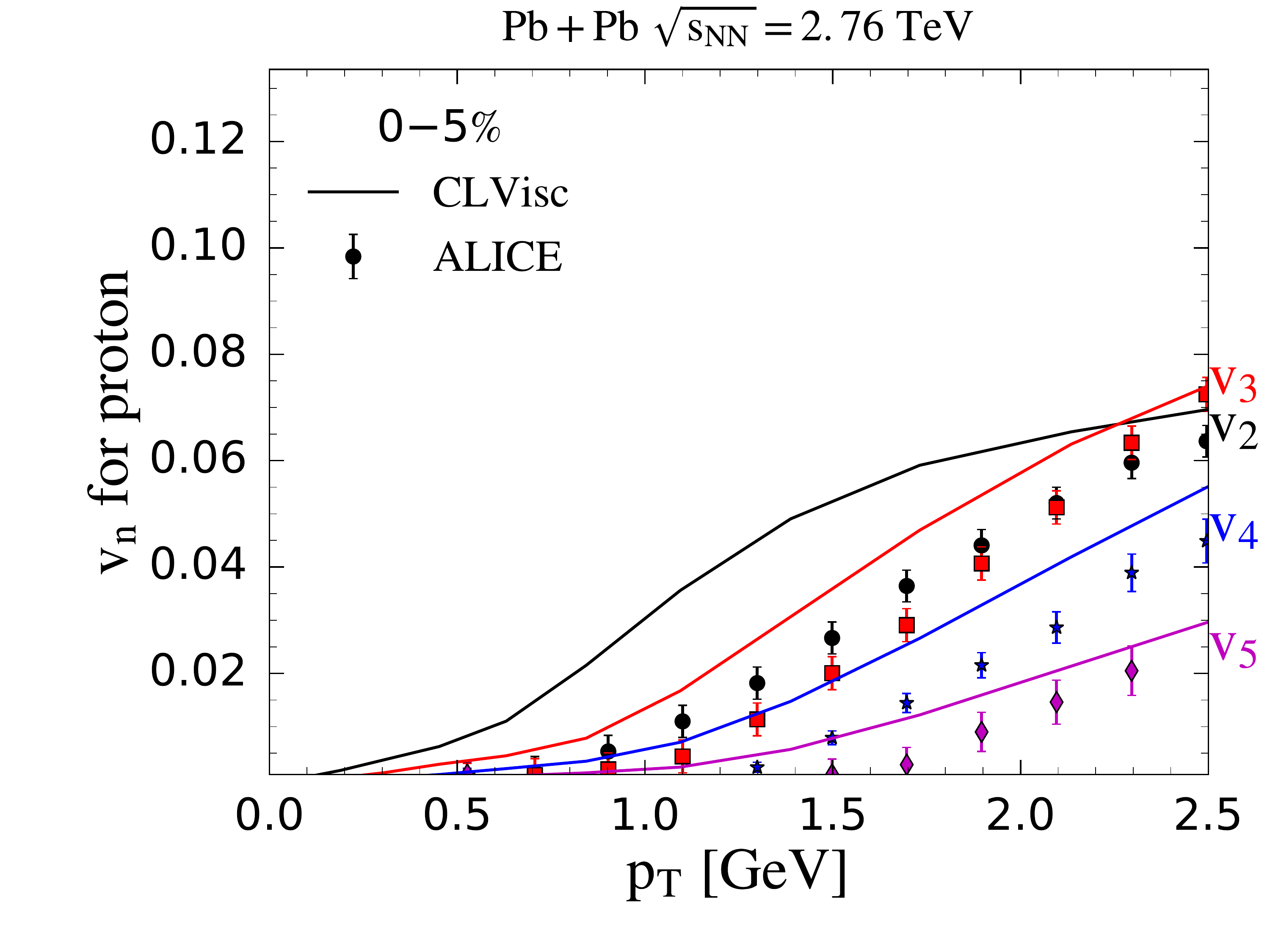} \\
    \includegraphics[width=0.32\textwidth]{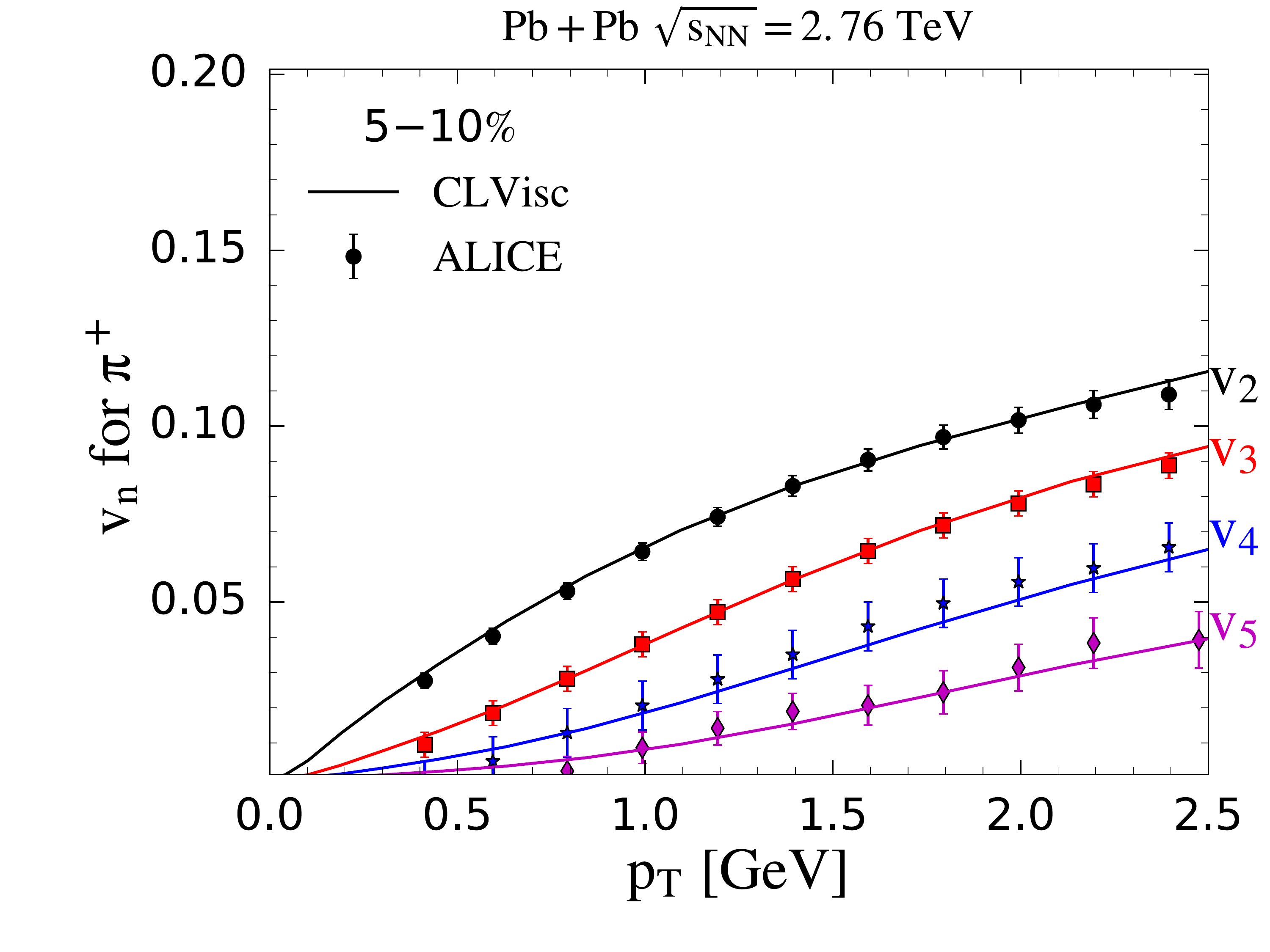} \includegraphics[width=0.32\textwidth]{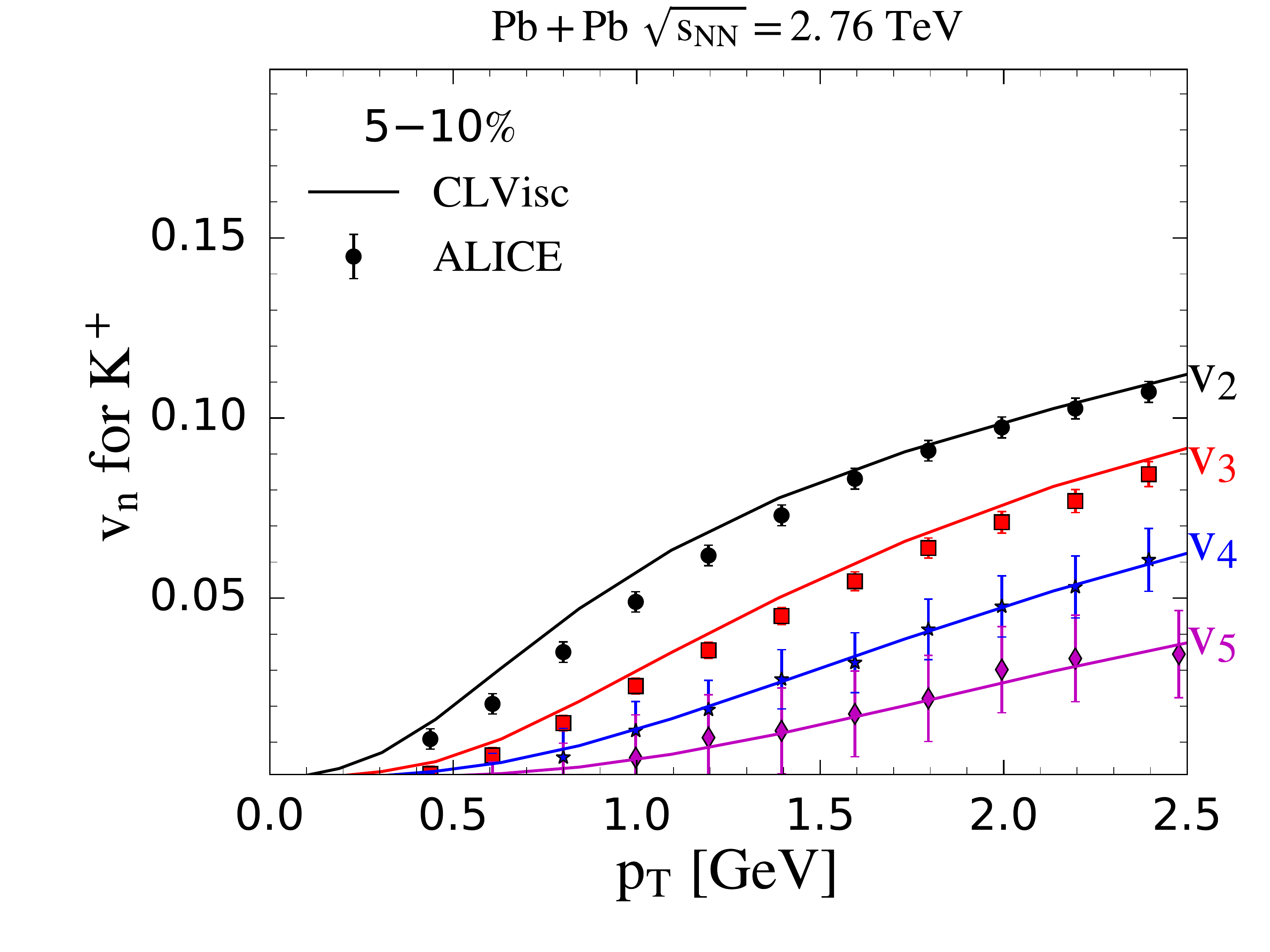} \includegraphics[width=0.32\textwidth]{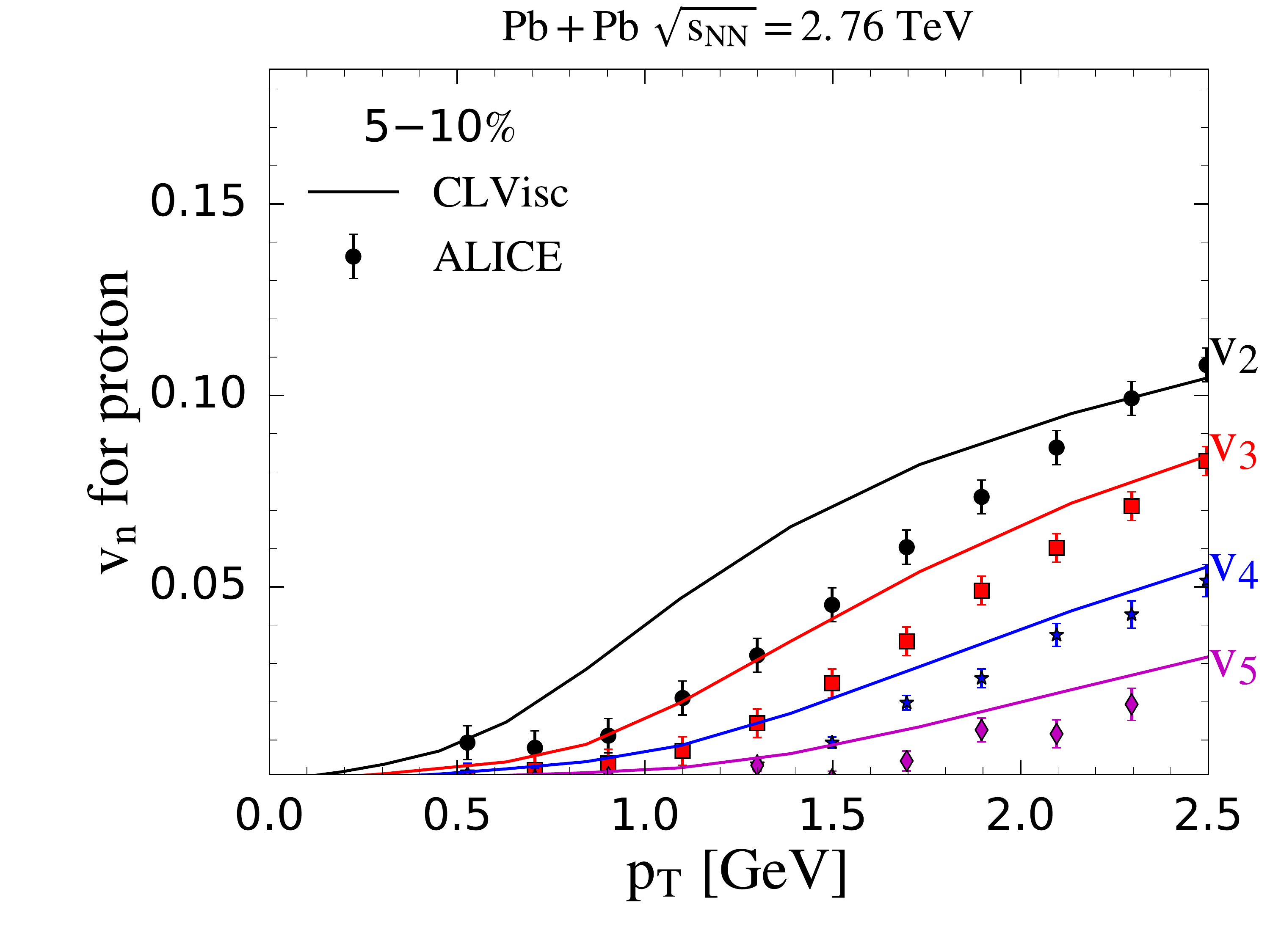} \\
    \includegraphics[width=0.32\textwidth]{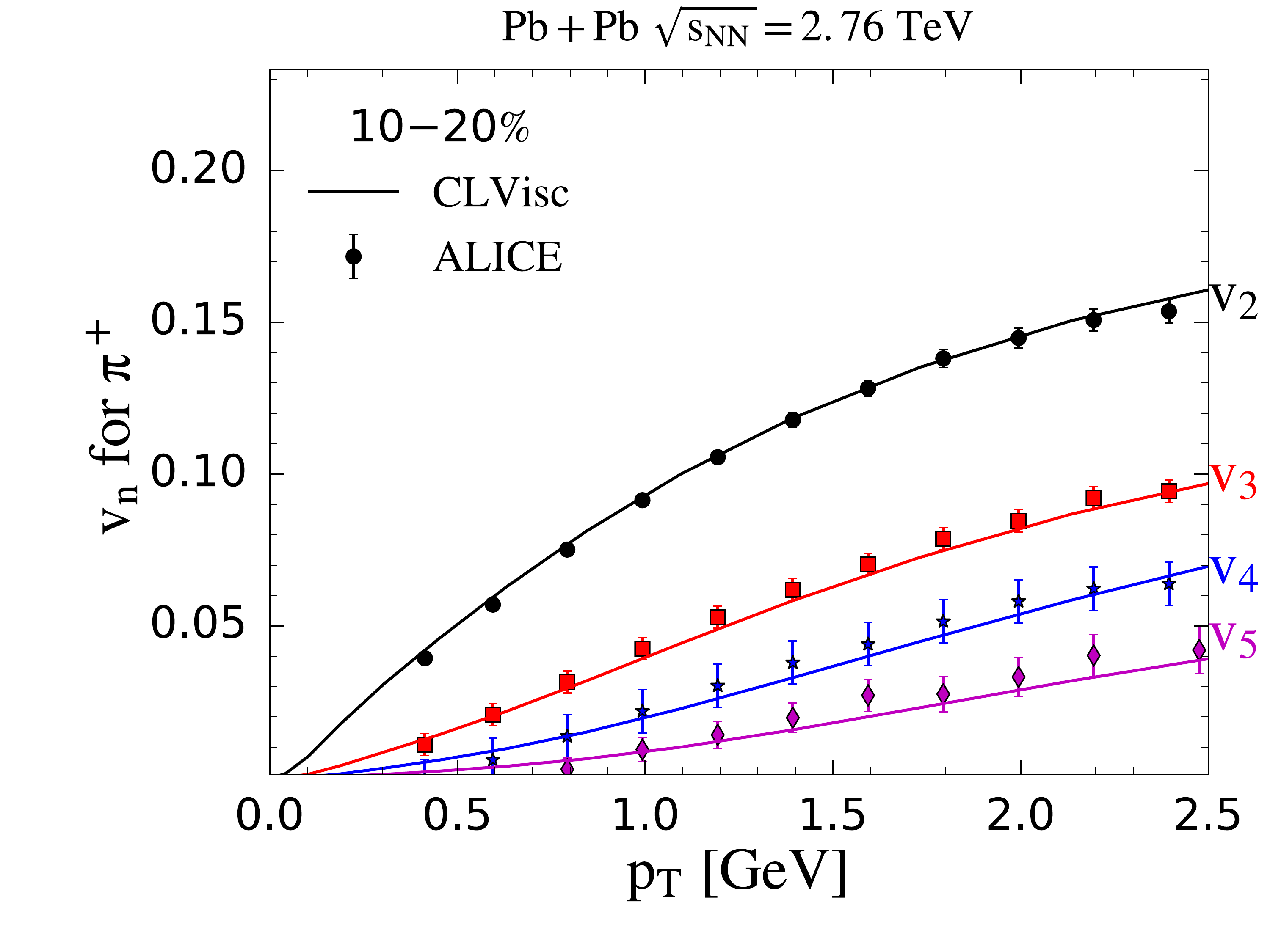} \includegraphics[width=0.32\textwidth]{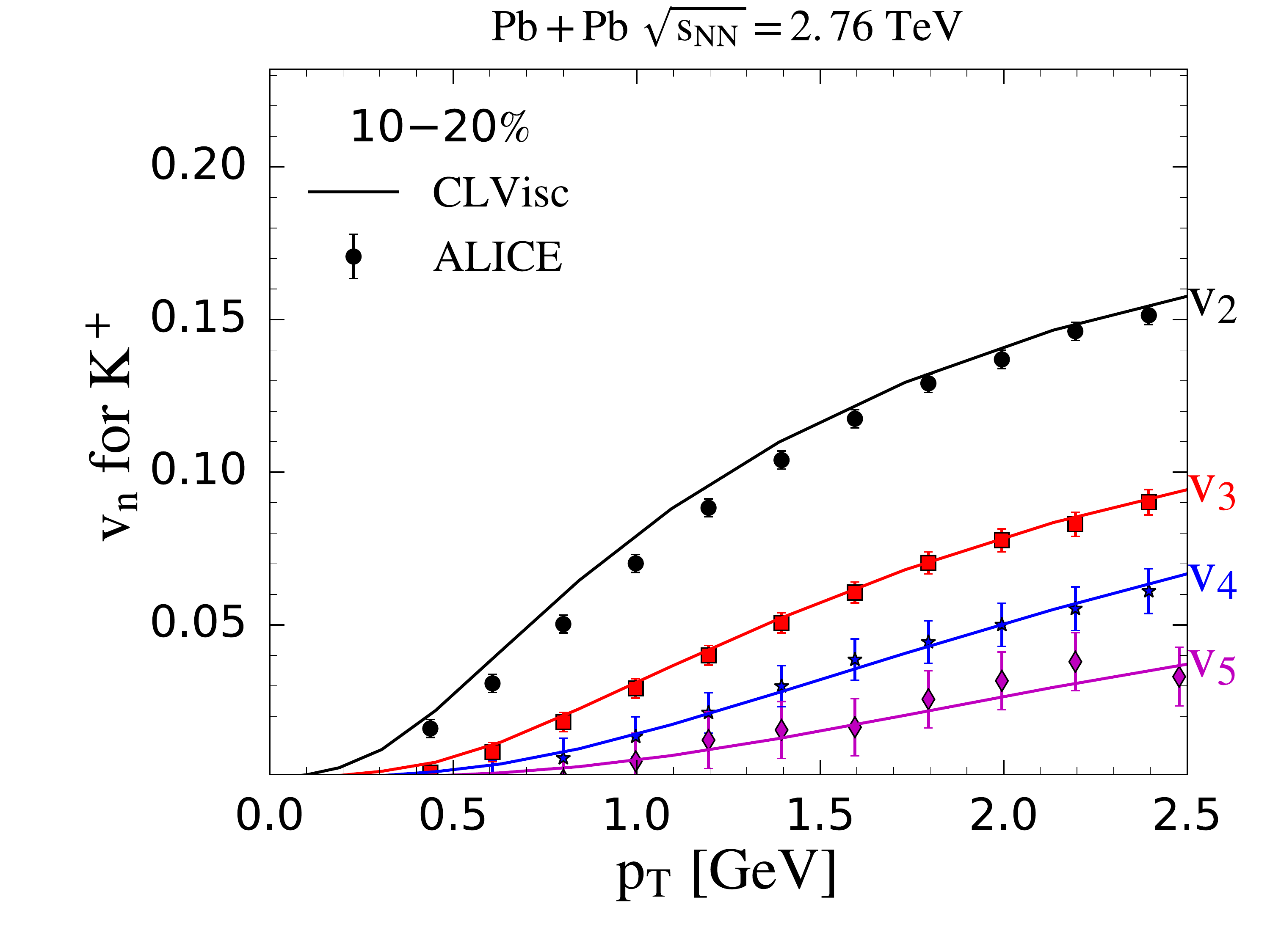} \includegraphics[width=0.32\textwidth]{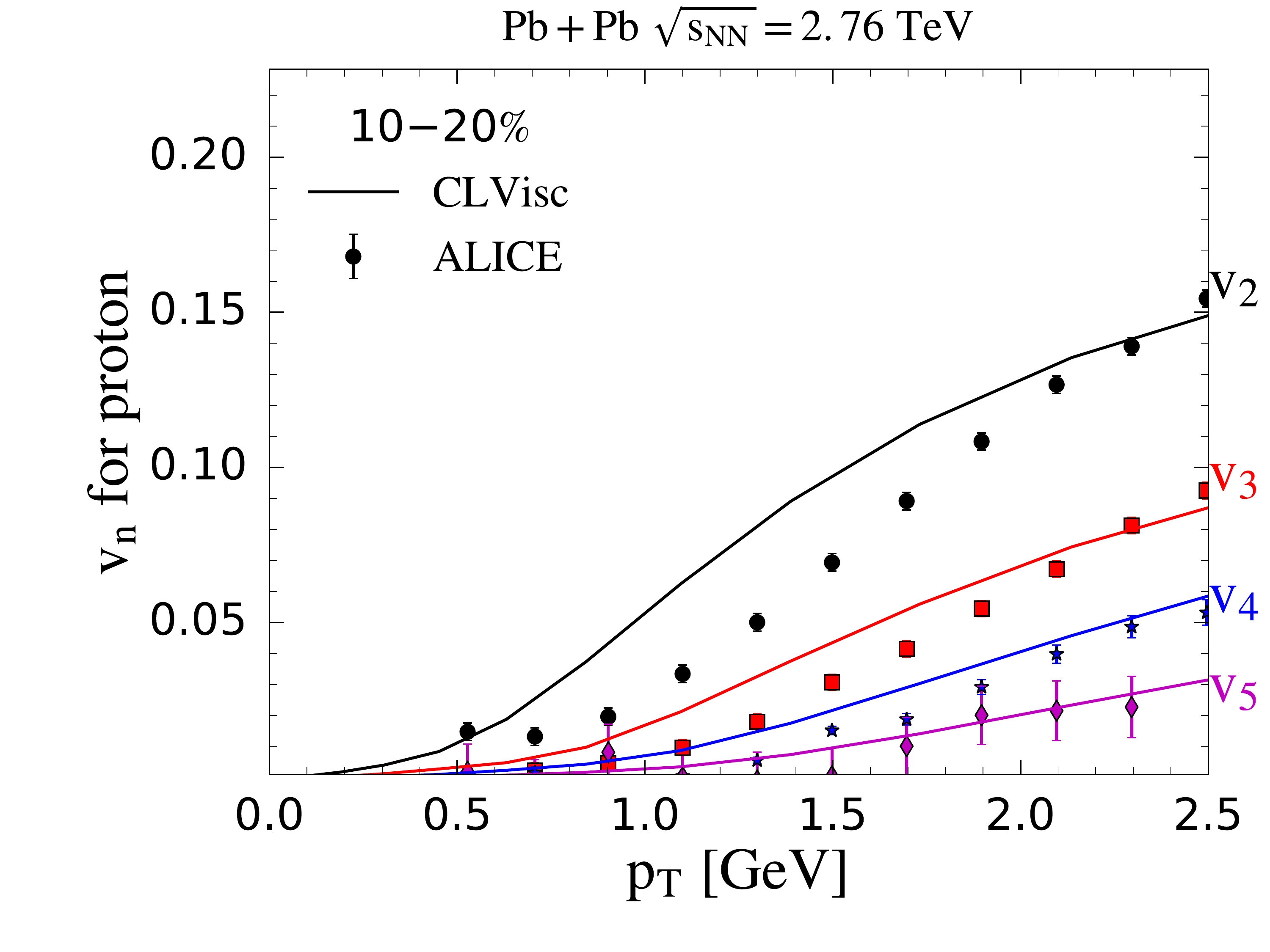} \\
    \includegraphics[width=0.32\textwidth]{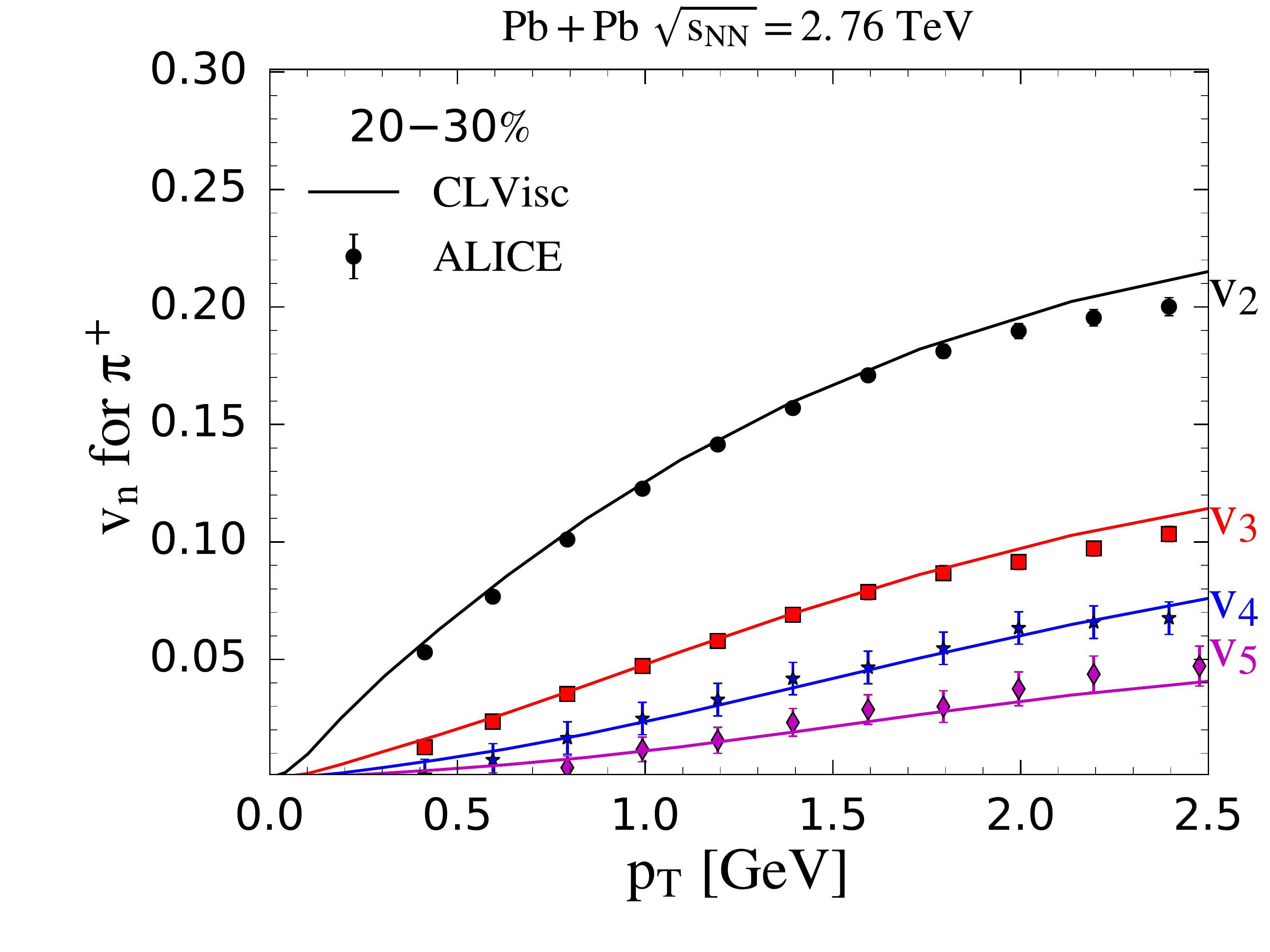} \includegraphics[width=0.32\textwidth]{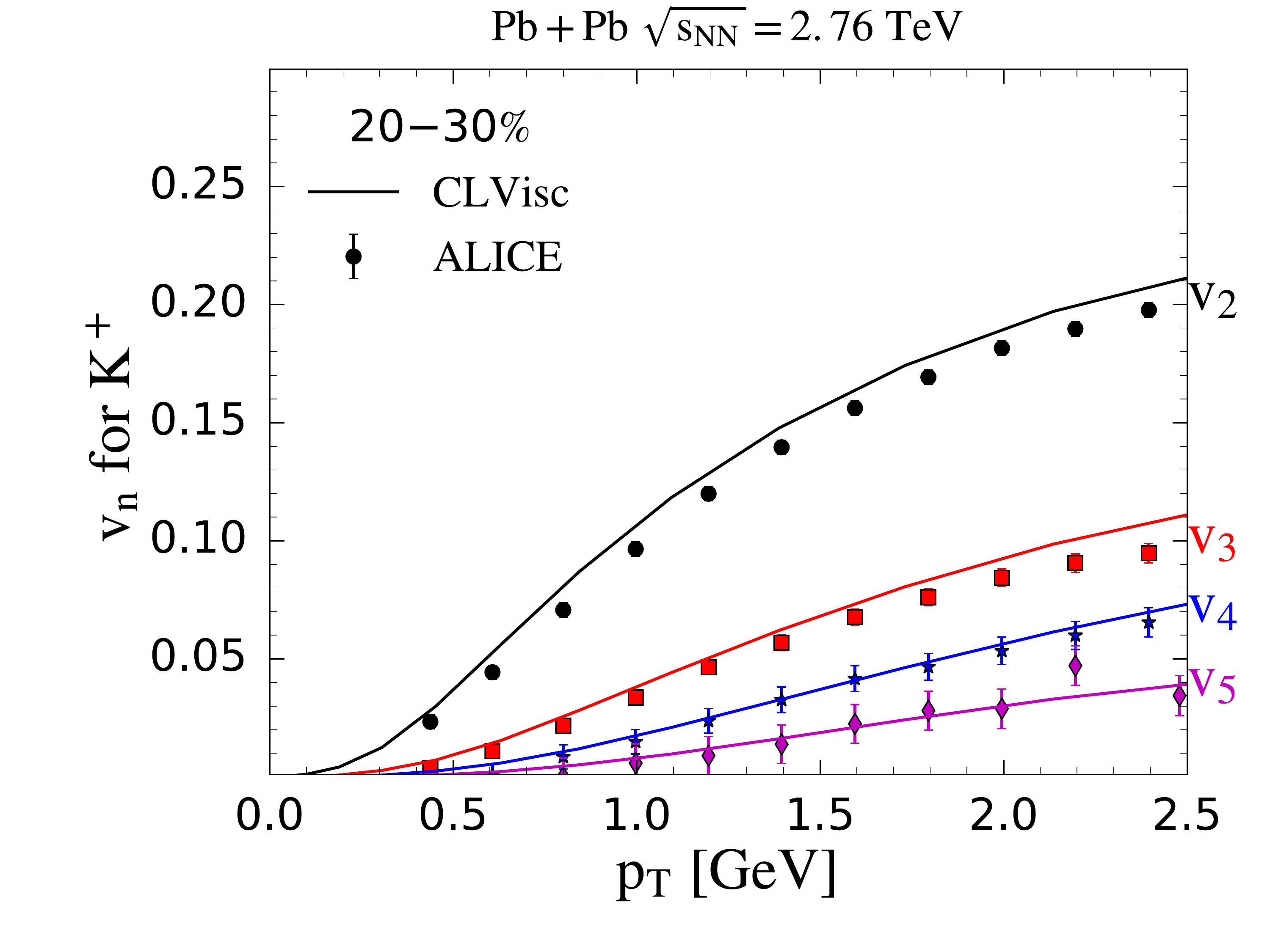} \includegraphics[width=0.32\textwidth]{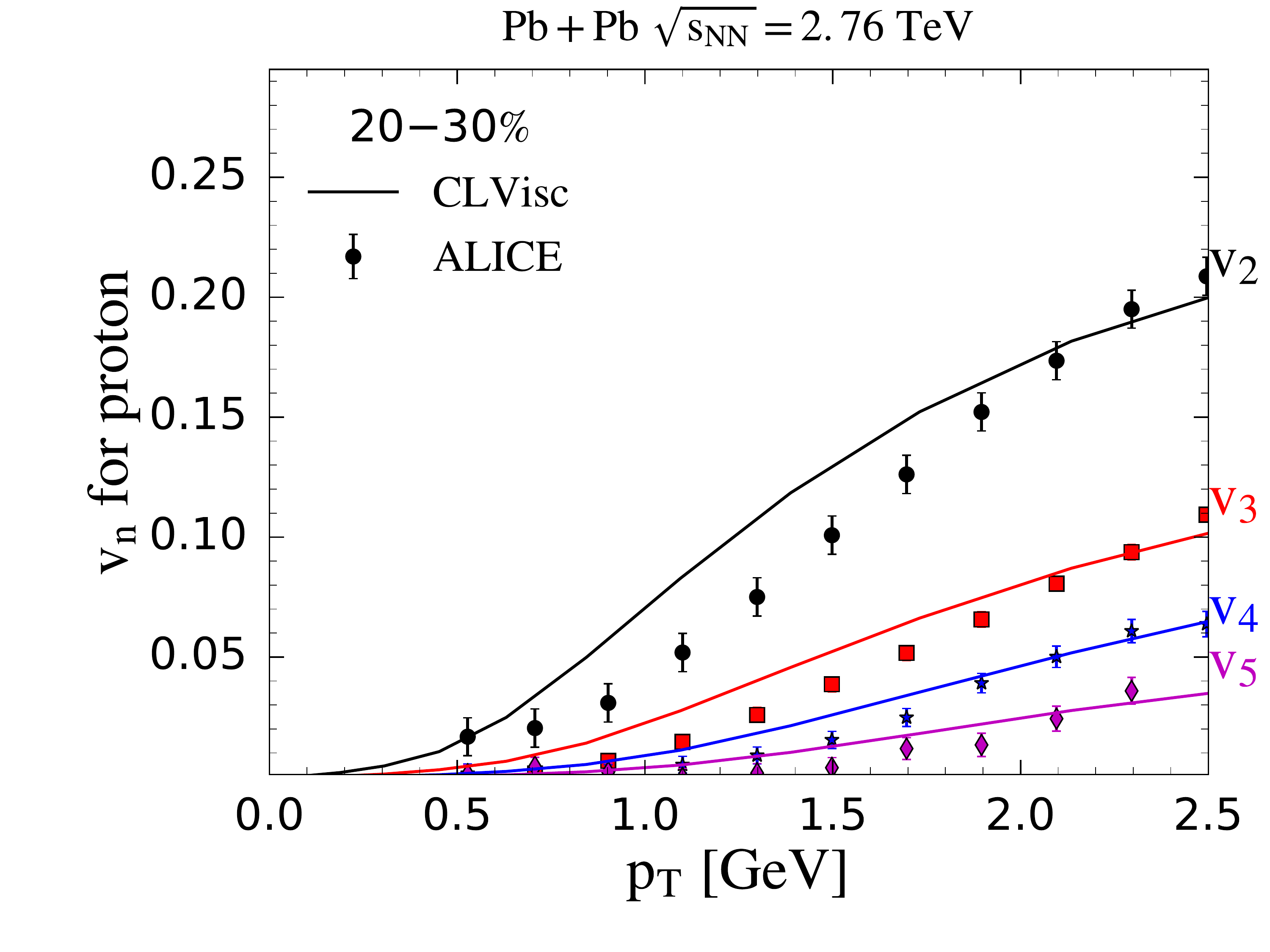}
    \protect\protect\caption{(color online) The centrality dependence of the anisotropic flows $v_2$, $v_3$, $v_4$ and $v_5$ from scalar-product method in Pb+Pb collisions at $\sqrt{s_{NN}}=2.76$ TeV with centrality ranges $0-5\%$, $5-10\%$, $10-20\%$ and $20-30\%$, from CLVisc (solid-lines) and LHC experimental data (markers) by ALICE collaboration \cite{Adam:2016nfo}.   \label{fig:pbpb2760_vn}}
\end{figure*}

CLVisc with Trento initial conditions and $T_{\mathrm{f}}=137$ MeV can reproduce experimental data on $v_2$, $v_3$, $v_4$ and $v_5$ for charged pions for all
available centralities as shown in Fig.~\ref{fig:pbpb2760_vn}. For pure relativistic hydrodynamic simulations without hadronic after-burner,
the $v_n$'s from CLVisc overshoot the experimental data by $5\%$ for $K^+$ and a large margin for protons. It has been shown that the $p_T$ differential elliptic flow of kaon and protons are boosted to higher $p_T$ in hydro-transport hybrid models by hadronic rescattering \cite{Song:2010aq}. On the other hand, the pion $v_n(p_T)$ is not very sensitive to hadronic afterburner and serves as a good measure of the QGP expansion. The consistency of freeze-out temperature best fitted to the transverse momentum spectra (100 MeV) and transverse momentum differential anisotropic flow (137 MeV) can also be resolved by matching hydrodynamic models with hadronic transport evolution in the final stage which will contribute to the further development of anisotropic flow. The range of freeze-out temperatures could also be used as a prior for Bayesian analysis. 

\section{The pseudo-rapidity dependence of anisotropic flow}
\label{sec:vneta}

To study the pseudo-rapidity dependence of anisotropic flow $v_2\{2\}$ and $v_3\{2\}$ of charged hadrons in this section and the longitudinal fluctuation and correlation in the next section, we need realistic and fluctuating longitudinal distributions of the initial entropy density.  For this purpose, the AMPT model is employed to generate event-by-event initial conditions 
that fluctuate both in the transverse plane and along the longitudinal direction. Notice that the $v_n\{2\}$ in this section are given by 2-particle cumulants method using sampled hadrons while the $v_n(p_T)$ in the previous section are given by scalar product method using smooth particle spectra.

\begin{figure*}[!htp]
    \includegraphics[width=0.33\textwidth]{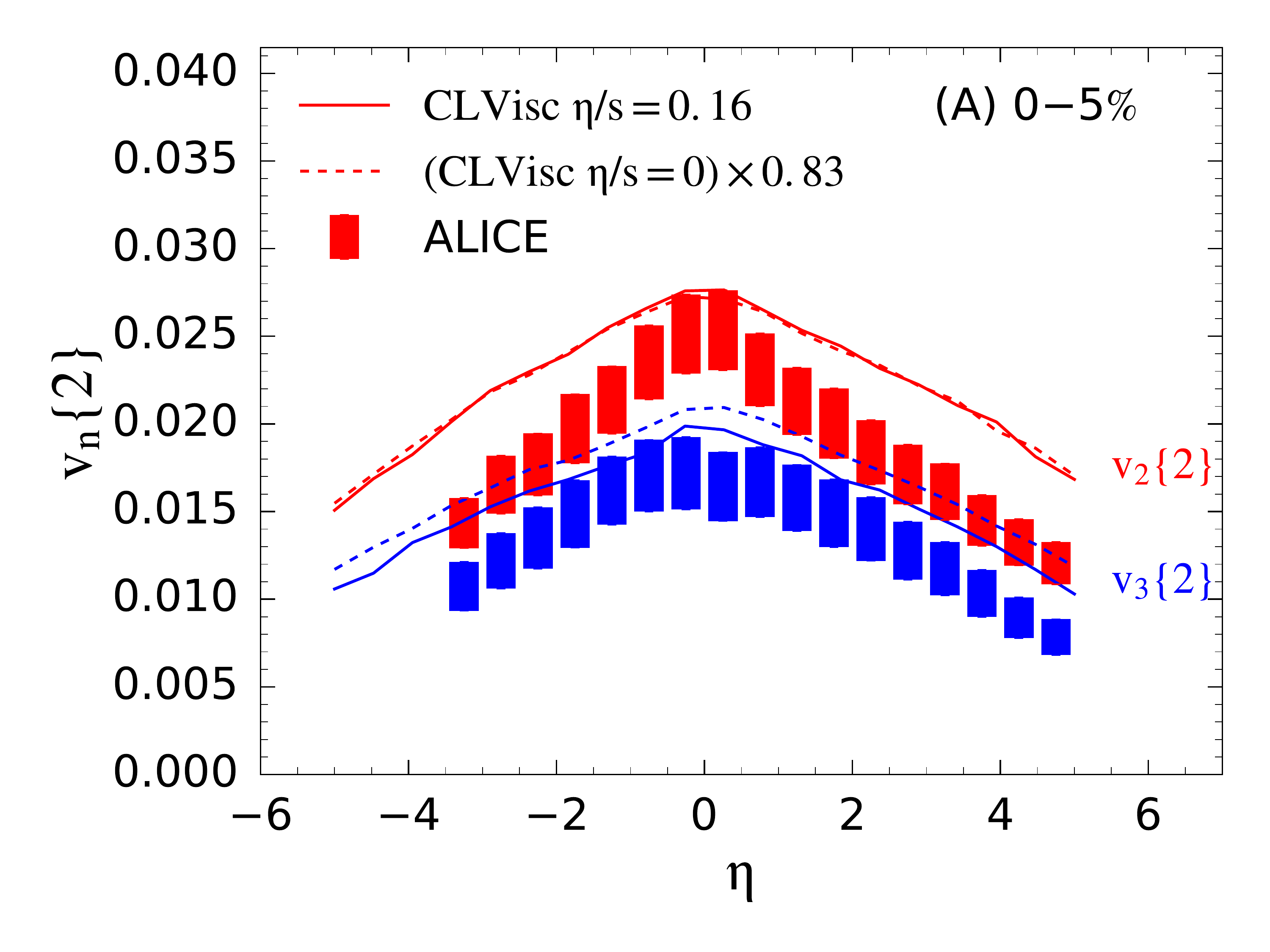} \includegraphics[width=0.33\textwidth]{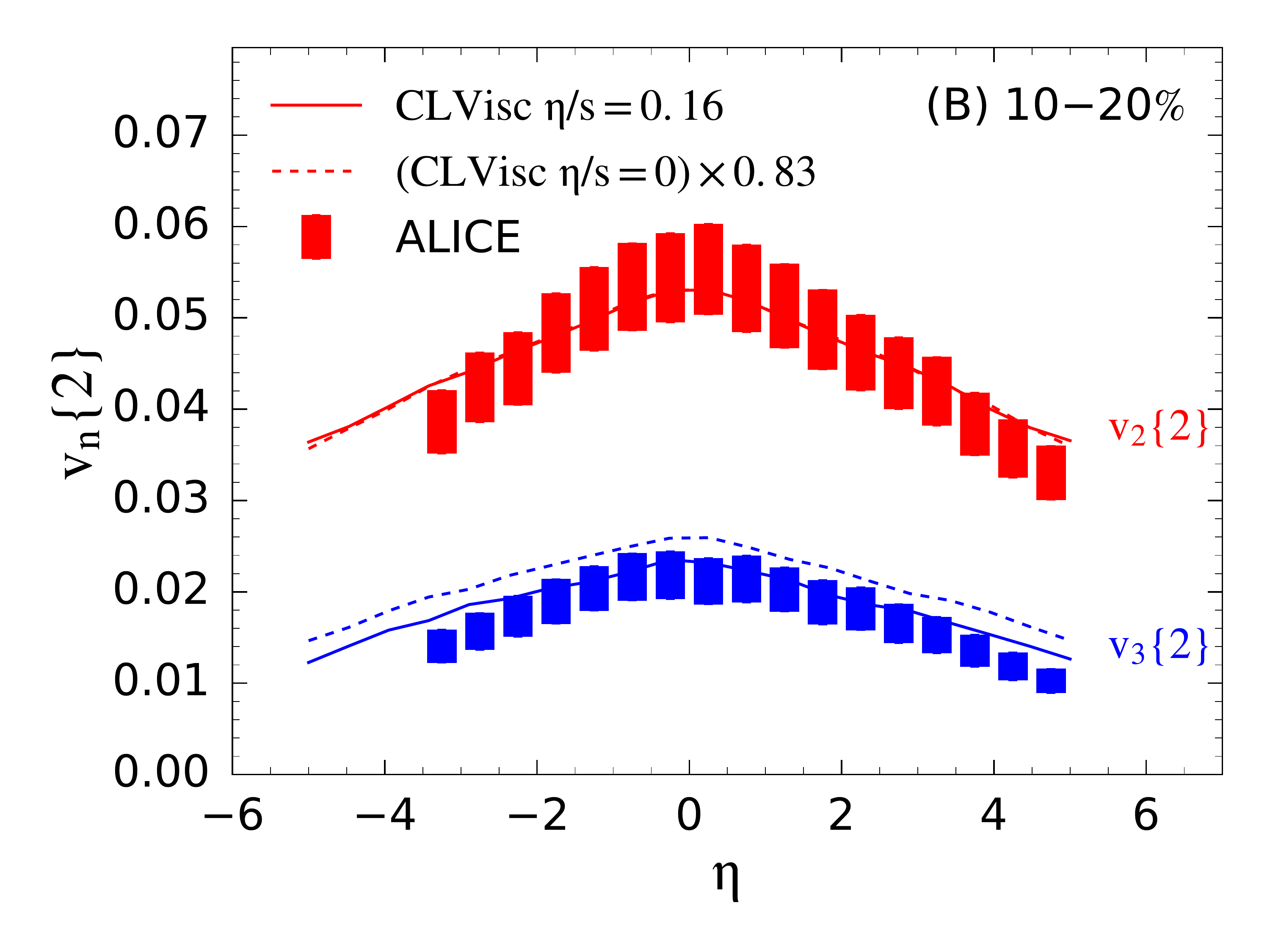}\includegraphics[width=0.33\textwidth]{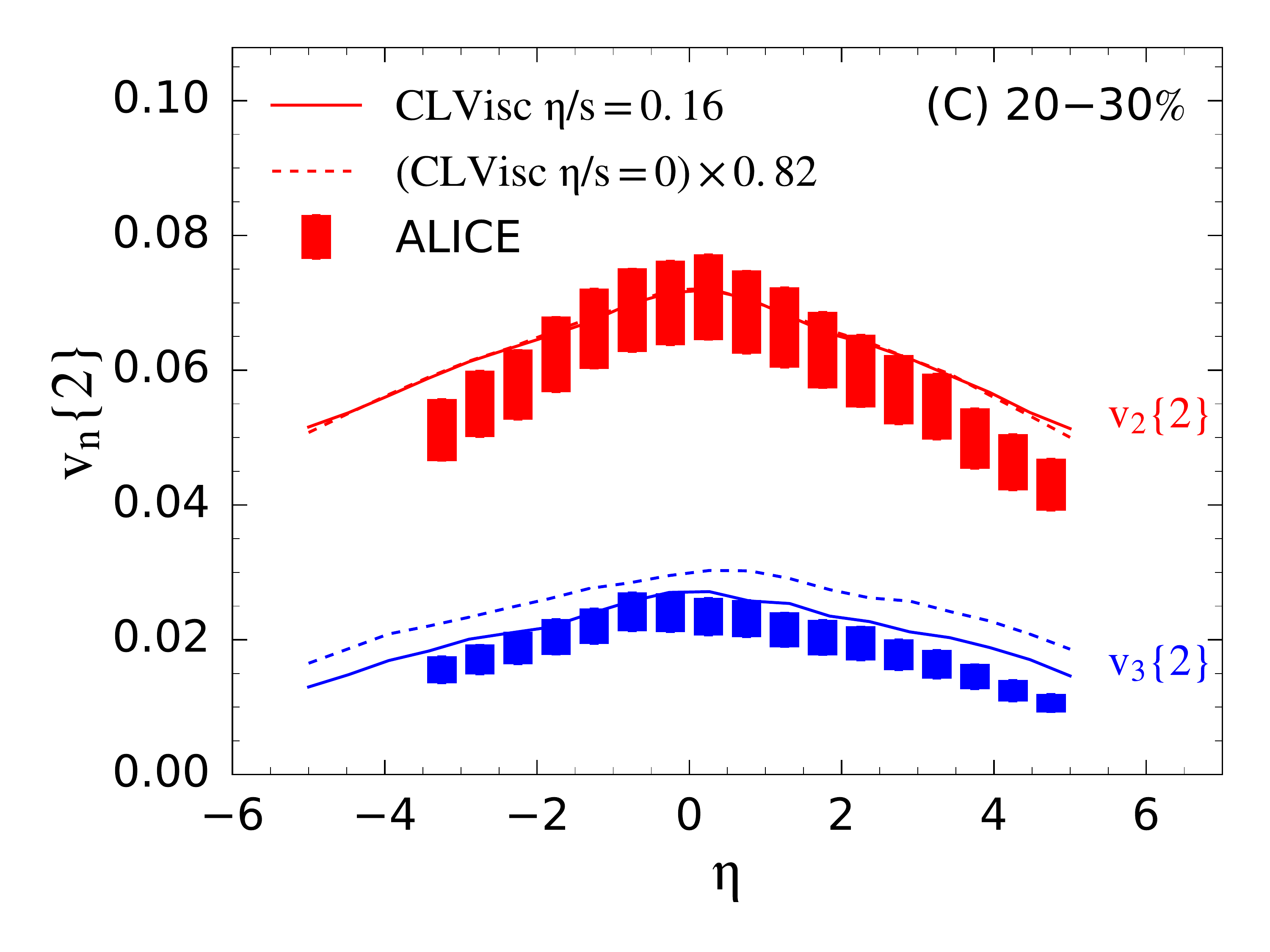} \\
    \includegraphics[width=0.33\textwidth]{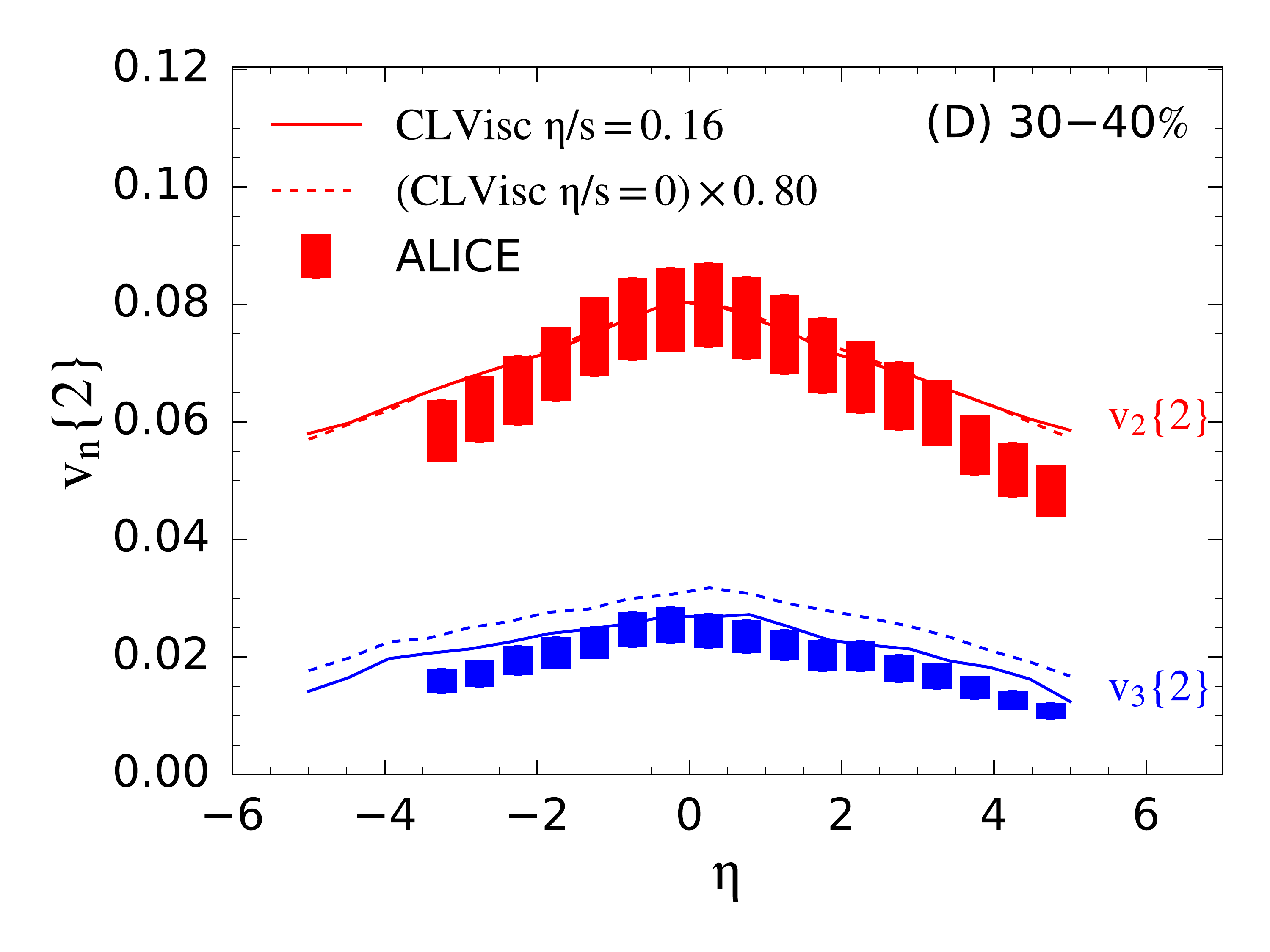} \includegraphics[width=0.33\textwidth]{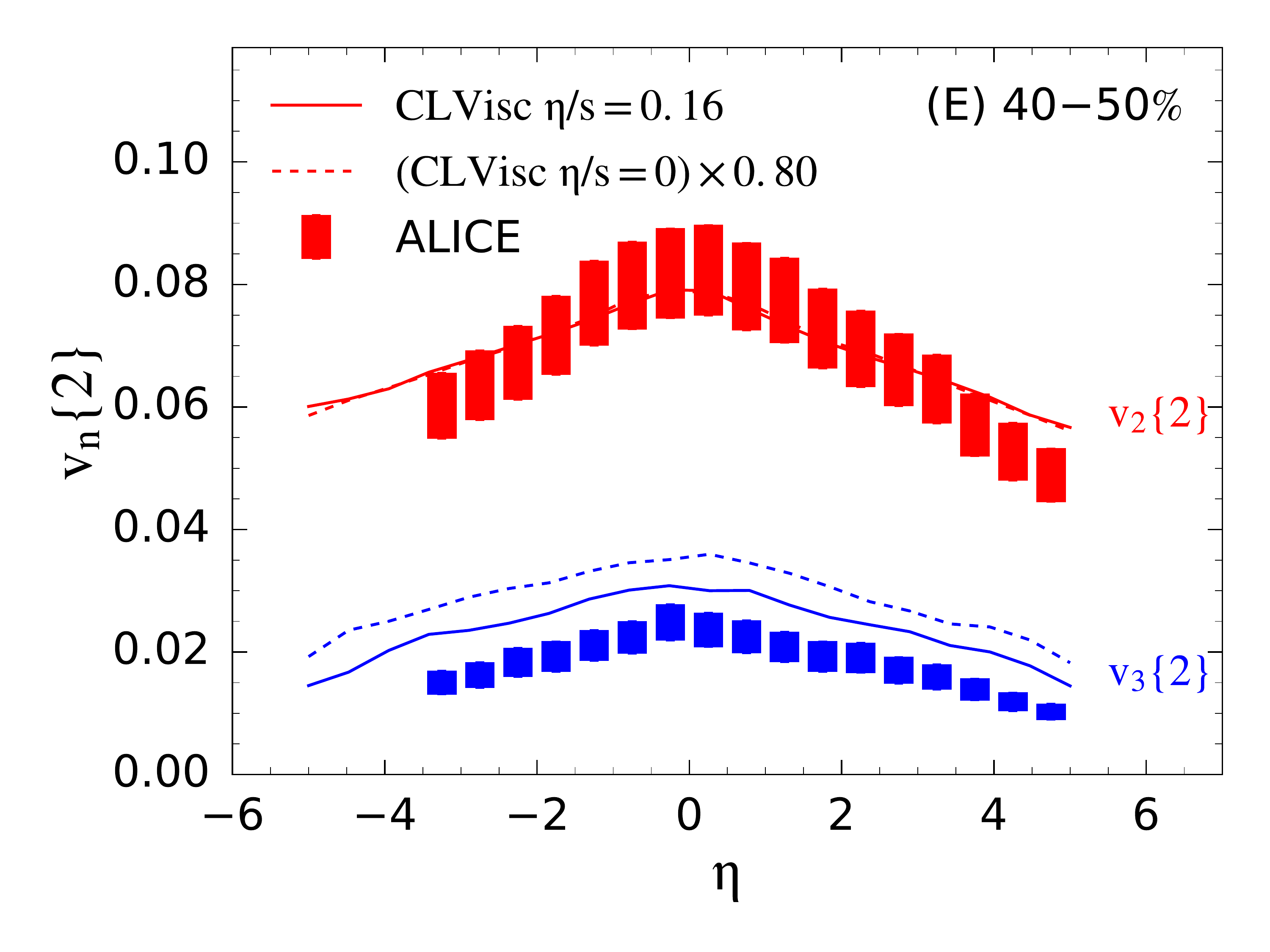}\includegraphics[width=0.33\textwidth]{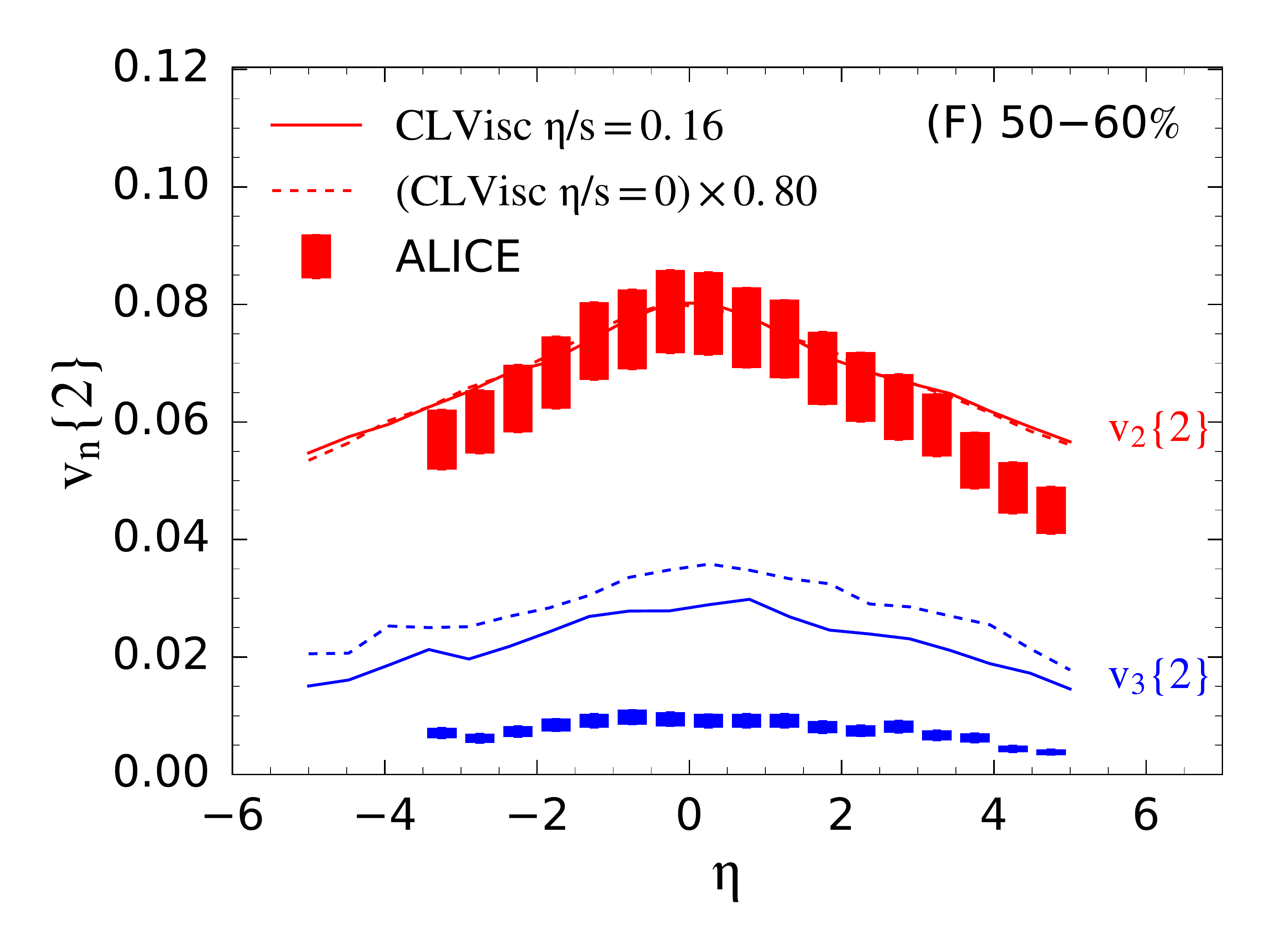}
    \protect\protect\caption{The pseudo-rapidity dependence of elliptic flow and triangular flow, for Pb+Pb $\sqrt{s_{NN}}=2.76$ TeV collisions with centrality range 0-5, 10-20, 20-30, 30-40, 40-50 and 50-60, from (3+1)D viscous hydrodynamic simulations starting from AMPT initial conditions without initial fluid velocity and evolve with $\eta_v/s=0.16$ as compared with LHC measurements from ALICE collaboration \cite{Adam:2016ows}.
    \label{fig:pbpb2760_vn_vs_eta}}
\end{figure*}

As shown in Fig.~\ref{fig:pbpb2760_vn_vs_eta},  $v_2\{2\}$ and $v_3\{2\}$ from CLVisc with $\eta_v/s=0.16$ in Pb+Pb collisions at $\sqrt{s_{NN}}=2.76$ TeV 
agree well with experimental data from the ALICE collaboration \cite{Adam:2016ows} for most of the centralities. The ratios between $v_2\{2\}$ and $v_3\{2\}$ are correctly reproduced for most central and semi-central collisions. The mean value of the ratio  $v_2\{2\}/v_3\{2\}$ increases as the system goes from most central to peripheral collisions. 
In most central collisions, both $v_2\{2\}$ and $v_3\{2\}$ from CLVisc+AMPT simulations are larger than experimental data.
For very peripheral collisions (e.g. $50-60\%$ centrality), the hydrodynamic simulations still produce reasonable $v_2\{2\}$ as a function of pseudo-rapidity while the $v_3\{2\}(\eta)$ is two times larger than the experimental data. For all centralities, the $v_n\{2\}(\eta)$ decreases faster at large rapidities in the experimental data than that given by the relativistic hydrodynamics with AMPT initial conditions. It was conjectured that temperature dependent $\eta_v/s$ may resolve this small overshoot of $v_n\{2\}$ at large rapidities \cite{Denicol:2015nhu}. In earlier works the rapidity dependence was reproduced by including the hadronic rescattering in 3+1 dimensional hydrodynamic calculations \cite{Hirano:2005xf,Nonaka:2006yn}. To investigate the sensitivity of the shape along rapidity, we show a calculation with $\eta_v/s=0$ that is scaled to match the $v_2(\eta)$ and see the same drop from middle to large rapidities. With the same scaling factor for $v_2\{2\}$ and $v_3\{2\}$ in ideal hydrodynamics, we see that the shape of $v_n\{2\}(\eta)$ from CLVisc is not sensitive to $\eta_v/s$ at all. The ratio $v_2\{2\}/v_3\{2\}$ is quite sensitive to $\eta_v/s$ since shear viscosity suppresses higher order harmonics stronger than lower order harmonics. As a result, the shape of the $v_n\{2\}(\eta)$ is only sensitive to the longitudinal distribution of initial entropy density but the ratios between different harmonic flows are good observables to constrain $\eta_v/s$. 

With constant $\eta_v/s$ and energy density fluctuations along the space-time rapidity in CLVisc,
the $v_n\{2\}(\eta)$ overshoots the experimental data at large rapidities.
It is not yet clear whether the temperature dependent $\eta_v/s(T)$ can fix the disagreement as suggested in \cite{Denicol:2015nhu} or if hadronic rescattering is necessary.
Furthermore, the net baryon density should become significant in the large  rapidity region, especially in low beam energy collisions at RHIC. 
One in principle has to take into account baryon chemical potential dependence of the EoS in the forward rapidity region \cite{Kapusta:2017ytu} in
order to describe the pseudo-rapidity dependence of $v_n\{2\}$.

\section{Longitudinal decorrelation of anisotropic flow}
\label{sec:rneta}

The decorrelation of anisotropic flow along the longitudinal direction has been computed in CLVisc with AMPT initial conditions and $\eta_v/s=0$ for the hydrodynamic evolution \cite{Pang:2015zrq}. In the current work, we focus on the effect of the shear viscosity and the initial fluid velocity on the longitudinal decorrelation observables.

The longitudinal decorrelation observable $r_n(\eta^a, \eta^b)$, which does not only capture the twist of event planes but also the anisotropic flow fluctuations along the longitudinal direction,  is defined as \cite{Khachatryan:2015oea},
\begin{equation}
    r_n(\eta^a, \eta^b) = \frac{\langle \vec{Q}_n(-\eta^a) \vec{Q}_n^*(\eta^b) \rangle}{\langle \vec{Q}_n(\eta^a) \vec{Q}_n^*(\eta^b) \rangle}
    \label{eq:rn_a_b}
\end{equation}
where $\eta^a$ and $-\eta^a$ are $16$ pseudo-rapidity windows each with size $\Delta \eta=0.3$ uniformly distributed in the range $[-2.4, 2.4]$ and 
$\eta^b$ are reference pseudo-rapidity windows to remove the effect of short range non-flow correlations, with the first reference window $\eta^b\in (3, 4)$  denoted as ``ref1'' and the second $\eta^b \in (4.4, 5.0)$ denoted as ``ref2''. The anisotropic flows and their orientation angles in a given pseudo-rapidity window are quantified by $\vec{Q}_n$,
\begin{equation}
    \vec{Q}_n \equiv Q_n e^{i n \Phi_n} = \frac{1}{N}\sum_{j=1}^{N} e^{i n \phi_j}
    = \frac{\int e^{i n \phi_j} \frac{dN}{d\eta dp_T d\phi}dp_T d\phi}{\int \frac{dN}{d\eta dp_T d\phi}dp_T d\phi},
    \label{eq:Qn_vec}
\end{equation}
where $\phi_j = \arctan{p_{yj}/p_{xj}}$ is the azimuthal angle of the $j$th particle in momentum space.
The smooth particle spectra are integrated over the azimuthal angle $\phi \in [0, 2\pi)$ and the corresponding transverse momentum $p_T$ ranges.
Following the CMS experimental setup \cite{Khachatryan:2015oea}, the $p_T$ range is $[0.3, 3.0]$ GeV/c for particles in $\eta^a$ and is $[0.0, \infty)$ for particles in $\eta^b$.
Since the Pb+Pb collisions are symmetric along the beam direction, by definition $r_n(\eta^a, \eta^b)$ should equal $r_n(-\eta^a, -\eta^b)$.
Following the suggestion through private communication with CMS collaboration, we use $\sqrt{r_n(\eta^a, \eta^b) r_n(-\eta^a, -\eta^b)}$ to improve statistics.
Let us note here once again, that the highly efficient GPU parallelized algorithm is crucial to obtain reliable results for correlation observables within reasonable computing time. 

\begin{figure*}[!htp]
    \includegraphics[width=0.49\textwidth]{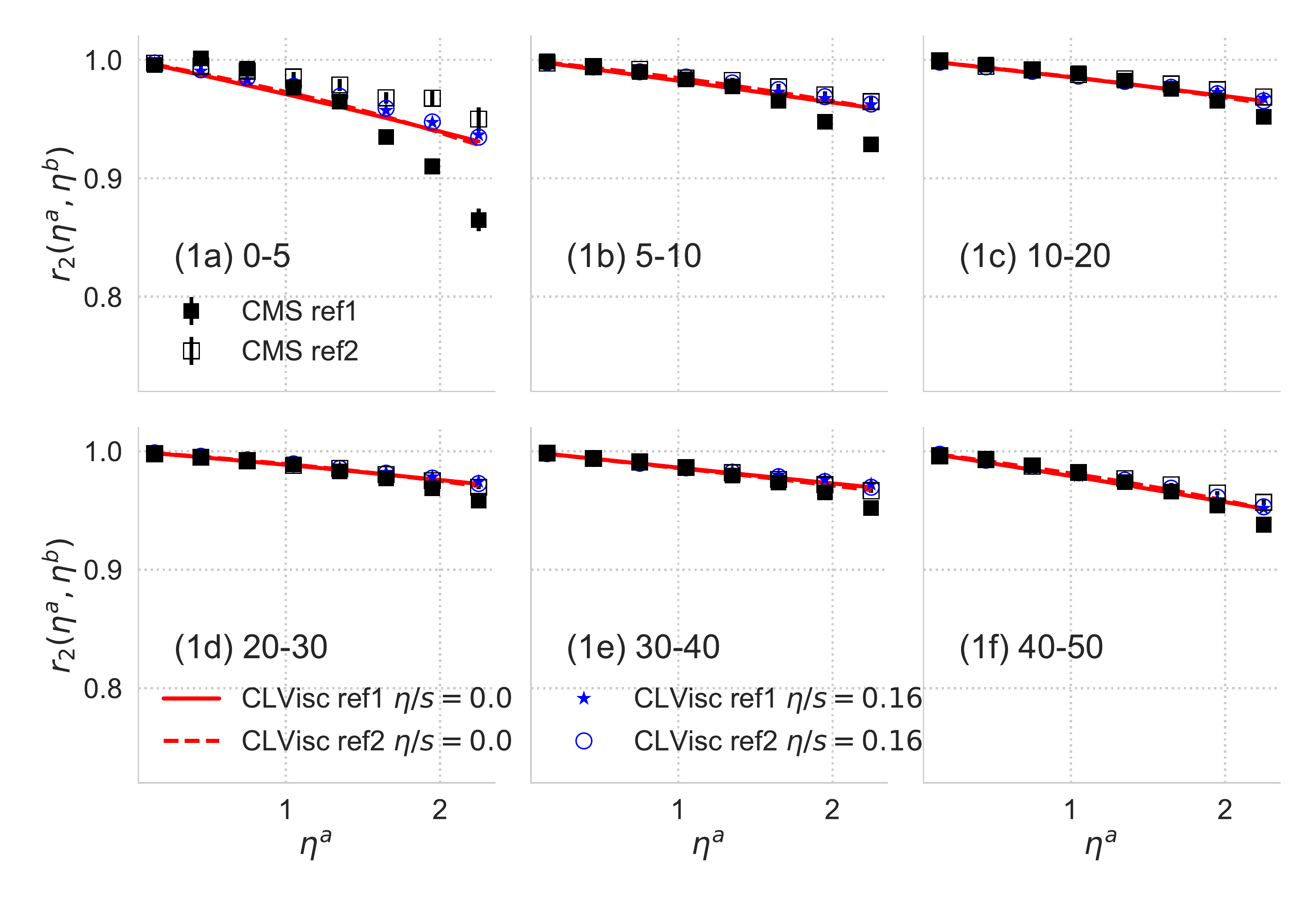} \includegraphics[width=0.49\textwidth]{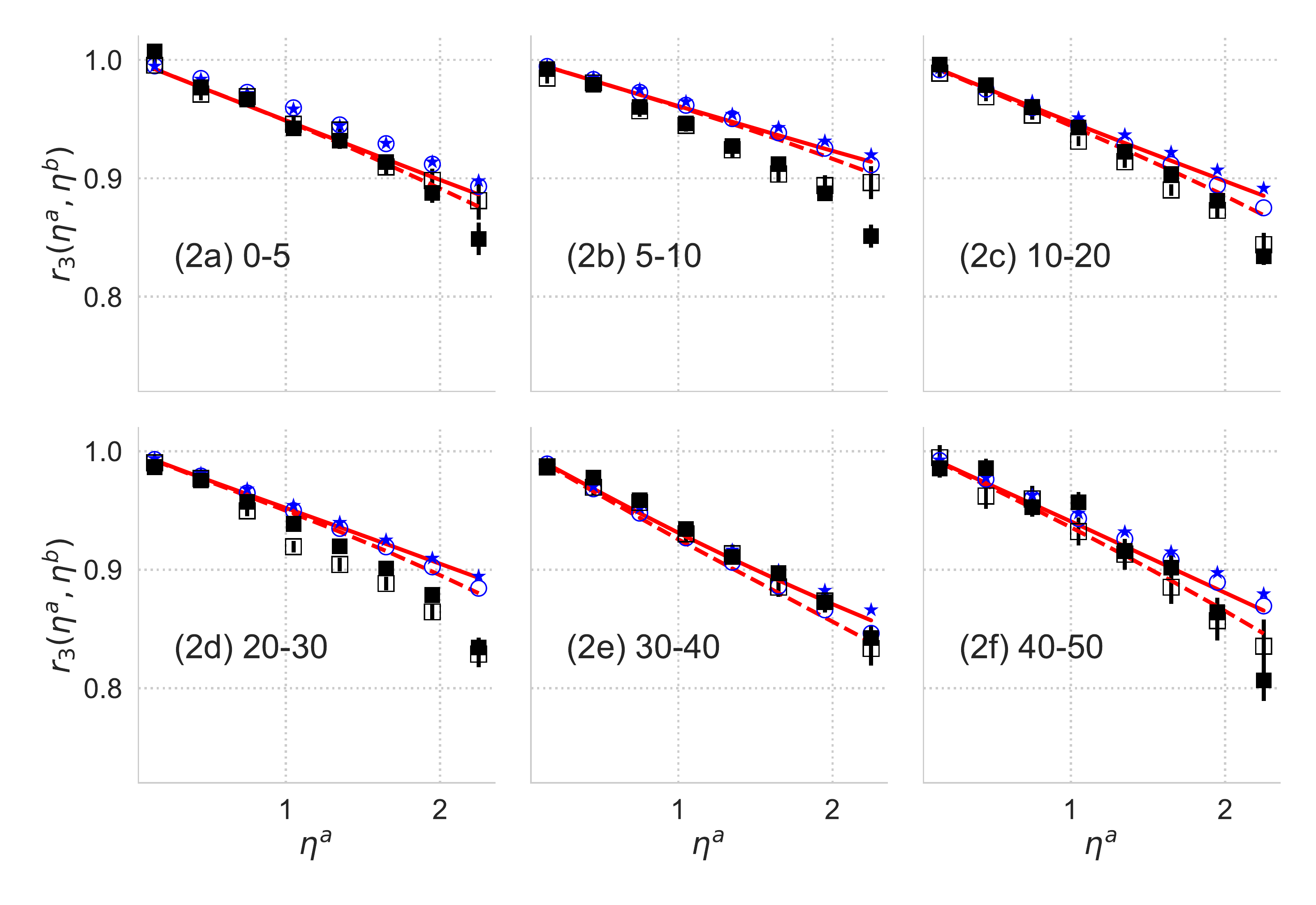}
    \protect\protect\caption{(color online) The decorrelation of elliptic flow (1a)-(1f) and triangular flow (2a)-(2f) along the pseudo-rapidity direction, for Pb+Pb $\sqrt{s_{NN}}=2.76$ TeV collisions with centrality range 0-5, 5-10, 10-20, 20-30, 30-40 and 40-50, from (3+1)D viscous hydrodynamic simulations starting from AMPT initial conditions without the initial fluid velocity ($\eta_v/s=0$ for red lines and $\eta_v/s=0.16$ for blue circles and stars) as compared with LHC measurements at CMS (black squares). The ``ref1'' denotes $3.0<\eta^b<4.0$ while ``ref2'' denotes $4.4<\eta^b<5.0$.
    \label{fig:pbpb2760_rn_noflow}}
\end{figure*}

\begin{figure*}[!htp]
    \includegraphics[width=0.49\textwidth]{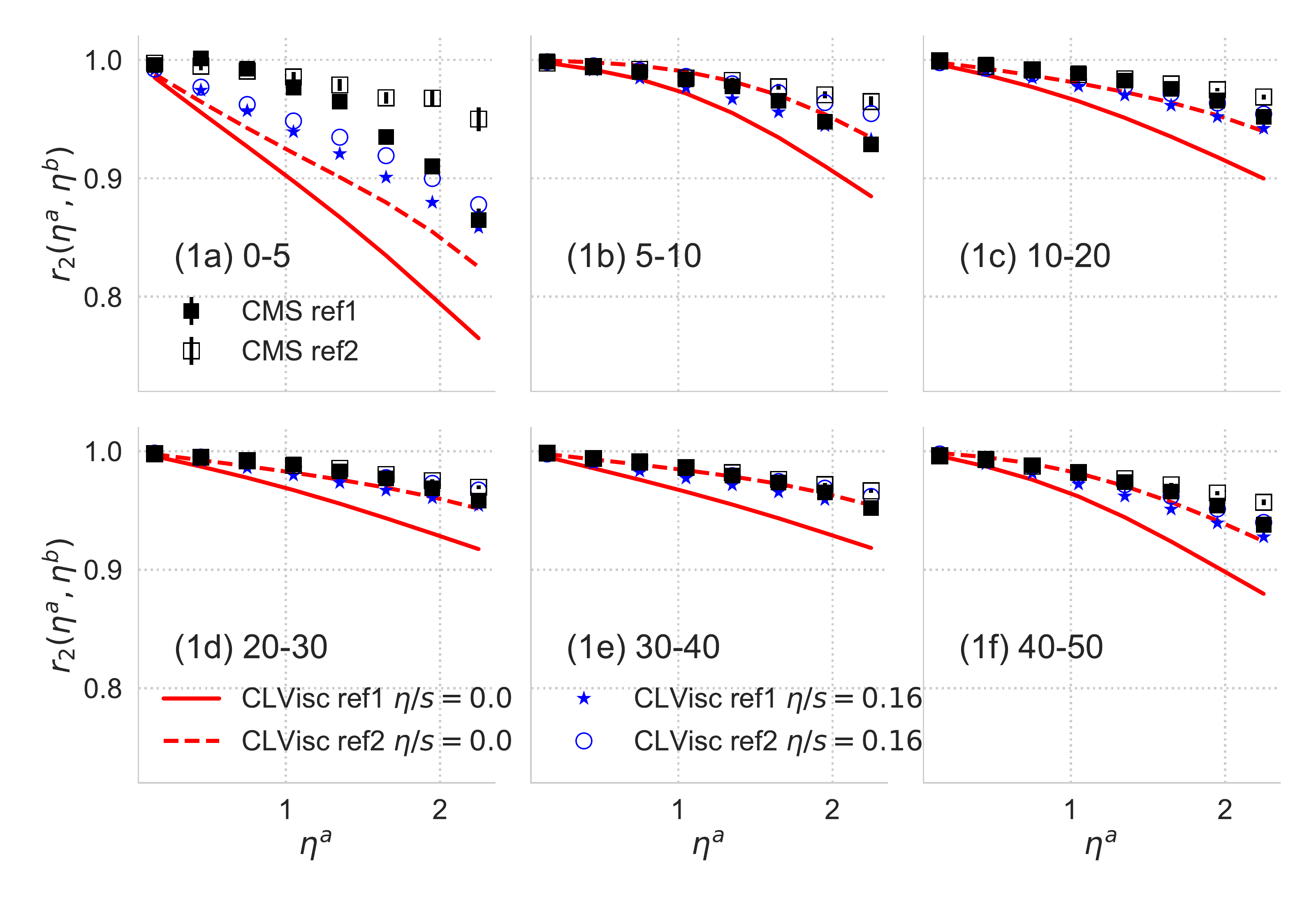} \includegraphics[width=0.49\textwidth]{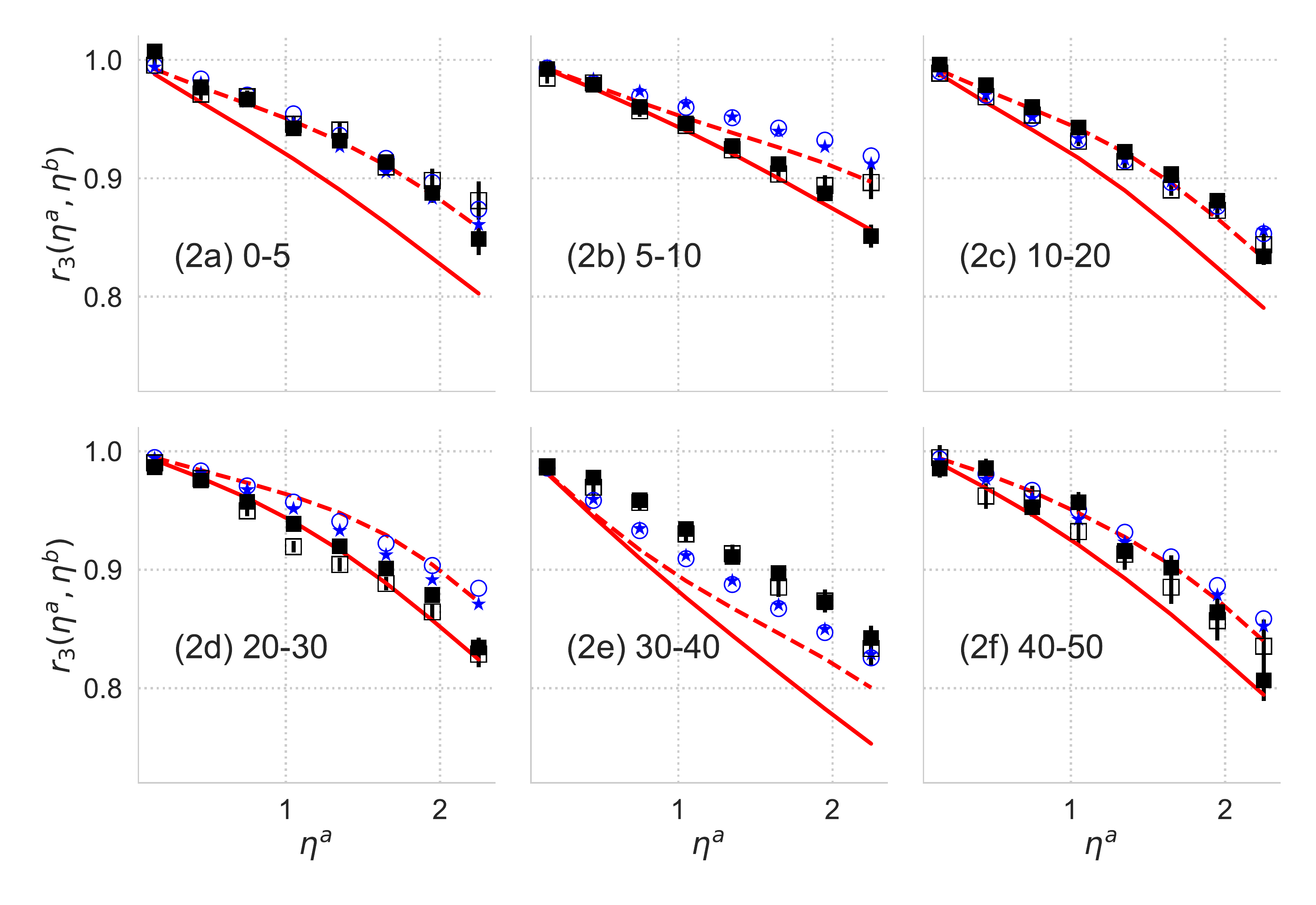}
    \protect\protect\caption{(color online) The decorrelation of elliptic flow (1a)-(1f) and triangular flow (2a)-(2f) along the pseudo-rapidity direction, for Pb+Pb $\sqrt{s_{NN}}=2.76$ TeV collisions with centrality range 0-5, 5-10, 10-20, 20-30, 30-40 and 40-50, from (3+1)D viscous hydrodynamic simulations starting from AMPT initial conditions with the initial fluid velocity ($\eta_v/s=0$ for red lines and $\eta_v/s=0.16$ for blue circles and stars) as compared with LHC measurements at CMS (black squares). The ``ref1'' denotes $3.0<\eta^b<4.0$ while ``ref2'' denotes $4.4<\eta^b<5.0$.
    \label{fig:pbpb2760_rn_withflow}}
\end{figure*}

We study the effect of the shear viscosity and the initial fluid velocity on $r_n(\eta^a, \eta^b)$ by comparing the results from CLVisc with $\eta_v/s=0.0$ and $\eta_v/s=0.16$,
starting from AMPT initial conditions with the initial state fluid velocity switched on and off.
Notice that in the comparison, parameters for ideal hydrodynamics are kept unchanged as given in the previous paper except that the freeze-out temperature is changed from $137$ MeV to $100$ MeV. In the viscous hydrodynamics simulation, the initial scaling factor is changed to $K=1.2$ to take into account the extra entropy production due to finite
shear viscosity in order to fit the charged multiplicity for $0-5\%$ central Pb+Pb collisions at $\sqrt{s_{NN}}=2.76$ TeV. 

Shown in Fig.~\ref{fig:pbpb2760_rn_noflow} are the decorrelation functions of elliptic flow (1a-1f) and triangular flow (2a-2f) from CLVisc with AMPT initial conditions and initial fluid velocity switched off as compared with CMS experimental data \cite{Khachatryan:2015oea} at the LHC. Both the decorrelations of elliptic flow and triangular flow agree with experimental data to a reasonable level. Two different values of $\eta_v/s$ used in CLVisc produce very similar longitudinal decorrelations. This indicates that the decorrelation observable is not sensitive to the value of $\eta_v/s$ used for the hydrodynamic evolution if there is no initial flow. For $r_{2}(\eta^a, \eta^b)$, the hydrodynamic results do not show difference for two different  $\eta^b$ reference windows. For $r_{3}(\eta^a, \eta^b)$, there is a very small splitting between two different $\eta^b$  reference windows. It is suggested that the non-flow short-range correlations in the denominator between particles in the window $[\eta^a -0.15, \eta^a+0.15]$ and the first reference window $3 < \eta^b < 4$ depress the value of $r_n(\eta^a, \eta^b)$. This is consistent with the negligible splitting from CLVisc with the zero-flow initial condition, since no near-side short-range correlations from jets are considered in the simulations.

The agreement between $r_{2}(\eta^a, \eta^b)$ and experimental data for all centralities are as good as 
our previously published results using ideal hydrodynamics with $T_{\mathrm{f}}=137$ MeV \cite{Pang:2015zrq}.
Moreover, the $r_{3}(\eta^a, \eta^b)$ with $T_{\mathrm{f}}=100$ MeV increases slightly as compared with $T_{\mathrm{f}}=137$ MeV.

With a finite ratio of shear viscosity over entropy density $\eta_v/s=0.16$, $r_2$ from CLVisc simulations fits the CMS data better, if the second reference window $\eta^b \in [4.4, 5.0)$ is chosen. For $r_n(\eta^a, \eta^b)$ computed with the first reference $\eta^b$ window, the shear viscosity decreases the decorrelation of elliptic flow slightly for zero-flow initial condition but strongly when initial fluid velocity is included in the initial condition. For $r_n(\eta^a, \eta^b)$ computed with the second reference $\eta^b$ window, the effect of the shear viscosity is very small. When there are longitudinal fluctuations, the non-Bjorken longitudinal expansion due to pressure gradients along the space-time rapidity is strong.
In ideal hydrodynamics, this longitudinal expansion decreases elliptic flow \cite{Pang:2012he}. However, in viscous hydrodynamics, the shear viscosity speed up the expansion along the transverse direction and slow down the expansion along the longitudinal (space-time rapidity) direction.
The anisotropic flow in viscous hydrodynamics with both transverse and longitudinal fluctuations are therefore affected by the entanglement between the accelerated transverse expansion and the decelerated longitudinal expansion.

When the initial fluid velocity computed from $T^{\tau \mu}$ is included in the initial condition, the short range ``non-flow'' correlations from mini-jets become stronger in ideal hydrodynamics. The short range correlations in the denominator between particles in the window $[\eta^a -0.15, \eta^a+0.15]$ and the first reference window $3 < \eta^b < 4$ suppress the value of $r_n(\eta^a, \eta^b)$. This is clearly seen in Fig.~\ref{fig:pbpb2760_rn_withflow} as the red-dashed line for $r_n(\eta^a, \eta^b=ref2)$ is always above the red-solid line for $r_n(\eta^a, \eta^b=ref1)$ from ideal hydrodynamic simulations.
For viscous hydrodynamics with initial fluid velocity, the splitting between two $\eta^b$ reference windows is much smaller than ideal hydrodynamics.
The comparisons between Fig.~\ref{fig:pbpb2760_rn_noflow} and Fig.~\ref{fig:pbpb2760_rn_withflow} shows that the decorrelation strength together with the splitting between two reference windows are sensitive to both the initial fluid velocity and shear viscosity. With shear viscosity constrained by other physical observables, the splitting between two reference windows for $0-5\%$ and $5-10\%$ central collisions might be a good observable to determine the initial fluid velocity.

\section{Summary}

We have developed a full (3+1)D viscous relativistic hydrodynamic model CLVisc in which both the hydrodynamic evolution with KT algorithm and Cooper-Frye particlization with integration on the freeze-out surface are parallelized on GPU using OpenCL. We achieved 60 and 120 times performance increase for the space-time evolution and Cooper-Frye particlization, respectively, relative to the performance of the code on a single core CPU. Such  increased performance makes many event-by-event studies of high-energy-heavy-ion collisions, such as the Coupled Linear Boltzmann Transport and hydrodynamics (CoLBT-hydro) model  \cite{Chen:2017zte} for jet propagation and medium response, possible. We have validated the CLVisc code with comparisons with several analytic solutions of ideal and viscous hydrodynamic equations such as Riemann, Bjorken and Gubser solutions as well as numerical solutions from VISH2+1. We have also compared results from CLVisc using the Trento Monte Carlo initial conditions with  experimental data 
on hadron spectra in heavy-ion collisions at both RHIC and LHC. We carried out a novel study with CLVisc on the pseudo-rapidity dependence and decorrelation of anisotropic flows in the longitudinal direction with initial conditions given by the AMPT model. We confirmed the observation that the magnitude and the relative ratio of anisotropic flows are sensitive to the shear viscosity to entropy density ratio $\eta_v/s$. We also found that the decorrelation of anisotropic flow along the pseudo-rapidity and the splitting between different reference rapidity window are sensitive both to the initial flow velocity and the shear viscosity to entropy density ratio.

In the comparisons to the experimental data on flavor dependence of the hadron spectra and anisotropic flows, CLVisc fails to describe the experimental data like all other pure hydrodynamic models. As illustrated by previous studies \cite{Song:2010aq,Ryu:2012at}, it is imperative to include non-equilibrium dynamics of hadronic scattering after the hadronization. CLVisc with the option of Monte Carlo sampling for Cooper-Frye particlization is well suited to work together with a hadronic transport model to account for this dynamic process. This will be investigated in the near future.

\begin{acknowledgments}
We thank Derek Teaney for helpful discussions on how to estimate the derivatives before each time step.
This work was supported in part by the National Science Foundation of China under grant No. 11521064 (L.-G.P. and X.-N.W),
National Science Foundation (NSF) within the framework of
the JETSCAPE collaboration, under grant number ACI-1550228 (L.-G.P. and X.-N.W.), the Director, Office of Energy
Research, Office of High Energy and Nuclear Physics, Division of Nuclear Physics, of the U.S. Department of
Energy under Contract Nos. DE-AC02-05CH11231 (X.N.W.),  funding of a Helmholtz Young Investigator Group VH-NG-822 from the
Helmholtz Association and GSI and the Helmholtz International Center for the Facility for Antiproton
and Ion Research (HIC for FAIR) within the framework of the Landes-Offensive
zur Entwicklung Wissenschaftlich-Oekonomischer Exzellenz (LOEWE) program launched by the State of Hesse (L.-G.P and H.P.). 
Computational resources have been provided by the GSI green cube and the GPU workstations at Central China Normal University.
\end{acknowledgments}

\section*{APPENDIX}
\label{sec:appendix}

\subsection{GPU architecture and the parallelization of the KT algorithm}
\begin{figure}
    \includegraphics[width=0.48\textwidth]{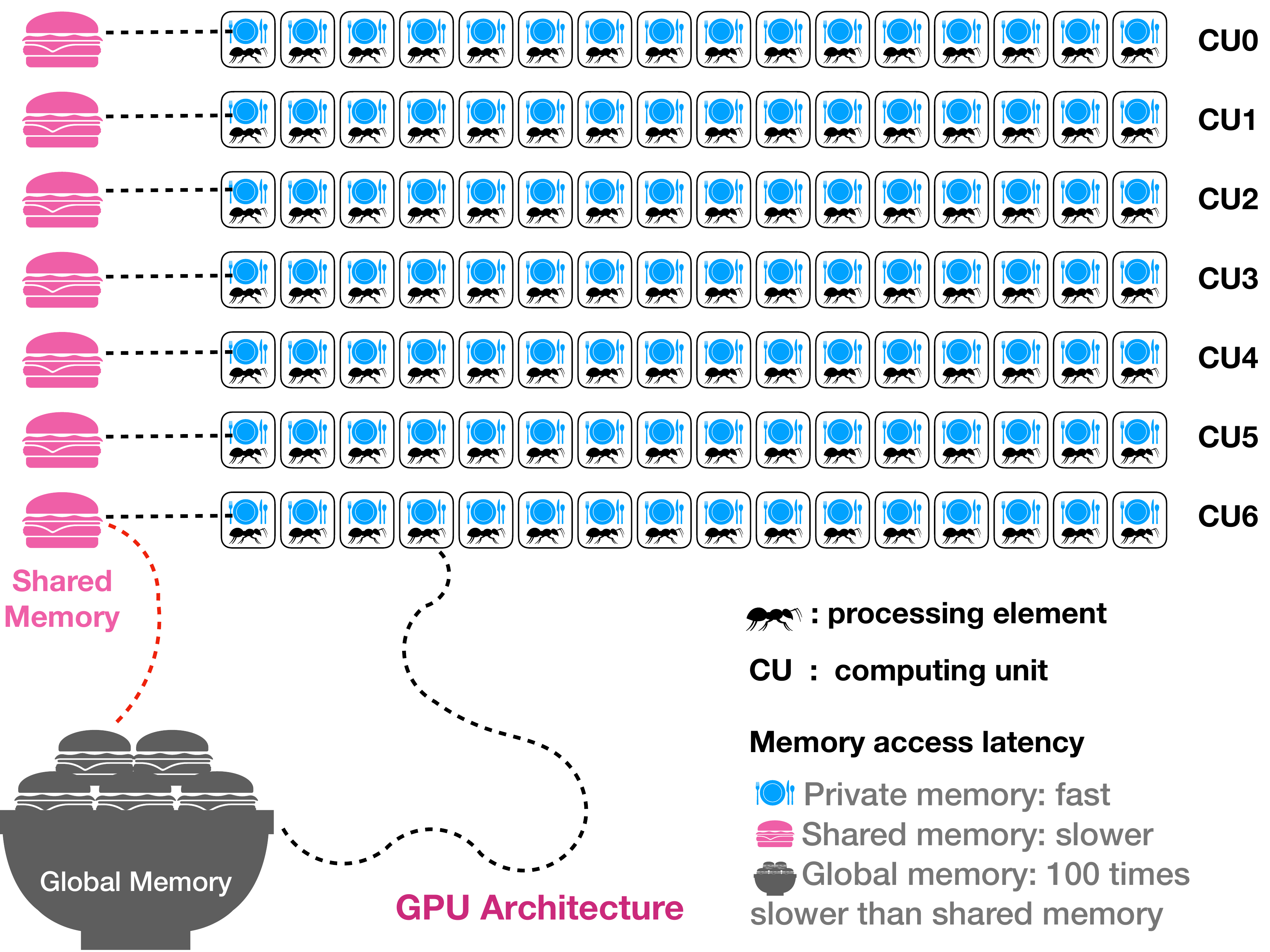}
    \caption{(color online) Cartoon diagram of the architecture of GPUs. \label{fig:gpu_structure}}
\end{figure}

Parallelization and optimization of relativistic hydrodynamic program on GPUs require expertise.
In this section we provide many technical details that are critical to GPU parallelization.
Shown in Fig.~\ref{fig:gpu_structure} is one cartoon diagram of the GPU architecture.
The smallest component of the GPU is the processing element (PE) which is comprised 
of a worker (the ant) that owns a very small piece of  private memory (the dish).
The accessing latency for the processing element to read data from the private memory is very low.
However, usually the private memory is so small that it is impossible
to store a big amount of data in private memory for processing at the same time. 
If more private memory is used than provided, the processing element will store data in global memory and read from there in each access.
This is not good practice, since there is a long distance between the global memory  (food source in the out environment) and
the private memory (the dish of the ant).
As a result, reading data directly from global memory to private memory has a large latency.
The clever ants decided to construct one granary (named as shared memory in CUDA and local memory in OpenCL) to store food that is fetched from out environment and will be shared by multiple ants.
The memory access from shared memory (the granary) to private memory (the dish) is more than 100 times faster than directly reading data
from global memory (out environment).
Pre-fetching data from global memory to shared memory for frequent accessing usually speeds up the program by a large margin.
Although the private memory and the shared memory have lower accessing latency than global memory, their capacities and horizons are much smaller.
The private memory (capacity = dozens of float numbers) can only be accessed by each processing element,
while the shared memory (capacity = 32KB -- 64KB) can be accessed by all the processing elements in the same computing unit.
As a comparison, the global memory (capacity = several GB) is large and can be accessed by all the processing elements.
If some data is shared by all the processing elements, a special region of the global memory -- ``constant memory'' can be used to balance the horizon and accessing latency.
Notice that all memories are located on the GPUs and transferring data from CPU memory to the global memory of GPUs also take time.
The good practice is to transfer data from CPU memory to the GPU global memory and performing all calculations before transferring back to CPUs for output.

In the 3D KT algorithm, the required data to update the source terms
$S_{\pi}$, $S_{N}$, $S_{T}$ and $S_{\Pi}$ at lattice $(i,j,k)$ are
$4$ components in $(\varepsilon,v_{x},v_{y},v_{\eta_s})$,
$10$ components in $\pi^{\mu\nu}$,
$2$ components in $N$ and $\Pi$,
on $13$ lattice grids.
As a result, at least $16 \times13=208$ float numbers are necessary to update one hydrodynamic cell.
Without using shared memory, there is too much redundant fetching
from global memory to private memory, which slows down the calculation.
In the beginning, a 3D stencil was used to fetch a 3D block of data to
shared memory, all the threads in the same work group read data
from shared memory. However, numerous halo cells are needed in each
direction in order to update the boundary cells in the local block.
In order to update one $7\times7\times7$ block,
one needs $7\times7\times4\times3$ halo cells.
The total shared memory used for the effective block and halo cells
in this simple case is $16\times7\times7\times(7+12)\times4/1024=56$ KB,
which already exceeds the maximum shared memory provided by the most
advanced GPUs on the market (typical size of shared memory is $32$ KB).
A trade off is to read halo cells directly from global memory instead
of storing them in shared memory, which reduces the shared memory
usage to $20$ KB.
On the other hand, concurrent reading from global memory is only
possible along one dimension,
depending on in which direction the data is stored continuously.
The data in one 3D array can only be stored continuously
in one direction, which makes concurrent reading impossible in
the other 2 directions.
For the 3D stencil, it is possible to store each block of data
$(7,7,7)$ continuously in global memory,
other than the common $(x,y,z)$ order for the whole $(\mathrm{nx,ny,nz})$ array.
It is also possible to construct the halo cells for each block
and store them continuously in global memory for concurrent
accessing.
One should keep in mind that constructing halo cells for the 3D block
is error-prone and asks for much more global memory.

\begin{figure}[!htp]
    \includegraphics[width=0.5\textwidth, trim={0cm 9cm 0cm 8cm}, clip]{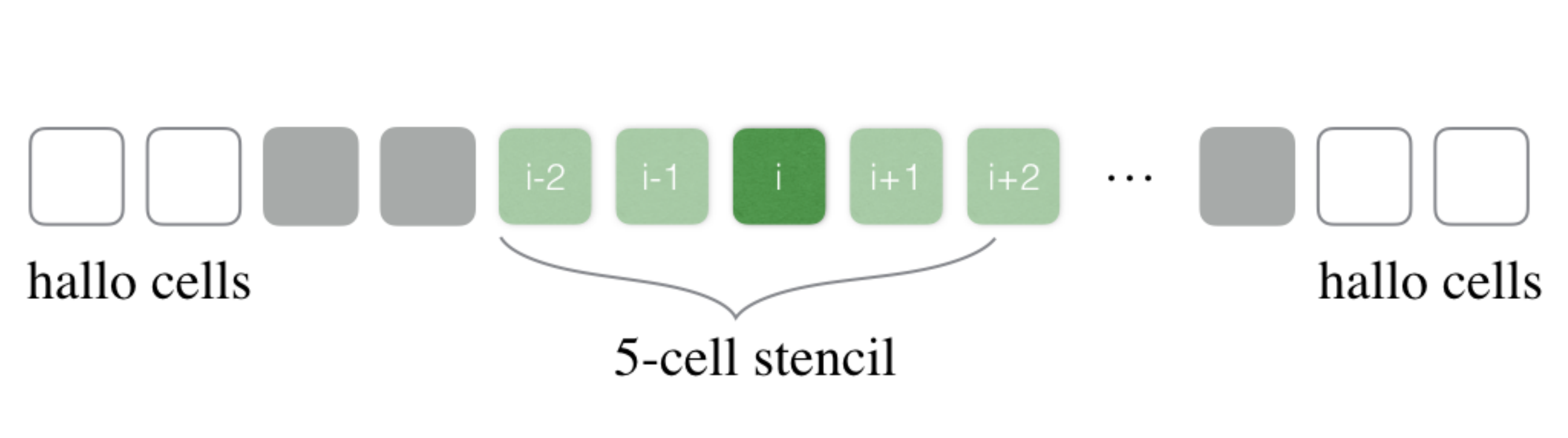}
    \protect\protect\caption{One strip of data stored in the shared memory for 5-cell stencil in KT algorithm.
       \label{fig:oned_stencil}}
\end{figure}

In the current version of CLVisc, the source terms are split into $3$ directions. The 1D data along each direction is put in the shared memory as shown in Fig.~\ref{fig:oned_stencil}.
The total shared memory used for one strip is $N\times16\times4/1024=32$ KB for $N=512$ lattices along the $x$ direction.
Each hydrodynamic cell shares $5\times 16$ single precision floating numbers along the $x$ direction and only $4$ halo cells at the boundary are needed.

\subsection{Parallelization of the smooth particle spectra calculation}
\begin{figure}[!htp]
    \includegraphics[width=0.5\textwidth, trim={3cm 3cm 3cm 3cm}, clip]{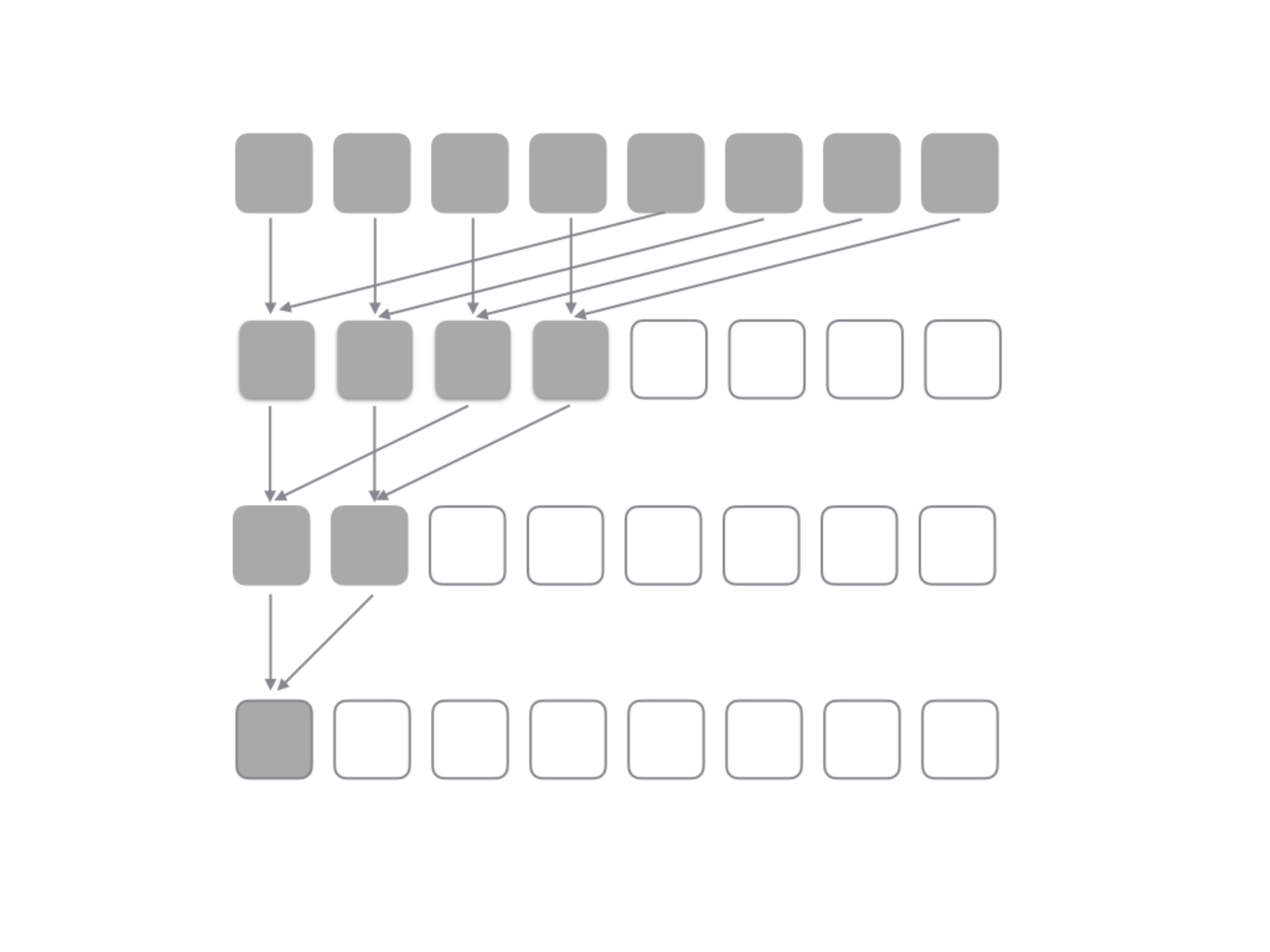}
    \protect\protect\caption{Parallel reduction used on GPU to compute the summation of particle spectra from millions of freeze-out hyper-surface elements.
    \label{fig:parallelreduction}}
\end{figure}

Since the integration kernel in Eq. (\ref{eq:cooper_frye}) is independently calculated for different
freeze-out hyper-surface elements before the summation, it is a perfect job to fit in GPU parallel computing.
If the Cooper-Frye integration is only needed once for all the hyper-surface, it can be
done efficiently using the two step parallel reduction algorithm as shown in Fig.~\ref{fig:parallelreduction} from Nvidia and AMD SDK.
In reality we need to do hyper-surface integration $308\times 41 \times 15 \times 48 $ times,
it is quite slow to load each hyper-surface element from global memory to private memory so many times.
In order to reduce the global memory access, we share the hyper-surface elements in one work group
for multiple $(pid, Y, p_T, \phi)$ combinations.
The computing time for $300$ resonances is reduced from 8 hours on a single core CPU to 3 minutes
on the modern GPUs like Nvidia K20 and AMD firepro S9150 for one typical hydrodynamic event.

Shown in Fig.~\ref{fig:parallelreduction} is one demonstration of parallel reduction. 
E.g., in order to sum all the numbers in one big array, one first put the numbers in many groups,
in each working group the working items iteratively add the second half of the sub-array to the first half in parallel.
After several iterations, the final result will be the value in the first working item.
Notice that the parallel reduction has not only been used in CLVisc to compute the summation
of particle spectra from the huge amount of freeze-out hyper-surface cells, but has also been used to compute the maximum energy density $\varepsilon_{\mathrm{max}}$ in the 
fluid field at each output time step. 
The $\varepsilon_{\mathrm{max}}$ is used to stop the time evolution of hydrodynamics when its value is smaller than the freeze-out energy density determined by 
the freeze-out temperature.
In order to find $\varepsilon_{\mathrm{max}}$ in the fluid field, one has to check $N_x \times N_y \times N_{\eta_s}$ fluid cells in the collision system with both 
transverse and longitudinal fluctuations.
This can be done easily in python, if the energy density values of the whole fluid field stay in the host memory (CPU memory).
However, transferring the values of a big 3D matrix from GPU to CPU at each output time step is very time consuming.
CLVisc uses parallel reduction to compute the maximum energy density of the fluid field on the GPU side and transfer a scalar $\varepsilon_{\mathrm{max}}$ back to the CPU side.
In order to avoid the data transfer between CPU and GPU memory, the freeze-out hyper-surface finding algorithm \cite{Pang:2012he} is also implemented on GPU.

\subsection{Profiling for the (3+1)D viscous fluid dynamic evolution}
\label{sec:profile}

In order to solve 3D partial differential equation, we need to update the values of  $N_{\mathrm{cells}} = NX\times NY \times NZ$ cells at each time step. Without parallel computing, there is only one computing element that updates these cells one after another. The modern GPUs have more than $N_{\mathrm{workers}} = 2500$ processing elements such that more than $2500$ cells can be updated simultaneously. In practice, the performance boost can not approach $2500$ for several reasons, (1) the computing power of each computing element on GPU is not as strong as CPU (2) reading data from global memory of GPU to the private memory of one computing element has big latency. The easiest optimization on GPU is to put the data shared by a block of processing elements on shared memory to reduce the global accessing latency. In the 5-stencil central scheme KT algorithm, the site information on each cell is shared 5, 9 and 13 times by its neighbors in 1-D, 2-D and 3-D respectively.

\begin{table}[htp]
\centering
\begin{ruledtabular}
\begin{tabular}{llllll}
block size    &  8       & 16        & 32        & 64        & 128      \\
Ideal(s)-GPU  &  0.37  & 0.218   & 0.178   & 0.155   & 0.157   \\
Visc(s)-GPU   &  3.12  & 1.65     & 1.17     & 1.01     & 1.17     \\
Visc(s)-CPU   &  6.64   & 6.45    & 6.63     & 7.0       & 7.58     \\
\end{tabular}
\end{ruledtabular}
\caption{Computing time for one time step on various computing devices for several different block sizes.}
\label{tab:block_size}
\end{table}

The optimal block size -- denotes the number of processing elements assigned to process one workgroup of cells, vary between different computing devices.
As shown in Table. \ref{tab:block_size}, we run (3+1)D viscous hydrodynamics with number of cells $N_{\mathrm{cell}}=385 \times 385 \times 115$ for 1600 time steps. Shown in the table are the mean time for one-step update on GPU AMD S9150 (2496 processing elements) and server CPU Intel Xeon 2650v2 (10 cores, 20 threads). The computing time for one-step update changes for different block sizes. For GPU AMD S9150, the optimal block size for this task is $64$ while for the CPU Intel Xeon 2650v2, the optimal block size is $16$. Running on GPU is about 6 times faster than running on a $10$ cores CPU with the same program. The (3+1)D ideal hydrodynamics with the same parallelization is about 6.5 times faster than the viscous version.

The performance can be further improved using deeper optimizations.
In the 1D-KT algorithm together with dimension splitting, each lattice point needs to be loaded $3$ times.
This is a trade off between implementation difficulty and efficiency.
However, it is already much better than independent fetching from global memory where the data on each lattice point are reloaded $13$ times.

\emph{Concurrent reading from global memory}

It is shown that the 1D KT algorithm is much faster along $\eta_s$ direction
than along $x$ and $y$ direction for $N_x=N_y=N_{\eta_s}=256$ grids.
The ratio of computing time along these three axis is $t_x : t_y : t_{\eta_s} = 38 : 28 : 1$.
There is the concurrent reading problem, since the data is only stored continuously in one direction.
Transposing the matrix in each time step is suggested by \cite{OpenCL_programing_guide} to increase the concurrent reading.
Another way is to use the native 3D image buffer, which provides a different storing order and constant extrapolation for boundary
cells.
We did not choose image buffer because it is read only or write only in one kernel in OpenCL version earlier than 2.0, and it does not support double precision.

\emph{Warp divergence}
Threads in the same workgroup are executed in warps of $32$ or $64$, with all the threads in one warp execute the same instruction at the same time.
If there is \emph{if/else} branching for two threads in the same warp, all the threads in the same warp will execute the instruction under both
of the two branches. This is called warp divergence. The root finding algorithm on each lattice cell needs different number of iterations to achieve the required precision, which will bring serious warp divergence. This should be kept in mind, but currently there is no way to tackle this problem.

\emph{Bank conflict}
On each computing unit there is one piece of shared memory whose size is around $32KB-48KB$.
Each work group occupy one piece of shared memory, the data in this piece of shared memory are stored in 32 banks with each bank holds many 32 bits data.
For example if we have one floats (32 bits) array A whose length is 500,
the first bank will store $A[0], A[32], \dots, A[32*n]$ and the second bank
will bank will store $A[1], A[33], \dots, A[32*n+1]$.
If multiple threads in the same warp read the same 32 bits data from one bank,
the data will be read only once and broadcast to all the requested threads,
there is no bank conflict in this case.
However, if $n$ threads in the same warp read $n$ different 32 bits data from 
the same bank, the operation is serialized and the program is slowed down,
this is called $n$-way bank conflict.
Bank conflict is also one way to slow down the program if the data is poorly structured.
For more details of GPU parallel computing, one can refer to \cite{OpenCL_programing_guide,OpenCL_in_action,OpenCL_standard}.

\subsection{Momentum sampling from Fermi-Dirac and Bose-Einstein distributions}
\label{subsec:draw_mom}

On the freeze-out hyper-surface, the baryons obey Fermi-Dirac distribution and mesons obey Bose-Einstein distribution. One needs to sample the momentum magnitude from these two distribution functions. The most straight forward method is native rejection sampling, which is not encouraged here due to too many rejections at large momentum when the probability is small. We introduce Scott Pratt's method and Adaptive Rejection Sampling (ARS) which are much faster to tackle this problem.

{\it Scott Pratt's method} There is a math trick to sample momenta from Juttner distribution function $f(p)=p^{2}\exp(-\sqrt{p^{2}+m^{2}}/T)$. The Fermion-Dirac distribution function can be approximated by Juttner distribution since $\exp(m/T) \gg 1$ even for the lightest baryon (e.g. proton with mass $m_p = 0.938$~GeV and freeze-out temperature $T\sim 0.2$~GeV gives out $\exp(m/T) \approx 90 \gg1$).

The Bose-Einstein distribution can be approximated using geometric sequence expansion with high precision, 
\begin{eqnarray*}
f(p) & = & \frac{p^2}{e^{E/T} -1 } = p^2 e^{-E/T} \frac{1}{1 - e^{-E/T}} \\
      & = & p^2 \left(e^{-E/T}  + e^{-2 E/T} + e^{- 3 E/T} + e^{-4 E/T} + ...\right),
\end{eqnarray*}
where $E=\sqrt{p^2 + m^2}$ is the energy of one particle in the co-moving frame of fluid. The problem is simplified to sampling from several Juttner distribution functions with effective freeze-out temperatures $T$, $T/2$, $T/3$, $T/4$ .... 

For massless particles whose distribution functions read $f(p)=p^{2}e^{-p/T}$, one uses the math trick: for probability distribution $x^{n-1}e^{-x}$,
one can draw $x$ by taking the natural log of $n$ random numbers $x=-\ln(r_{1}r_{2}...r_{n})$ with $r_{i}$ uniformly distributed between zero and one. It is easy to draw the momentum magnitude, polar and azimuthal angles in 3-dimensions, from Juttner distribution function,
\begin{eqnarray*}
p & = & -T\ln(r_{1}r_{2}r_{3}),\\
\cos\theta & = & \frac{\ln(r_{1})-\ln(r_{2})}{\ln(r_{1})+\ln(r_{2})},\\
\phi & = & \frac{2\pi\left[\ln(r_{1}r_{2})\right]^{2}}{\left[\ln(r_{1}r_{2}r_{3})\right]^{2}}.
\end{eqnarray*}

By checking the Jacobian, indeed,
\begin{eqnarray*}
dpd\cos\theta d\phi & = & |J|\ dr_{1}dr_{2}dr_{3}\\
 & = & \frac{8\pi T}{r_{1}r_{2}r_{3}\left[\ln(r_{1}r_{2}r_{3})\right]^{2}}dr_{1}dr_{2}dr_{3}\\
 & = & \frac{8\pi T}{e^{-p/T}p^{2}/T^{2}}dr_{1}dr_{2}dr_{3},
\end{eqnarray*}
and $dr_{1}dr_{2}dr_{3}=\frac{1}{8\pi T^{3}}p^{2}e^{-p/T}dpd\cos\theta d\phi$.

For massive hadrons, 
\[
p^{2}e^{-(E-\mu)/T}=p^{2}e^{-p/T}e^{(p-E+\mu)/T}.
\]
One first draws $p$ from $p^{2}e^{-p/T}$, then accept or reject with weight function $\omega(p)=e^{(p-E)/T}=e^{(p-\sqrt{p^{2}+m^{2}})/T}$.
For heavy hadrons $\omega(p)\ll1$, too many rejections slows down the sampling. Scott Pratt introduces a numerical trick,
\begin{eqnarray}
 p & = & \sqrt{E^{2}-m^{2}},\ dp=E/pdE\\
 dpp^{2}e^{-E/T} & = & dE\frac{E}{p}p^{2}e^{-E/T}\\
 & = & dEpEe^{-E/T}\\
 & = & dk\frac{p}{E}(k+m)^{2}e^{-k/T}e^{-m/T}\\
 & = & dk(k+m)^{2}e^{-k/T}\omega(p)\\
 & = & dk(k^{2}+2mk+m^{2})e^{-k/T}\omega(p)
\end{eqnarray}
where $k=E-m$ and $\omega(p)=\frac{p}{E}e^{-m/T}$ are weight functions that satisfy $E-m>0$ and $p/E<1$.
The $e^{-m/T}$ and $e^{-\mu/T}$ terms are not important and can be discarded.
The upper distribution is split into 3 parts and their discrete probabilities are determined by the k-integration,
\begin{eqnarray}
\int dkk^{2}e^{-k/T} & = & 2T^{3}\\
\int dk2mke^{-k/T} & = & 2mT^{2}\\
\int dkm^{2}e^{-k/T} & = & m^{2}T
\end{eqnarray}
Using this method, the sampled $k$ is accepted with very high probability $p/E$.

{\it Adaptive Rejection Sampling (ARS)} can not only be used to sample the Juttner, Fermion-Dirac and Boson-Einstein distribution, but also Woods-Saxon distribution and any distribution functions that are log-concave ($h''(x)<0$ for any $x$ where $h(x)= \log f(x)$).  ARS is very useful in nuclear physics and high energy physics. The philosophy of ARS is to generate a piecewise exponential upper bound $q(x)$ for $f(x)$ and refine this bound with rejected points. Notice that $q(x)\propto\exp(g(x))$ is constructed from $g(x)$ which is the piecewise linear upper bound of $\log f(x)$ -- whose existence requires the log-concave property. The ordered change points are $z_{0}<z_{1}<z_{2}...<z_{n}$ and $g(x)$ has slope $m_{i}$ in $(z_{i-1},z_{i})$. The area under each piece of exponential segment $\exp(g(x_{i}))$ is,

\[
A_{i}=\int_{z_{i}-1}^{z_{i}}e^{g(x)}dx=\frac{1}{m_{i}}\left(e^{g(z_{i})}-e^{g(z_{i-1})}\right)
\]
First sample $j$ from discrete\_distribution(\{$A_{i}$\}), then
sampling $x\in(z_{j-1,}z_{j})$ from distribution function $q(x)=\exp(a+m_{i}x)$.
By inversely sampling uniform distribution $r\in[0,1]$ from the cumulative
probability 

\[
Q(x)=\int_{z_{i-1}}^{x}q(y)dy=\frac{q(x)-q(z_{i-1})}{q(z_{i})-q(z_{i-1})}=r
\]
we get $x$ from the exponential distribution,

\[
x=\frac{1}{m_{i}}\ln\left(re^{m_{i}z_{i}}+(1-r)e^{m_{i}z_{i-1}}\right)
\]

With this $x$ we can do rejection test: $ran()<\frac{f(x)}{q(x)}=\exp(h(x)-g(x))$.
If a point is rejected, it will be used to refine the upper bound
which will make the upper bound closer to $f(x)$.
In squeezing test step, lower bound is also needed which we call $l(x)$.
Squeezing test is true if $ran()<\frac{l(x)}{q(x)}$. 
The ARS method can be extended to arbitrary distributions
by isolating the distribution function into concave and convex
parts with different upper bounds.

\subsection{Code structure}

This section describes the software aspect of the GPU parallelization and the code structure of CLVisc. 
Programming on GPUs usually uses two levels of language, one for the host side to read configurations, query devices, dispatch jobs to different computing devices and transferring data between host and devices,
the other is on the device side to do the real calculation using CUDA or OpenCL.
The CLVisc is comprised of several modules with two modules that provide examples for both Python--OpenCL and C++--OpenCL combinations.

\begin{itemize}
    \item[--] The relativistic hydrodynamic module which solves the partial differential equations and finds the freeze-out hyper-surface uses Python for the host side and OpenCL for the device side.
    \item[--] The smooth particle spectra calculation and resonance decay program use C++ for the host side and OpenCL for the device side.
    \item[--] Sampling hadrons from freeze-out hyper-surface and forcing resonance decay uses C++.
\end{itemize}

In CLVisc, the computing kernels are written in OpenCL and the host side for fluid dynamics is in Python.
Employing python as the host side language for the main CLVisc program has several benefits.
Comparing the host side language in C++ (used in smooth spectra calculation) and that is given in python by PyOpenCL, 
we found that the python version is much more compact and easier to program.
The built-in modules \verb!argparse!, \verb!logging!, \verb!unittest! together with \verb!PyOpenCL! make the host side programming in Python a
much better experience than using C++.
The kernels written in OpenCL can be directly used in a program whose host side language is C++ without any changes.
It is also much easier to connect to the later data analysis using \verb!numpy!, \verb!scipy!, \verb!pandas! and \verb!matplotlib!.
All the popular modern machine learning and deep learning libraries use Python as their user interface, 
which can also be easily connected to the CLVisc output.

\subsection{Code Availibility}

The CLVisc code is publicly available from \url{https://gitlab.com/snowhitiger/PyVisc}. In the package, there are example codes to run event-by-event hydrodynamics 
with optical Glauber, Trento initial condition or AMPT initial conditions.

\bibliographystyle{unsrt}
\bibliography{inspire,not_inspire}

\end{document}